%% file: ms.tex
\documentclass[journal]{IEEEtran}
\usepackage{graphicx}
\pdfoutput=1
\usepackage{tikz}
\usepackage{comment}
\usepackage{pgfplots}
\usepackage{amsmath,amssymb,mathtools, commath}
\usepackage{multicol}
\usepackage{booktabs}
\usepackage{tabularx}
\usepackage{multirow}
\usepackage{url}
\usepackage{balance}
\usepackage{ulem}
\usepackage[noadjust]{cite}
\usepackage[caption=false,font=normalsize,labelfont=sf,textfont=sf]{subfig}
\pgfplotsset{compat=newest}
\usetikzlibrary{positioning,spy,backgrounds,calc,arrows}
\DeclareMathOperator*{\argmin}{arg\,min}

\newcommand\crule[3][black]{\textcolor{#1}{\rule{#2}{#3}}}

\begin{document}
\bstctlcite{IEEEexample:BSTcontrol}

\title {J-MoDL: Joint Model-Based Deep Learning  for Optimized Sampling and Reconstruction  }

\author{Hemant~Kumar~Aggarwal,~\IEEEmembership{Member,~IEEE,}
        Mathews~Jacob,~\IEEEmembership{Senior Member,~IEEE,}%
\thanks{Hemant~Kumar~Aggarwal (email: hemantkumar-aggarwal@uiowa.edu) and Mathews Jacob (email: mathews-jacob@uiowa.edu) are with the Department
of Electrical and Computer Engineering, University of Iowa, IA, USA, 52242.
}% <-this % stops a space
\thanks{Manuscript received Month day, year; revised Month day, year.}
\thanks{This work is supported by 1R01EB019961-01A1. This work was conducted on an MRI instrument funded by 1S10OD025025-01 }
}

\markboth{Journal of Selected Topics in Signal Processing,~Vol.~14, No.~6, 2020}%
{Shell \MakeLowercase{\textit{et al.}}: Bare Demo of IEEEtran.cls for IEEE Journals}

\maketitle

\begin{abstract}
  Modern MRI schemes, which rely on compressed sensing or deep learning algorithms to recover MRI data from undersampled multichannel Fourier measurements, are widely used to reduce the scan time. The image quality of these approaches is heavily dependent on the sampling pattern. We introduce a continuous strategy to optimize the sampling pattern and the network parameters jointly. We use a multichannel forward model, consisting of a non-uniform Fourier transform with continuously defined sampling locations, to realize the data consistency block within a model-based deep learning image reconstruction scheme.  This approach facilitates the joint and continuous optimization of the sampling pattern and the CNN parameters to improve image quality. We observe that the joint optimization of the sampling patterns and the reconstruction module significantly improves the performance of most deep learning reconstruction algorithms.   The source code of the proposed joint learning framework is available at \url{https://github.com/hkaggarwal/J-MoDL}.
\end{abstract}


\section{Introduction}

\IEEEPARstart{M}R imaging offers several benefits, including good soft-tissue contrast, non-ionizing radiation, and the availability of multiple tissue contrasts. However, its main limitation is the slow image acquisition rate. The last decade has witnessed several approaches, including parallel MRI and compressed sensing, to recover the images from undersampled k-space measurements. Recently, deep learning methods are emerging as powerful algorithms for the reconstruction of undersampled k-space data; they offer significantly improved computational efficiency and higher image quality than classical methods. Several direct-inversion methods including  \cite{wangCTtmi2017,jong2019kspace,dagan,gan_cyclic,sigmanet,dar2018,dar2017transfer,zhu2018} use a convolutional neural network (CNN) to recover the images from the undersampled data directly. Another family of methods pose the image recovery as an optimization problem involving a physics-based forward model and a deep-learned regularization prior \cite{roth,admmnet,istanet,casecadeDynamic,modl,hammernik,zhang2017magazine,omodl,modlmussels,mardaniGANCS}. These model-based methods can be thought of as learning based variants of earlier plug-and-play methods \cite{venkatakrishnan2013plug,ahmad2020plug}, which used off-the-shelf denoisers as regularization penalties. In this work, we will focus on our implementation \cite{modl}, which is termed as model-based deep learning (MoDL). We refer the reader to~\cite{modl} for the details of MoDL, including its benefits over (a) direct-inversion based methods, (b) similar unrolled architectures and learned plug-and-play priors, (c) the use of conjugate gradients in contrast to steepest descent update to enforce data consistency, (d) as well as its ability to work with smaller CNN modules that allows it to learn from smaller datasets. The image quality offered by all of the above methods heavily depends on the sampling pattern. Early parallel MRI hardware~\cite{smash} was designed to eliminate the need to sample adjacent k-space samples, making uniform undersampling of k-space a desirable approach. By contrast, compressed sensing~\cite{candes2007sparsity,lustig2008compressed} advocates for the sampling pattern to be maximally incoherent. Since the k-space center is associated with high energy, variable density schemes that sample the center with a higher density are preferred by practitioners. Many of the current methods rely on the Poisson-disc variable density approach, which is a heuristic that combines the above intuitions~\cite{vasnawalaPoissonDisc}. Early empirical studies in the context of deep learning suggest that incoherent sampling patterns, which are widely used in compressed sensing, may not be necessary for good reconstruction performance in this context~\cite{hammernik}. Computational methods were introduced as a systematic approach to design the sampling patterns for each setting. 

The computational design of sampling patterns has a long history in MRI. Current solutions can be broadly classified as algorithm-dependent and algorithm-agnostic. The algorithm-agnostic approaches such as \cite{Reeves2000,xu,haldar2019oedipus,levine2017,senel2019} consider specific image properties and optimize the sampling patterns to improve the measurement diversity for that class. Image properties, including image support \cite{Reeves2000}, parallel acquisition using sensitivity encoding~(SENSE) \cite{xu,samsonov,levine2017}, and sparsity constraints \cite{haldar2019oedipus}, have been introduced.  These experiment design strategies often rely on the Cramer-Rao (CR) bound, assuming the knowledge of the image support or location of the sparse coefficients. Algorithm-dependent schemes such as \cite{sherry2019,gozcu2018learning} optimize the sampling pattern, assuming specific reconstruction algorithms (e.g., TV or wavelet sparsity). These approaches~\cite{sherry2019,gozcu2018learning} only consider single-channel settings with undersampled Fourier transform as a forward model. They utilize a subset of discrete sampling locations using greedy or continuous optimization strategies to minimize the reconstruction error. The main challenge with the above computational approaches is the significantly high computational complexity. The main contributor to the complexity is the evaluation of the loss associated with a specific sampling pattern. For instance, algorithm-dependent schemes need to solve the compressed sensing problem for each image in the dataset to evaluate the loss for a specific sampling pattern. The design of sampling pattern thus involves a nested optimization strategy; the optimization of the sampling patterns is performed in an outer loop, while image recovery is performed in the inner loop to evaluate the cost associated with the sampling pattern. The use of deep learning methods for image reconstruction offers an opportunity to speed up the computational design. Specifically, deep learning inference schemes enables the fast evaluation of the loss associated with each sampling pattern. In addition, these methods also facilitates the evaluation of the gradients of the cost with respect the sampling pattern. Unlike classical methods that rely on specific image properties (e.g., sparsity, support-constraints), the non-linear convolutional neural networks (CNN) schemes exploit complex non-linear redundancies that exist in images. This makes it difficult to use the algorithm-agnostic computational optimization algorithms discussed above in this setting. In addition, these learning-based methods often learn representations that may be strongly coupled to the specific sampling scheme. A joint strategy, which simultaneously optimizes for the acquisition scheme as well as the reconstruction algorithm, is necessary to obtain the best performance.

Most of the current sampling pattern optimization schemes for deep learning relies on a binary sampling mask  \cite{weiss,pilot,loupe}, which chooses a subset of the Cartesian sampling pattern. For instance, the recent LOUPE algorithm~\cite{loupe} jointly optimizes the sampling density in k-space and the reconstruction algorithm. It assumes each binary sampling location to be an independent random variable. The independence assumption makes it difficult of LOUPE to account for dependencies between sampling locations. We note that the popular Poisson disc sampling strategy~\cite{vasnawalaPoissonDisc} assumes
the sampling locations to be separated by a minimum distance~\cite{lustig2010spirit}, in addition to following a density. This separation is vital for exploiting the redundancies resulting from multichannel sampling with smooth coil sensitivities as described in~\cite{smash}. The PILOT approach \cite{weiss,pilot} instead relies on a relaxation of the binary mask to make the cost differentiable. A challenge with this scheme is the large number of trainable parameters, which often translate to convergence issues \cite{pilot}. In our own settings, a non-parametric strategy that aimed to optimize for all the sampling locations failed to converge, especially when large training datasets are not used. We note that another class of deep learning solutions involve active strategies \cite{jin,Zhang}, where a neural network is used to predict the next k-space sample to be acquired based on the image reconstructed from the current samples. We do not focus on such active paradigms in this work. We also note that similar work involving the optimization of the forward model have been also explored in the context of optical imaging \cite{metzler2019deep,muthumbi2019,horstmeyer2017,cheng2019,chakrabarti2016}.

The main focus of this work is to jointly optimize the sampling pattern and the deep network  parameters for parallel MRI reconstruction. We rely on an algorithm-dependent strategy to search for the best sampling pattern. The main contributions of this work are
\begin{enumerate}
	\item Unlike previous methods \cite{weiss,loupe,sherry2019,gozcu2018learning} that constrain the sampling pattern to be a subset of the Cartesian sampling pattern, we assume the sampling locations to be continuous variables. The earlier methods  \cite{weiss,loupe,sherry2019,gozcu2018learning} rely on relaxations or approximations of the discrete mask to make the cost function differentiable. The proposed scheme does not need any approximations since the derivatives with respect to the sampling locations are well-defined. 
	\item Unlike \cite{pilot}, we solve for the sampling pattern rather than the sampling density. Hence, our approach can account for complex dependencies between k-space sampling locations, which may be difficult for a density-based approach.
	\item Unlike the previous optimization strategies \cite{weiss,pilot,loupe,sherry2019,gozcu2018learning} that were only restricted to the single-channel setting, we extend the scheme to the multichannel setting where there is the most gain. 
	\item We introduce a parametric representation of the sampling patterns to reduce the degrees of freedom of the sampling pattern. The reduced search space improves the ability to learn the sampling pattern even from smaller datasets. 
\end{enumerate}
 The main objective of the proposed work is to optimize the sampling pattern for a specific anatomy (e.g., knee, brain) and protocol, rather than optimizing it for each subject. We note that the earlier optimization strategies in MRI are also designed for similar settings \cite{knollsampling,pilot}. Our experiments show that most of the deep learning algorithms significantly benefit from sampling pattern optimization, which is a relatively under-explored area compared to reconstruction network architecture and training. Our experiments involving the fastMRI knee dataset~\cite{fastmri}, acquired from multiple sites and scanners, demonstrate the robustness of the approach.

\section{Method}

\subsection{Image Formation}
We consider the recovery of the complex image $\boldsymbol \rho \in \mathbb{C}^{M\times N}$ from its non-Cartesian Fourier samples: 
\begin{equation}
\label{measurement}
  b[i,j]= \sum_{\mathbf m \in \mathbb{Z}^2} s_{j}[\mathbf m] ~\rho[\mathbf m] ~e^{-j \mathbf k_i^T \mathbf m} + n[i,j], \mathbf k_i \in \boldsymbol \Theta.
\end{equation}
Here, $\boldsymbol \Theta$ is a set of sampling locations and $n[i,j]$ is the noise process. $s_{j}; j=1,..,J$ corresponds to the sensitivity of the $j^{\rm th}$ coil, while $\mathbf k_i$ is the $i^{\rm th}$ sampling location. The above mapping can be compactly represented as  $\mathbf b = \mathcal A_{\boldsymbol \Theta}(\boldsymbol \rho) + \mathbf n$. The measurement operator $\mathcal A_{\boldsymbol \Theta}$ is often termed to as the forward model. It captures the information about the sampling pattern as well as the receive coil sensitivities. We note that the forward model is often modified to include additional information about the imaging physics, including field inhomogeneity distortions and  relaxation effects~\cite{doneva}.

\subsection{Regularized Image recovery}
Model-based algorithms are widely used for the recovery of images from heavily undersampled measurements, such as \eqref{measurement}. These schemes pose the reconstruction as an optimization problem of the form
\begin{equation}
\label{modelbased}
  \widehat{\boldsymbol \rho}_{\{\boldsymbol \Theta,\Phi\}}= \argmin_{\boldsymbol \rho} \|\mathbf b-\mathcal A_ {\boldsymbol \Theta}(\boldsymbol \rho) \|_2^2 + ~ \mathcal R_{\Phi}(\boldsymbol \rho).
\end{equation}
Here, $\mathcal R_{\Phi}$ is a regularization penalty. Regularizers include transform domain sparsity~\cite{ista2003wavelet}, total variation regularization~\cite{shiqianma2008}, and structured low-rank methods~\cite{jacobspmag}. For instance, in transform domain sparsity, the regularizer is chosen as $\mathcal R(\boldsymbol \rho) = \lambda \|\mathbf T \boldsymbol \rho \|_{\ell_1}$, with $\Phi = \{\lambda, \mathbf T\}$ denoting the parameters of the regularizer and the transform. We rely on the notation $ \widehat{\boldsymbol \rho}_{\{\boldsymbol \Theta,\Phi\}}$ for the solution of~\eqref{modelbased} to denote its dependence on the regularization parameters as well as the  sampling pattern. 

\subsection{Deep learning based image recovery}

\begin{figure}
  \centering
  \subfloat[J-UNET architecture]{
  \includegraphics[width=.6\linewidth]{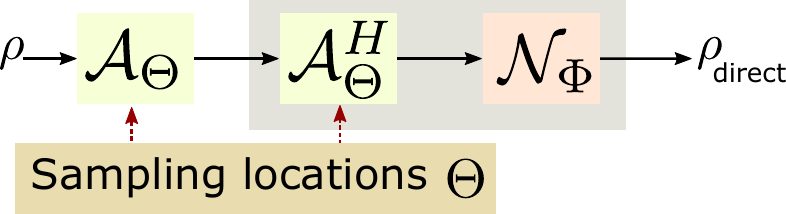}
}

  \subfloat[proposed J-MoDL architecture]{
    \includegraphics[width=.99\linewidth]{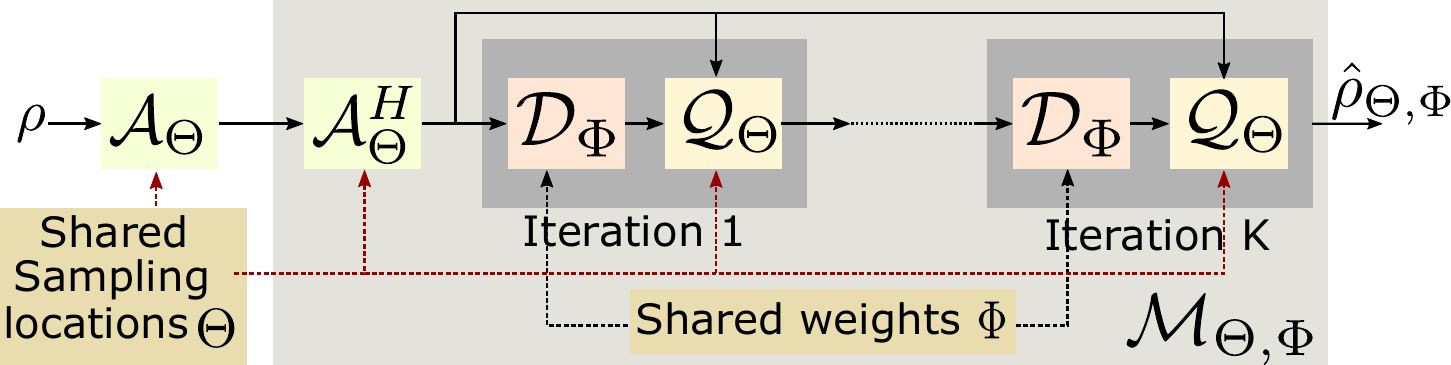}    
   }
  \caption{Illustration of the simultaneous sampling and reconstruction architectures. (a) The direct-inversion (J-UNET) architecture described by \eqref{direct}, where a CNN $\mathcal N_{\Phi}$ is used to recover the images from $\mathcal A_{\Theta}^H \mathbf b$. As discussed previously, the CNN parameters are closely coupled with the specific sampling pattern, making joint optimization challenging. (b) corresponds to the J-MoDL architecture, described by \eqref{modldc} and \eqref{denoised12}. Each iteration alternates between the CNN denoiser  $\mathcal D_{\Phi}$ and the data consistency block  $\mathcal Q_{\Theta}$. The data consistency block $\mathcal Q_{\boldsymbol \Theta}$ inverts the measured Fourier samples assuming $\mathbf z_n$, while $\mathcal D_{\Phi}$ acts as a \emph{denoiser} of the current iterate. The blocks $\mathcal D_{\Phi}$ and $\mathcal Q_{\boldsymbol \Theta}$ are relatively independent of $\Theta$ and $\Phi$, respectively.  }
  \label{fig:jmodl}
\end{figure}

Deep learning methods are increasingly being investigated as alternatives for regularized image reconstruction. Instead of algorithms that rely on the hand-crafted priors discussed above, these schemes learn the parameters from exemplar data. Hence, these schemes are often termed as data-driven methods. 
\subsubsection{Direct-inversion schemes}
Direct-inversion approaches~\cite{wangCTtmi2017,jong2019kspace} rely on a deep CNN $\mathcal N_{\Phi}$ to recover the images from undersampled gridding reconstruction $\mathcal A_{\boldsymbol \Theta}^H(\mathbf b)$ as 
\begin{equation}\label{direct}
\boldsymbol{\rho}_{\rm direct} = \mathcal N_{\Phi}\left( \mathcal A_{\boldsymbol \Theta}^H\mathbf b\right).
\end{equation}
Here $\Phi$ denotes the learnable parameters of the CNN $\mathcal N_{\Phi}$ (see Fig.~\ref{fig:jmodl}(a)). 
\subsubsection{Model-based deep learning}
Several unrolled approaches, which combine physics-based priors with learned priors, have been introduced for image recovery \cite{roth,admmnet,istanet,casecadeDynamic,modl,hammernik,zhang2017magazine,omodl,modlmussels,mardaniGANCS}.
In this paper, we will focus on the model-based deep learning (MoDL) ~\cite{modl} framework, where image recovery is formulated as
\begin{equation}
\label{modl}
\widehat{\boldsymbol \rho}_{\{\boldsymbol \Theta,\Phi\}}= \argmin_{\boldsymbol \rho} \|\mathbf b-\mathcal A_ {\boldsymbol \Theta}(\boldsymbol \rho) \|_2^2 + ~  \|\boldsymbol \rho - \mathcal D_{\Phi}(\boldsymbol \rho)\|_F^2,
\end{equation}
where $\mathcal D_{\Phi}$ is a residual learning-based CNN that is designed to  extract the noise and alias terms in $\boldsymbol \rho$.  The optimization problem specified by~\eqref{modl} is solved using an iterative algorithm, which alternates between a denoising step and a data consistency step: 
\begin{eqnarray}\label{modldc}
\boldsymbol\rho_{n+1} &=& \left(\mathcal A_{\Theta}^{H}\mathcal A_{\Theta} + 
\mathcal I\right)^{-1} \left(\mathbf z_n + \mathcal A_{\Theta}^{H}\mathbf b\right)\\\label{denoised12}
\boldsymbol z_{n+1} &=& \mathcal D_{\Phi}(\boldsymbol \rho_{n+1} ).
\end{eqnarray}
Here, \eqref{modldc} is implemented using a conjugate gradient algorithm. This iterative algorithm is unrolled  to obtain a deep recursive network $\mathcal M_{\boldsymbol \Theta,\Phi}$, where the weights of the CNN blocks and data consistency blocks are shared across iterations, as shown in Fig.~\ref{fig:jmodl}(b). Specifically, the solution to \eqref{modl} is given by 
\begin{equation}\label{modlnw}
\widehat{\boldsymbol \rho}_{\boldsymbol \{\Theta,\Phi\}} = \mathcal M_{\boldsymbol \Theta,\Phi} \left(\mathcal A_{\boldsymbol \Theta}(\boldsymbol \rho)\right).
\end{equation}
 Note that once unrolled, the image reconstruction algorithm is essentially a deep network, shown in Fig.~\ref{fig:jmodl}(b). Thus, the main distinction between MoDL and direct-inversion scheme is the structure of the network $\mathcal M_{\Theta,\Phi}$.  Please see \cite{modl} for details.

\subsection{Optimization of sampling patterns and hyperparameters}

The focus of this work is to optimize the sampling pattern specified by $\Theta$ in \eqref{measurement} and the parameters $\Phi$ of the reconstruction algorithm \eqref{modelbased} to improve the quality of the reconstructed images. Conceptually, the regularization priors encourage the solution to be restricted to a family of feasible images (e.g., sparse wavelet representation). The objective is to optimize the sampling pattern to capture information that is maximally complementary to the image representation. 

Early approaches that rely on compressed sensing algorithms  \cite{sherry2019,gozcu2018learning} optimize the sampling pattern $\Theta$ such that 
\begin{equation}
\label{algdep}
\{\boldsymbol \Theta^*\} =  \argmin_{\boldsymbol \Theta} \sum_{i=1}^{N} \|   \widehat{\boldsymbol \rho}_{i,\{\boldsymbol \Theta,\Phi\}} -\boldsymbol \rho_i \|_2^2,
\end{equation}
is minimized. Here $\rho_i; i=1,..,N$ are the different training images used in the optimization process and $\widehat{\boldsymbol \rho}_{i,\{\boldsymbol \Theta,\Phi\}}$ are the corresponding reconstructed images, recovered using \eqref{modelbased}. Greedy \cite{gozcu2018learning} or continuous optimization schemes \cite{sherry2019} are used to solve \eqref{algdep}. However, the main challenge associated with these schemes is the high complexity of the optimization algorithm used to solve \eqref{modelbased}.  Note that the optimization scheme \eqref{modelbased} is in the inner loop; for each sampling pattern, the $N$ images have to be reconstructed using computationally expensive CS methods to compute the loss in \eqref{algdep}. This makes it challenging to train the pattern using a large batch of training images. In addition, the hyperparameters of the algorithm denoted by $\Phi$ are assumed to be fixed during this optimization. 

Recent schemes such as LOUPE~\cite{loupe} and PILOT~\cite{pilot} exploit the fast deep-learned reconstruction algorithms to optimize for the sampling pattern. Instead of directly solving for the k-space locations, the LOUPE approach optimizes for the sampling density \cite{loupe}. Specifically, they assume the k-space sampling locations that are acquired to be binary random variables $\mathbf k_i \sim \mathcal B(p_i)$ and optimize for the probabilities $p_i$. They rely on several random realizations of $\mathbf k_i$ and the corresponding reconstructions to perform the optimization.

\subsection{Proposed Joint Optimization Strategy}
 This work proposes a joint model-based deep learning~(J-MoDL) framework  to jointly optimize both the $\mathcal D_{\Phi}$ and  $\mathcal Q_{\boldsymbol \Theta}$ blocks in the MoDL framework~\eqref{modl}  with the goal of improving the  reconstruction performance. Specifically, we propose to jointly learn the sampling pattern $\boldsymbol \Theta$ and the CNN parameters $\Phi$ from training data using
\begin{equation}
\label{joint}
\{\boldsymbol \Theta^*,\Phi^*\} =  \argmin_{\boldsymbol \Theta ,\Phi} \sum_{i=1}^{N} \| \mathcal M_{\boldsymbol \Theta,\Phi} \left(\mathcal A_{\boldsymbol \Theta}(\boldsymbol \rho_i)\right) -\mathbf x_i \|_2^2.
\end{equation}
We note that $\mathcal M_{\boldsymbol \Theta,\Phi} $ denotes a general deep learning network architecture that includes direct-inversion schemes denoted by \eqref{direct} as well as unrolled architectures denoted by \eqref{modlnw}.

While the proposed J-MoDL  framework (in Fig.~\ref{fig:jmodl}(b)) can be generalized to other error metrics such as perceptual error, we focus on the $\ell_2$ error in this work.

\subsection{Forward model and parametrization of the sampling pattern}
We represent the forward model as 
\begin{equation}\label{fwd}
\mathbf b_{i} = \mathcal F_{\Theta} (\mathbf s_i\cdot \rho); ~~i=1,..,Nc,
\end{equation}

where $s_i; i=1,..,N_c$ denotes the coil sensitivities of the $i^{\rm th}$ channel to compactly represent \eqref{measurement}. Here, $\mathcal F_{\Theta}(\rho)$ denotes the Fourier transform of $\rho$ evaluated at the continuous sampling locations $\mathbf k_i$, whose set is denoted by $\Theta$.

\begin{figure}
\input{fig_sampling.tex}  
  \caption{Illustration of the proposed sampling parameterization to acquire $M$ samples of an $N$ dimensional signal such that the acceleration factor~$=N/M$. (a) The sampling operator $\mathcal F_{\Theta}$ is an $M \times N$ matrix that can capture $M$ samples from possibly non-integer locations $k_1, \cdots, k_M$. These $M$ locations are real-valued trainable parameters constrained between $[0,1]$. (b) In the 2-D case, we utilize two sampling operators, $\mathcal F_{\Theta_h}$ and $\mathcal F_{\Theta_v}$, in the horizontal and the vertical directions, respectively.  $\mathcal F_{\Theta_h}$ acquires $m_h$ samples, whereas $\mathcal F_{\Theta_v}$ acquires $m_v$ samples such that total $M=m_h \times m_v$ samples are acquired from $N=P\times Q$ dimensional image. }
  \label{fig:sampling}
\end{figure}
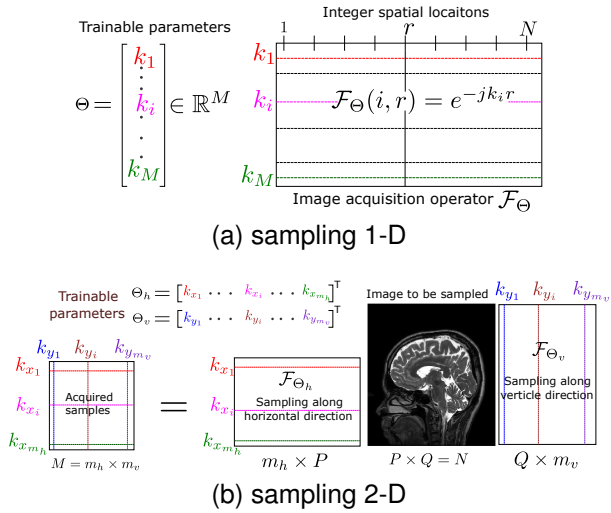

We found it challenging to directly optimize for the large number of trainable sampling locations which would require a huge amount of training data.
Hence, we propose to reduce the dimension of the search space by a parametrization of the sampling pattern as illustrated in Fig.~\ref{fig:sampling} for 1-D and 2-D case. Specifically, we assume that the sampling pattern to be the union of transformed versions of a template set $\Gamma$:
\begin{equation}\label{constrained}
\boldsymbol{\Theta} = \bigcup_{i=1}^{P} \mathcal T_{\theta_i}\left(\Gamma\right).
\end{equation}

Here, $\mathcal T_{\boldsymbol\theta_i}$ is a transformation that is dependent on the trainable parameters $\theta_i$. For example, one may consider the optimization of the phase encoding locations in MRI, while  the frequency encoding direction is fully sampled. Specifically, we choose $\Gamma$ as samples on a line and $\mathcal T_{\boldsymbol\theta_i}$ are translations orthogonal to the line. Here, $\theta_i; i=1,..,P$ are the phase encoding locations. In the 2-D setting,  we also consider sampling patterns of the form 
\begin{equation}\label{2d}
\Theta = \Theta_{\rm v} \cap \Theta_{\rm h},
\end{equation}
where $\Theta_{\rm v}$ and $\Theta_{\rm h}$ are 1-D sampling patterns in the vertical and horizontal directions, respectively. Here, we assume that the readout direction is orthogonal to the scan plane and is fully sampled. An example sampling pattern in this setting is shown in  Fig.~\ref{fig:2d8xloupe}. Specifically, the locations $k_{x_i}; i=1,..,h$ and $k_{y_i}; i=1,..,v$ are the unknowns that the algorithm optimizes for. We note that this approach reduces the number of trainable parameters from $hv$ to $h+v$.

In addition to reducing the parameter space, the above approaches also simplifies the implementation. We focus on this setting because the forward model in \eqref{fwd} can be implemented  in terms of the 1-D Fourier transforms as 
\begin{equation}\label{key}
\mathbf B = \mathbf F_h ~\mathbf X \;\mathbf F_v^H.
\end{equation}
Here, $\mathbf X$ is the 2-D image and $F_h$ and $F_v$ are 1-D discrete Fourier transform operators as described in Fig.~\ref{fig:sampling}(b). By eliminating the need for non-uniform Fourier transform (NUFT) operators, this approach accelerates the training and inference. 

\subsection{Architecture of the networks used in joint optimization}

Figure~\ref{fig:jmodl}(b) shows the proposed J-MoDL framework. The framework alternates between data consistency blocks $\mathcal Q_{\Theta}$, that depend only on the sampling pattern, and the CNN blocks $\mathcal D_{\Phi}$. We unrolled the MoDL algorithm in Fig.~\ref{fig:jmodl}(b) for K=5 iterations (i.e., five iterations of alternating minimization) to solve Eq.~\eqref{modl}.  The forward operator $\mathcal A_{\Theta}$ is implemented as a 1-D discrete Fourier transform to map the spatial locations to the continuous domain Fourier samples specified by $\Theta$, following the weighting by the coil sensitivities, as described by \eqref{measurement} and Fig.~\ref{fig:sampling}. The data consistency block $\mathcal Q_{\Theta}$ is implemented using 10 iterations of the conjugate gradient algorithm.  The CNN block $\mathcal D_{\Phi}$ is implemented as a UNET with four pooling and unpooling layers with $3 \times 3$ trainable filters as in the UNET model~\cite{ronneberger2015unet}.  The parameters of the blocks $\mathcal D_{\Phi}$ and $\mathcal Q_{\Theta}$ are optimized to minimize~\eqref{joint}. We relied on the automatic differentiation  capability of TensorFlow to evaluate the gradient of the cost function with respect to $\Theta$ and $\Phi$. 

We also study the optimization of the sampling pattern in the context of direct-inversion (i.e., when a UNET is used for image inversion). A UNET with the same number of parameters as the MoDL network considered above was used to facilitate fair comparison. This optimization scheme,  where both sampling parameters and the UNET parameters are learned jointly, is termed as J-UNET (Fig.~\ref{fig:jmodl}(a)).  Since MR images are inherently complex valued, all the networks were trained using complex k-space data as input and the training loss was calculated on the complex images. The complex data was split into real and imaginary parts, which were fed into the neural networks. The data consistency steps explicitly worked with the complex data type. 

\subsection{Proposed continuous optimization training strategy }
\label{sec:strategies}
 We first consider a collection of variable density random sampling patterns with 4\% fully sampled locations in the center of the k-space, and train only the network parameters $\Phi$. This training strategy  is referred to as $\Phi$-alone optimization. Once this training is completed, we fixed the trained network parameters and optimize for the sampling locations alone. Specifically, we consider the sampling operator $\mathcal A_{\Theta}$ and its adjoint as layers of the corresponding  networks. The parameters of these layers are the location of the samples, denoted by $\Theta$. We optimize for the parameters using stochastic gradient descent, starting with random initialization of the sampling locations $\Theta$. The gradients of the variables are evaluated using the automatic differentiation capability of TensorFlow. This  strategy, where only the sampling patterns are optimized,  is referred to as the $\Theta$-alone optimization; the parameters of the network derived from the $\Phi$-alone optimization are held constant. The third strategy, we refer as $\Theta,\Phi$-Joint or just Joint, simultaneously  optimizes for both, the sampling parameter $\Theta$ as well as the network parameters $\Phi$. The $\Phi$-alone optimization strategy take 5.5~hours to train in single-channel settings as described in section~\ref{sec:sc}. The $\Theta$-alone and $\Phi, \Theta$-joint strategies only take 1~hour to train with an initialization from  $\Phi$-alone model.

\section{Experiments and Results}
\subsection{Datasets}

We relied on three datasets for comparison. 

\subsubsection{Single-channel  knee data from fastMRI database} \label{fastmri}
  We used the data from the NYU fastMRI Initiative database~\cite{fastmri} (\url{fastmri.med.nyu.edu}) in this section. As such, NYU fastMRI investigators provided data, but did not participate in analysis or writing of this article. The primary goal of fastMRI is to test whether machine learning can aid in the reconstruction of medical images. We relied on a PCA-based complex combination of the multichannel images from 
  the database to obtain  single-coil images. The k-space data of  these images, computed using the forward model in~\eqref{measurement} with $J=1$ and $s_1(\mathbf x)=1$, are the input to the networks, while the corresponding complex images are used as the ground truth for training. We chose three subsets of the fastMRI dataset, consists of 100 training, 50 validation, and 100 test subjects. Unlike the other datasets considered in this work, this data was acquired on multiple scanners at different institutions, thus exhibiting significant diversity in the measurement settings. This dataset thus enables the evaluation of the scheme in a multi-site setting.

\subsubsection{Multichannel knee dataset} \label{florianknoll}We used a publicly available parallel MRI knee dataset as in~\cite{hammernik}. The training data constituted of 381 slices from ten subjects, whereas test data had 80 slices from two subjects. Each slice in the training and test dataset had different coil sensitivity maps that were estimated using the ESPIRIT~\cite{espirit2014} algorithm. Since the data was acquired by using a 2-D Cartesian sampling scheme, we relied on a 1-D undersampling of this data.

\subsubsection{Multichannel brain dataset} \label{brain}

We consider a parallel  MRI brain data using a 3-D~T2~CUBE sequence with Cartesian readouts using a $12$-channel head coil at the University of Iowa on a 3T GE MR750w scanner. The data was acquired according to the approved IRB protocol. Written consent was obtained from all subjects prior to the scan. 
The matrix dimensions were $256\times232\times 208$ with a $1$~mm isotropic resolution.  Fully sampled multi-channel brain images of nine  volunteers were collected, out of which data from five subjects were used for training, while the data from two subjects were used for testing and the remaining two for validation. Since the data was acquired with a 3-D sequence, we used this data to determine the utility of 1-D and 2-D sampling in parallel MRI settings.  Specifically, we performed a 1-D inverse Fourier transform along the readout direction and considered the recovery of each slice in the volume. Since the undersampling was performed on the phase encoding directions, these simulation studies are realistic. Following the image formation model in~\eqref{measurement}, additive white Gaussian noise of standard deviation $\sigma=0.01$ was added in k-space in all the experiments.

\subsection{Single-Channel Results}
\label{sec:sc}
\begin{table} \centering
	\caption{Single-channel settings: The mean $\pm$ std values of PSNR~(dB) and SSIM  over the test data of hundred subjects using different optimization strategies at 4x acceleration. }
	\label{tab:1d_sc}
	\input{tab_1d_sc.tex}
\end{table}

\begin{figure}
  \input{fig_1d_single_channel.tex}
\caption{The visual comparisons of different optimization strategies, described in section~\ref{sec:sc}, on a test slice in single-channel settings at 4x acceleration. The numbers in the subcaption show the PSNR values. The red arrow points to thin vertical features sharply captured by J-MoDL as compare to MoDL or J-UNET.}
\label{fig:1d_sc}
\end{figure}
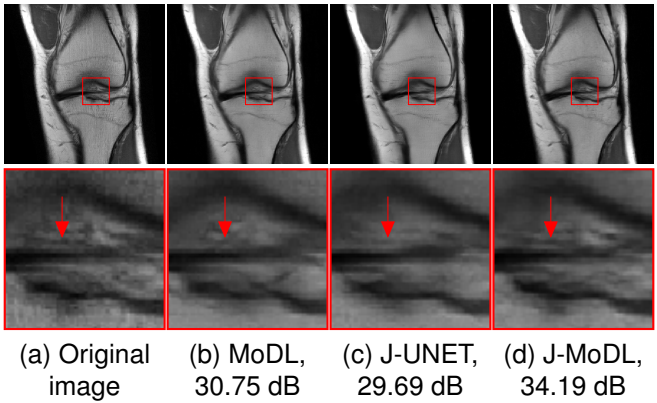

We first consider the single-channel experiments using the fastMRI data, as described in Section \ref{fastmri}. We note that almost all sampling pattern optimization schemes have considered  the single-channel settings~\cite{weiss,pilot,loupe,sherry2019,gozcu2018learning}, where an undersampled Fourier sampling forward operator is considered. Unlike the discrete optimization schemes that rely on relaxations of a discrete sampling mask  \cite{weiss,pilot,loupe,sherry2019,gozcu2018learning}, we  consider the optimization of the continuous values of the  phase encoding locations $k_1,\cdots,k_M$, as shown in Fig.~\ref{fig:sampling}(a). We consider an undersampling factor of four in this experiment.

Table~\ref{tab:1d_sc} reports the average PSNR and SSIM values obtained on the test data from 100 subjects. The top row corresponds to the optimization of the network parameters $\Phi$-alone, assuming the random variable density undersampling patterns with 4\% fully sampled center of the k-space. Each training slice had a different sampling pattern, whereas during testing each subject had a different sampling pattern; all slices of a subject had same sampling pattern. This approach made the network relatively insensitive to the specific sampling pattern, compared to the learning with a single pattern.  We note that the higher complexity of the MoDL framework translated to an approximate 3.5 dB improvement in performance over a UNET scheme in the $\Phi$-alone setting, even though both methods had the the same number of parameters. This observation is in line with the experiments in \cite{modl}. The second row in Table~\ref{tab:1d_sc} reports the result of only optimizing the sampling parameters $\Theta$, while keeping the reconstruction network fixed as the one trained in the first row ($\Phi$-alone). The last row of Table~\ref{tab:1d_sc} corresponds to the joint optimization scheme, where both $\Theta$ and $\Phi$ are trained with the initial sampling pattern used in the top row. The resulting J-MoDL scheme offers a 3.13 dB improvement in performance over the case where only the network is trained. The J-MoDL scheme is also better by 1.32 dB compared to only optimizing the sampling pattern. By contrast, the J-UNET approach provided only a 1.05 dB improvement over the initialization. The results demonstrate the benefit of the decoupling of the sampling pattern and CNN parameters offered by MoDL.

The visual comparisons of these strategies are shown in Fig.~\ref{fig:1d_sc}. The proposed J-MoDL method provides significantly improved results over the MoDL scheme, as highlighted by the zoomed region. The red arrows clearly show that the proposed J-MoDL architecture preserves the high-frequency details better than the MoDL architecture. The optimization of the sampling patterns also improved the UNET performance.

\subsection{Parallel Imaging (Multichannel)  with 1-D sampling}
\label{sec:mc}

\begin{table} \centering
	\caption{ Impact of optimization strategies for parallel MRI recovery of knee images using 1-D sampling. The results correspond to two subjects with a total of 80 slices.  }
	\label{tab:1d_mc}
	\input{tab_1d_mc.tex}
\end{table}

\begin{figure}
 \input{fig_1d4x_multi_channel}
	\caption{ Comparison of joint and  network-alone optimization in parallel imaging settings, described in section~\ref{sec:mc} ,with a 1-D sampling mask. The numbers in subcaptions are showing the PSNR (dB) values. (a) shows a fully sampled image from the test dataset. (b) shows the reconstructed image with a pseudo-random 4x acceleration mask using the MoDL approach. (c,d) shows joint optimization of sampling as well as network parameters using direct-inversion and model-based techniques, respectively.   The zoomed areas clearly show that joint learning better preserves the fine details. }
	\label{fig:1d_mc}
\end{figure}
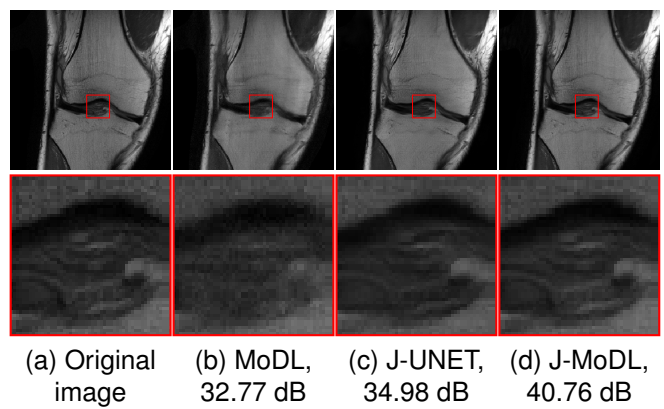

\begin{figure*}
  \input{fig_landscape.tex}
  \caption{This figure compares the optimization landscape of the MoDL (K=1) and UNET architecture for 1-D multichannel data. These plots show the mean squared error (MSE$\times 1000$) between the reconstructions and the corresponding  original images. The $n1$ and $n2$ axes represent continuous valued sampling locations around the ones marked on the mask. (a) and (b) show the landscape plot for MoDL architecture trained with a single sampling pattern and multiple sampling patterns, respectively. Similarly (c) and  (d) shows corresponding plots for the UNET architecture. These plots (a)-(d) are plotted at high-frequency values around locations 6 and 15, as marked with green in the mask. Similarly, (e)-(h) show landscape plots at relatively low frequencies around locations 135 and 167. From this controlled experiment, we observe that MoDL results in a smoother landscape as compared to UNET both at low and high frequencies. In addition, the UNET landscapes become comparatively smoother with the sampling pattern augmentation strategy, which makes the approach relatively insensitive to small differences in sampling pattern, as seen from (c) to (d) and (e) to (f).  }
\label{fig:landscape}
\end{figure*}
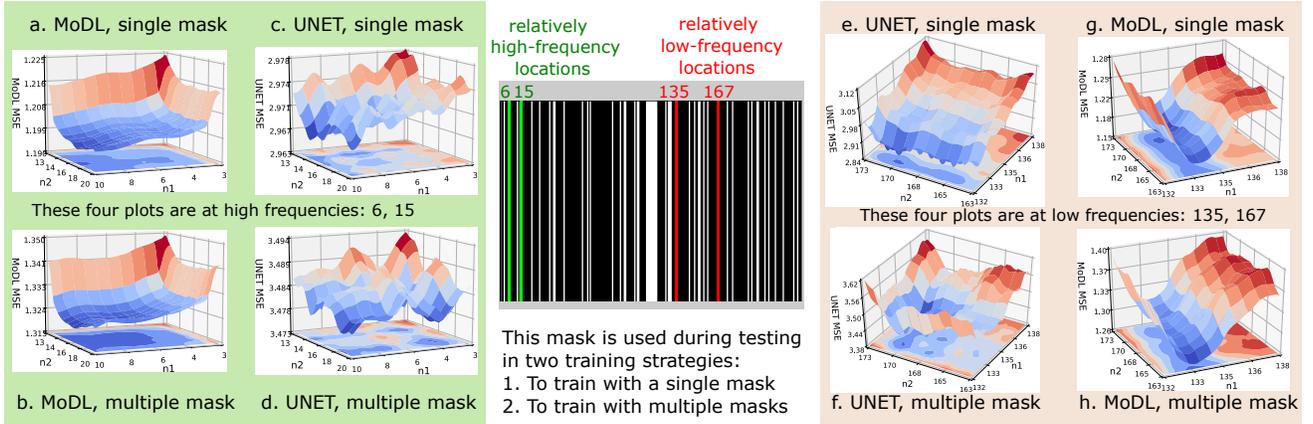

Table~\ref{tab:1d_mc} summarizes the results in the 1-D parallel MRI setting on knee images, described in Section~\ref{florianknoll}. The first row denoted as $\Phi$-alone in Table~\ref{tab:1d_mc} corresponds to optimizing the network parameters alone without optimizing the sampling mask. Unlike the setting in Section~\ref{sec:sc}, the network was not trained with different sampling masks. We choose the sampling mask as a single pseudo-random pattern for all the slices. In the second row, denoted as $\Theta$-alone optimization, only the sampling mask is optimized, while keeping the reconstruction parameters fixed to optimal values as derived in the first row. Here, the network parameters were initialized with the ones derived from the $\Phi$-alone optimization. Unlike the trend in Table~\ref{tab:1d_sc}, we observe that the performance of the UNET scheme dropped slightly, while the MoDL scheme that was trained with the same setting provided improved results. The last row compares joint optimization using direct-inversion and model-based techniques. We observe that both methods improved in this case. The J-MoDL provides around 7~dB improvement over $\Phi$-alone in the 4x setting and 3.5 dB in the 6x settings.

Figure~\ref{fig:1d_mc}.(a) shows an example slice from the test dataset that illustrates the benefit of jointly optimizing both the sampling pattern and the network parameters (Fig.~\ref{fig:1d_mc}.(c)), compared to the network-alone in the model-based deep learning framework~(Fig.~\ref{fig:1d_mc}.(b)). The zoomed image portion shows that joint learning using J-MoDL better preserves the soft tissues in the knee at the four-fold acceleration case in parallel MRI settings.

	To understand the drop in performance of the UNET scheme  during the $\Theta$-alone optimization, we compare the optimization landscape of the two schemes (MoDL and UNET in $\Theta$-alone settings) in Fig.~\ref{fig:landscape}. Since this is a large dimensional problem, we plot the variation in MSE with respect to two variables (sampling locations) at a time.
	 
	As described above, a single sampling pattern, shown in Fig.~\ref{fig:landscape}, was used to train UNET and MoDL architectures on the parallel imaging knee dataset.  We then computed the loss of the networks for perturbations of the sampling locations around the  sampling pattern shown in Fig.~\ref{fig:landscape}. Specifically, the trained models were used to reconstruct the test dataset, while two of the original sampling locations (denoted by the green and red lines in Fig.~\ref{fig:landscape}) are perturbed from their original values. The loss evaluated for each of the perturbations are plotted in (a)-(d) and (e)-(h), respectively. Specifically, (a)-(d) corresponds to perturbations around the green locations 6 and 15 from the high-frequency samples, while (e)-(h) correspond to the samples 135 and 167, closer to the k-space center. The losses of the networks are plotted in Fig.~\ref{fig:landscape}. The $n1$ and $n2$ axes on these four plots correspond to the sampling locations, while the vertical z-axis shows the  mean squared error~(MSE~$\times 1000$) between the predicted and original test image.  Each of the $n1$ and $n2$ axis were varied for 100 points around them, thus resulting in a total of 10,000 MSE evaluations on each of the plots.
	The plots show that the MoDL network exhibits a smoother cost landscape around its minimum, while the UNET, which was trained using the same settings and initialization, resulted in highly oscillatory landscape.

        We note that the proposed sampling pattern optimization scheme relies on stochastic gradient descent. The high gradients resulting from the oscillatory landscape, as well as the randomness in the gradient updates, likely resulted in the UNET converging to a bad local minimum. As shown in Table \ref{tab:1d_sc}, this problem can be mitigated by sampling pattern augmentation. However, we note that the MoDL scheme does not require the network to be trained with multiple sampling patterns to have a smoother optimization landscape, which explains its improved performance in this setting.

\subsection{Parallel Imaging (Multichannel) with 2-D sampling}
\label{sec:mc2d}
\begin{table} \centering
	\caption{Impact of optimization strategies for parallel MRI recovery of the brain images using 2-D sampling. The  PSNR and SSIM values are reported for the average $\pm$ std.~of 200 test slices at 6x acceleration factor.}
	\label{tab:2d_mc}
	\input{tab_2d_mc.tex}
      \end{table}
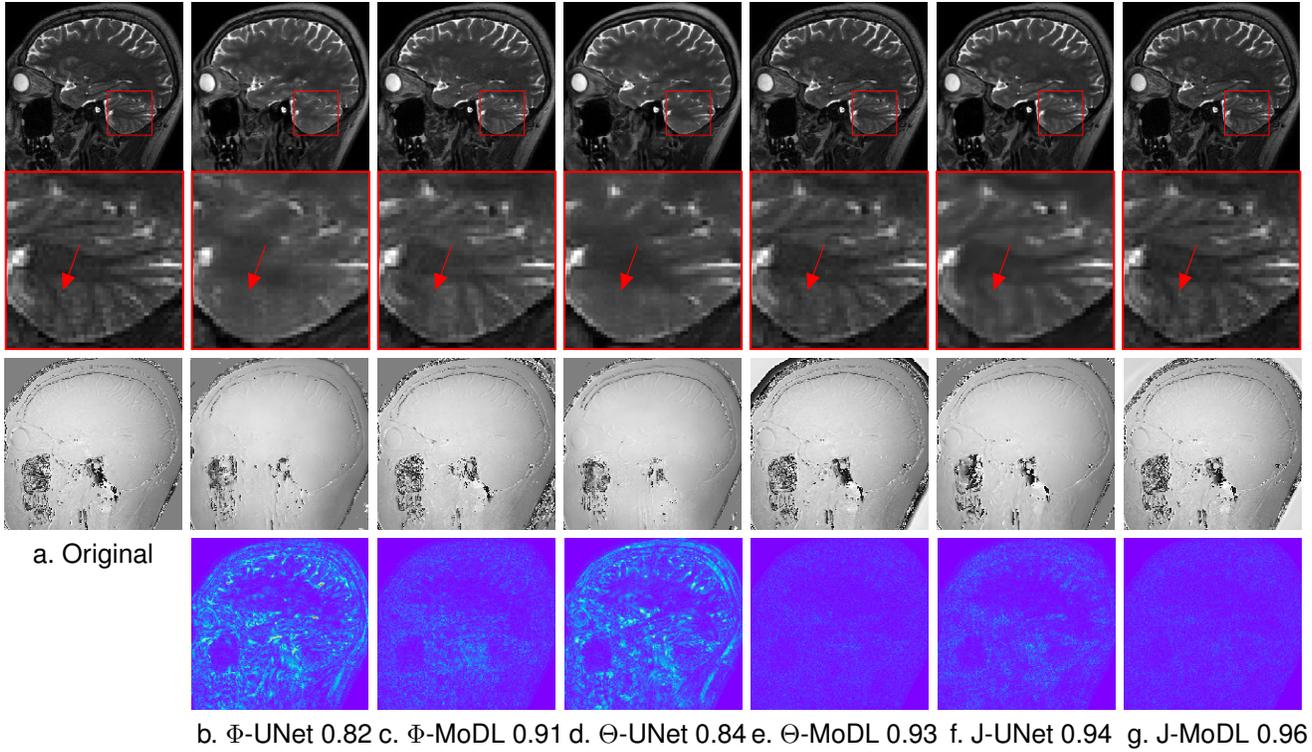
\begin{figure*}
  \input{fig_2d6x_mc.tex}
  \caption{This figure compares the different schemes in parallel imaging settings with a 2-D sampling mask at 6x acceleration, as described in section~\ref{sec:mc2d}. Row~1 and row~2 show magnitude images and a zoomed cerebellum region, respectively. Row~3 shows reconstructed phase images, while row~4 shows error maps of the reconstructed images with respect to the original image. $\Phi$-Unet~(b) and $\Phi$-MoDL~(c) optimize only the network parameters~$\Phi$. $\Theta$-UNet and $\Theta$-MoDL optimize only the sampling parameters~$\Theta$ with initialization from respective $\Phi$-alone models. The sub-captions denote the SSIM values. Finally, J-UNET and J-MoDL are the proposed joint optimization models. The J-MoDL approach preserves the fine features in the cerebellum region, as shown by the zoomed area. The red arrows in the zoomed area point to a feature that is well preserved by the joint optimization techniques versus the results of the networks that optimized only network parameters.  }
  \label{fig:2d6x_mc}
\end{figure*}

Table~\ref{tab:2d_mc} summarizes the comparison results
in the multichannel setting with 2-D sampling  patterns, as described by \eqref{2d} on the brain data described in Section \ref{sec:mc2d}. Both the direct-inversion based framework (UNET) and the model-based framework~(MoDL) are compared in Table~\ref{tab:2d_mc} at three different optimization strategies at 6x acceleration factor. As described in Section \ref{sec:sc}, the $\Phi$-alone network was trained with multiple sampling patterns to reduce its sensitivity to sampling patterns. The trends of the different methods continue to be the same as in Section \ref{sec:sc}.

The improved performance offered by the optimization of the sampling pattern in 2-D parallel imaging settings can be appreciated from Fig.~\ref{fig:2d6x_mc}. The zoomed portion in Fig.~\ref{fig:2d6x_mc} shows the cerebellum region in which all the fine features are reconstructed well by the proposed J-MoDL approach at 6x acceleration. The red arrows are pointing to a  high-frequency feature that is not recovered by the joint learning in the direct-inversion framework (J-UNET). This feature is also not recovered by the fixed model-based deep learning framework without joint optimization~(see Fig.~\ref{fig:2d6x_mc}(b) and (c)). Fig.~\ref{fig:2d6x_mc} also shows that proposed method can reconstruct the phase of the MR images. The error maps in Fig.~\ref{fig:2d6x_mc} shows that the proposed J-MoDL approach has the least error in reconstruction among competing methods.

\subsection{Comparison with classical sampling patterns}
\label{sec:classical}
\begin{figure}
	\input{fig_2d10x_vd_rand_learn.tex}
	\caption{This figure compares the reconstruction quality obtained by different sampling masks in 2-D parallel imaging settings at 10x acceleration as described in section~\ref{sec:classical}. Rows one, two, and three shows masks, $A^Hb$,  and reconstruction outputs, respectively. Two $\Phi$-alone models using the MoDL approach were trained with random masks as well as random variable-density~(VD) masks. It can be observed that the  2-D mask learned using the J-MoDL approach outperforms the reconstruction using fixed random and variable-density masks. The numbers in sub-captions are showing PSNR (dB) values. }
	\label{fig:2d10x_vd}
\end{figure}
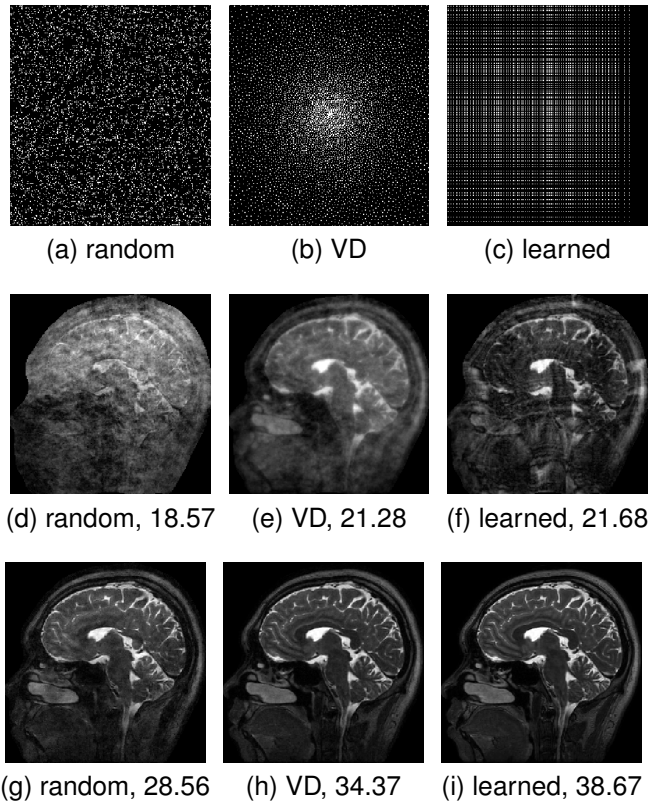

Figure~\ref{fig:2d10x_vd} shows the visual comparison of reconstruction quality obtained with three different sampling patterns at the same 10x acceleration in 2-D parallel MRI settings with the model-based deep learning framework. Figure~\ref{fig:2d10x_vd}(a) and (b) are showing pseudo-random and  variable density~(VD) masks, respectively, while Fig.~\ref{fig:2d10x_vd}(c) shows the  2-D mask learned using joint learning with J-MoDL. These masks result in gridding reconstructions, as shown in Fig.~\ref{fig:2d10x_vd}(d), (e), and~(f). It can be observed from Fig.~\ref{fig:2d10x_vd}(f) that learned mask results in a gridding reconstruction with comparatively fewer artifacts. Figures~\ref{fig:2d10x_vd}(g) and~\ref{fig:2d10x_vd}(h) are the reconstructed images using the  $\Phi$-alone optimization, whereas Fig.~\ref{fig:2d10x_vd}(i) corresponds to the reconstruction using joint learning.

\subsection{Impact of noise}
\label{sec:noise}
\begin{figure}
  \input{fig_2d8x_high_noise.tex}
  \caption{This figure demonstrates the impact of adding a high amount of noise in the k-space samples in 2-D parallel MRI settings at 8x acceleration, as described in section~\ref{sec:noise}. The first row shows different masks learned with the J-MoDL approach when the Gaussian noise of standard deviation $\sigma$ is added in the k-space samples. The second row shows corresponding reconstructions (Recon.). Subcaptions of (d), (e), and (f) are showing PSNR~(dB) values. As expected, higher noise levels promote the algorithm to learn the sampling parameters that sample more of the low-frequency components  from the center of k-space, leading to low-resolution reconstructions.}
  \label{fig:2d8xnoise}
\end{figure}
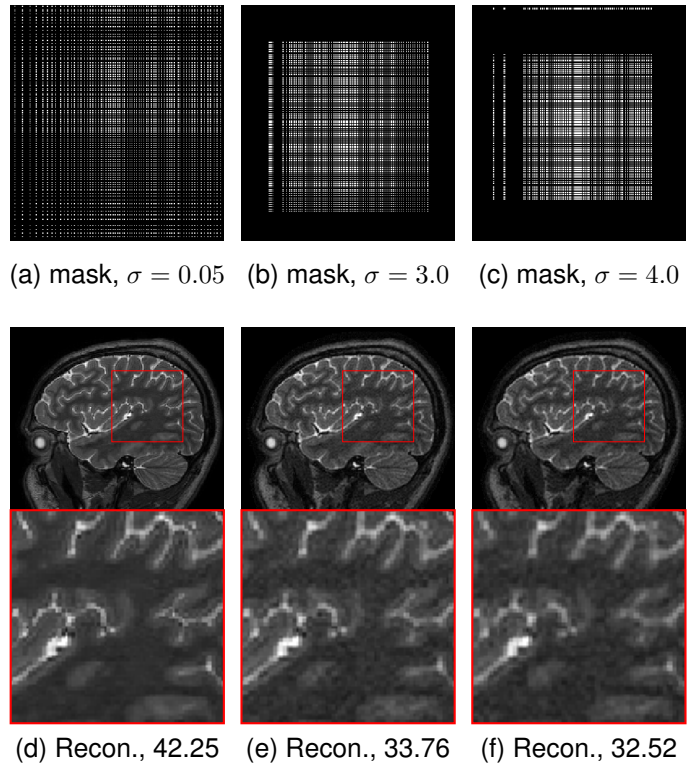

We study the impact of noise on the learned optimal sampling pattern and the reconstruction performance in Fig.~\ref{fig:2d8xnoise}. 
 We note that the data was already corrupted by noise. We further added complex Gaussian noise with different standard deviations to the  8x undersampled k-space measurements. The results show that as the noise standard deviation increases, the optimal sampling patterns get concentrated to the center of k-space. This is expected since the energy of the Fourier coefficients in the center of k-space is higher. As the standard deviation of the noise increases, the outer k-space regions become highly corrupted with noise and hence sampling them does  not aid  the reconstruction performance. As expected,  the restriction of the sampling pattern to the center of k-space results in image blurring. It can be noted that during training with different noise levels  no extra constraints were imposed  to promote a low-frequency mask. This behavior can be attributed to the explicit data consistency step in the model-based deep learning framework.  This experiment empirically shows  that the proposed J-MoDL  technique indeed conforms with classical model-based techniques while retaining the  benefits of deep learning methods.

\subsection{Convergence of the sampling pattern optimization scheme}
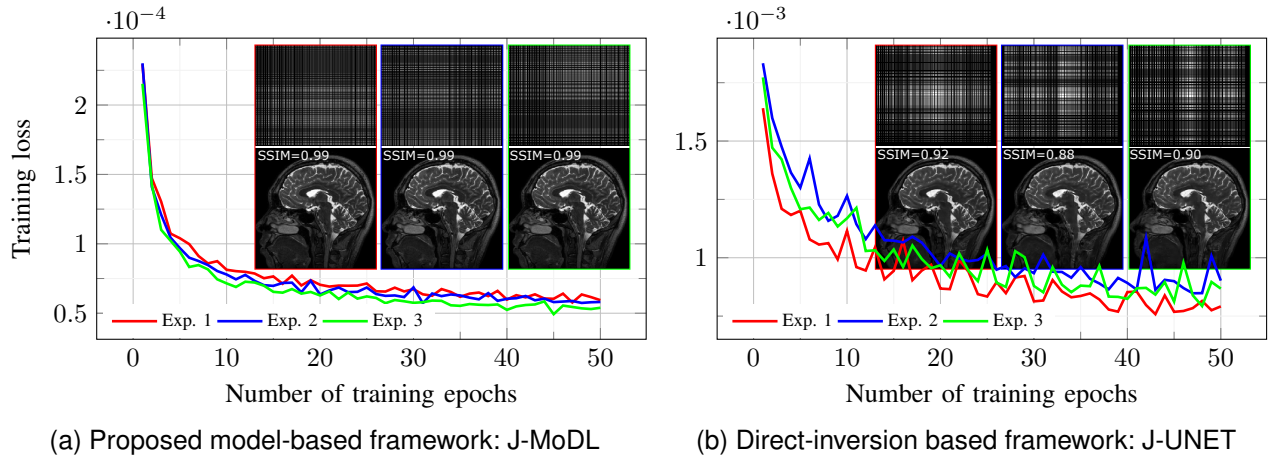
\begin{figure*}
\input{fig_converge.tex}
\caption{This figure compares the convergence of training loss in joint optimization with direct-inversion and model-based deep learning frameworks. We performed three independent experiments for each of the two frameworks.  The experimental setup for all the three experiments was identical except for the initialization of sampling parameters. The learned masks, reconstructed images, as well as the training loss, are plotted for each of the three experiments. We observe that J-MoDL convergence is relatively smoother compared to the J-UNET approach; all three initializations resulted in results with somewhat similar image quality. We note that the cost function may have multiple global minima. The J-UNET convergence was slower compared to the J-MoDL setting, likely because of the non-smooth optimization landscape as seen from Fig. \ref{fig:landscape}. }
\label{fig:2d6x_conv}
\end{figure*}

We empirically study the convergence of the sampling pattern optimization schemes in Fig.~\ref{fig:2d6x_conv}. We consider three different initial pseudo-random sampling patterns, each with  fully sampled center having 4\% lines as initialization.
  In each experiment, training is performed for 50 epochs. Figure~\ref{fig:2d6x_conv}(a) shows the decay of training loss with the proposed J-MoDL scheme, while Fig.~\ref{fig:2d6x_conv}(b) correspond to the J-UNET scheme. We observe that despite the highly non-convex optimization scheme the J-MODL network was able to converge to solutions with almost the same cost.  We also note that the images reconstructed with the final network are similar, even though the sampling patterns are different. We observe that the convergence of the J-UNET scheme was relatively less smooth, likely because of the non-smooth optimization landscape. The image quality of the final reconstructions are also more variable in this case. By contrast, the J-UNET scheme exhibit more variability in the final results.

\subsection{Comparison with other sample optimization schemes}
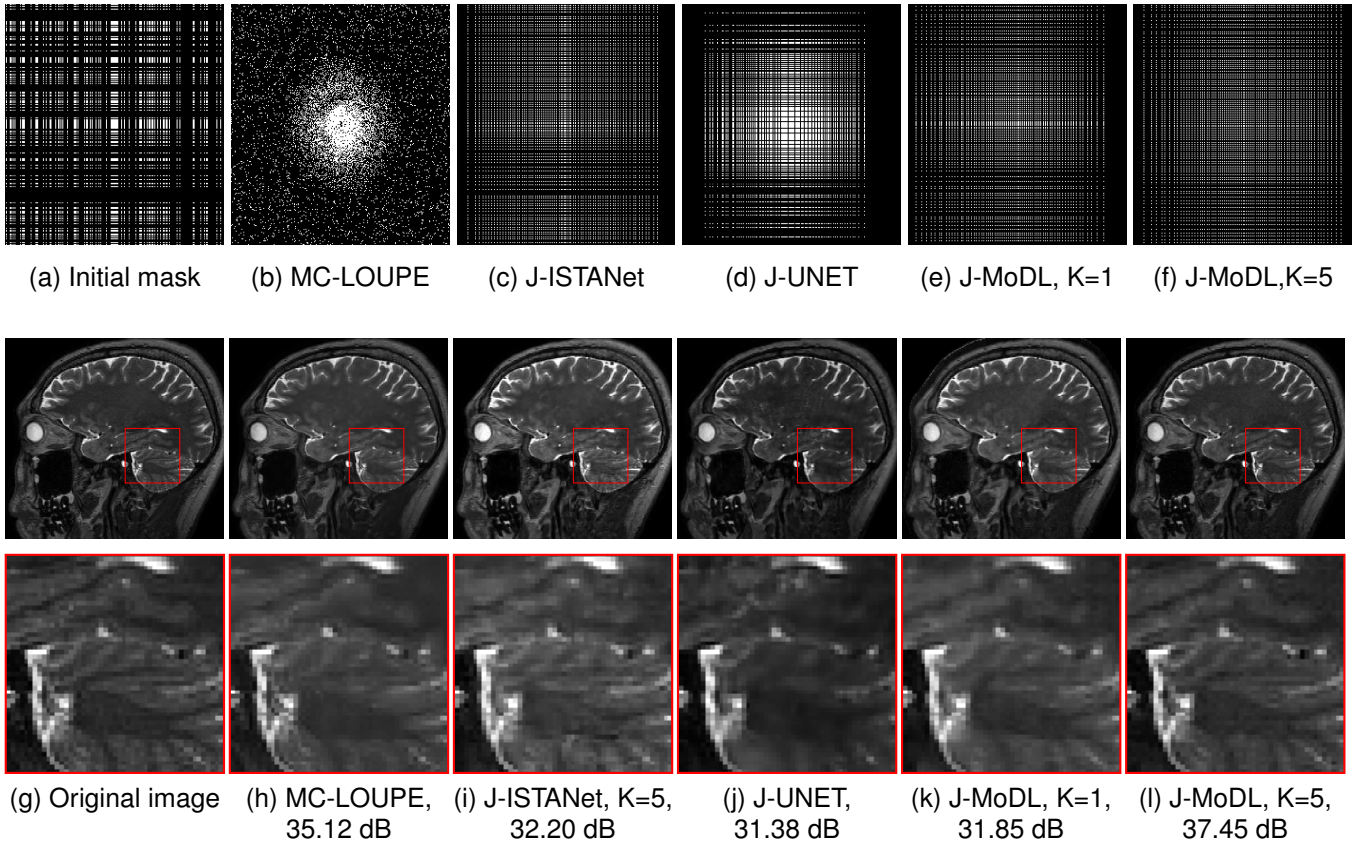
\begin{figure*}
  \input{fig_2d8x_loupe.tex}
  \caption{This figure shows comparative results between the proposed J-MoDL approach and existing algorithms at 8x acceleration in  multichannel~(MC)settings.  (a) shows the 8x sampling mask used for the initialization of the J-ISTANet, J-UNet, and J-MoDL approaches. Row~1 shows the learned masks by the respective algorithms. The learned masks by both J-MoDL approaches takes samples from all the locations in the k-space that help in preserving the high-frequency details in the reconstructed images as shown by the zoomed area near cerebellum. }
  \label{fig:2d8xloupe}
\end{figure*}

We compare the proposed J-MoDL scheme against other sampling pattern optimization methods. In particular, we compare J-MoDL against LOUPE \cite{loupe}, which is a discrete optimization strategy. We extended the original LOUPE algorithm to the multichannel setting to compare with the proposed scheme. We also study the proposed continuous sampling pattern optimization scheme for a range of network architectures, including ISTANet, UNET, and MoDL with one iteration.  These three methods were trained with identical initialization of the sampling mask as shown in Fig.~\ref{fig:2d8xloupe}(a).  See Fig.~\ref{fig:2d8xloupe} for the visual comparison of reconstructed images and learned masks. 

Table~\ref{tab:2d_loupe} summarizes the quantitative comparative results at 8x acceleration in the multichannel settings. The columns $\Phi$-alone denotes the network-alone optimization, when a variable density 2D pattern is used. The results of the joint optimization of the sampling pattern and network is shown in $\Theta~\Phi$ Joint columns. Since the LOUPE implementation available from the authors cannot be run without joint optimization, the results for the $\Phi$-alone case are not reported. We observe that the image quality of all methods improved significantly with the optimization of the sampling pattern.  
 	
 We note that all of the architectures in the above study have roughly the same number of trainable parameters. The ISTANet and MoDL ($K=5$) approaches  repeat the UNET five times, and hence have higher computational complexity over the UNET network and MoDL ($K=1$) network. We observe that MoDL ($K=1$) and UNET differ mostly in the addition of a data consistency step at the end. The results show that this gives around 3 dB improvement in performance during the network-alone training. We observe that the MoDL ($K=5$) network provided an additional 2dB improvement in performance over the one-iteration MoDL with a fixed sampling pattern, which is consistent with our earlier findings \cite{modl}. However, the performance improvement offered by this scheme with joint optimization is even more significant. 

\begin{table} \centering
\caption{ The average PSNR and SSIM values along with standard deviation are shown for different algorithms at 8x acceleration in the multichannel settings. The LOUPE algorithm is  for joint optimization therefore its results are not available~(NA) for the network alone ($\Phi$-alone) case.}
\label{tab:2d_loupe}
\input{tab_2d_loupe.tex}
\end{table}

\section{Discussion and Conclusion}
We introduced an approach for the joint optimization of the continuous sampling locations and the reconstruction network for parallel MRI reconstruction. Unlike past schemes, we consider a Fourier operator with continuously defined sampling locations, which facilitated the optimization of the sampling pattern without approximations. Our experiments show the benefit of the joint optimization strategy. We relied on a parametric sampling pattern with few parameters, which improved the convergence of the network with limited data. The experimental results demonstrate the significant benefits in the joint optimization of the sampling pattern in the proposed model-based framework. 

We note that the continuous optimization problem is highly non-convex with potentially many local minima and global minima. Specifically, any permutation of the optimal sampling pattern would be associated with the minimal cost. We note that similar symmetries do exist in the weights of neural networks. Fortunately, the stochastic gradient descent scheme is able to provide good solutions, despite the challenges in optimization.  

We note that the MoDL scheme relies on end-to-end training of the network parameters. This training approach is different from plug-and-play methods, where the network parameters are pre-trained. We refer the interested readers to~\cite{modl}, where the benefit of end-to-end training over pre-training is demonstrated. Similarly,~\cite{modl} also shows the benefit of using conjugate gradients in the data consistency blocks over steepest descent updates as in ISTANet. Further, a detailed comparisons between direct-inversion schemes and model-based schemes are covered in~\cite{modl}. We omit such comparisons in this work for brevity.  We note that the proposed sampling pattern optimization framework can also be utilized along with GAN-based reconstruction networks such as~\cite{gan_cyclic,dagan}.  

In this work, we used 10 iterations of the conjugate gradient in the data consistency step. Both the UNET and the J-MoDL unrolled for 10 iterations are trained on a 12 GB TitanV or any similar graphics card.     The offline training time for MoDL is almost 5 times longer than that of basic UNET. During inference, the basic UNET reconstructs 110 slices per second whereas MoDL reconstructs only 18 slices per second. We note that the UNET is around six times faster than the MoDL framework. However, we believe that the MoDL scheme is considerably faster than compressed sensing methods, and the improved image quality justifies its use in many applications.

We note that the MSE between the final reconstructions and original images was chosen as the loss function for both MoDL and UNET. However, we note that the final images in MoDL are obtained as the minimization of the cost function in Eq.~(\ref{modl}). Thus, one may view the MoDL training of the network parameters as consisting of two loss terms, one corresponding to the comparison in the image domain, and one corresponding to comparison with the  measured noisy samples. Since this training is more fine-tuned to the measurement process, the optimization is expected to yield improved results than the UNET  approach, which is confirmed by our experimental findings. We have reported  the standard deviation across slices in  our multichannel experiments. We understand that this might be an under-estimate since the slices across a single subject may be correlated. A larger study involving more testing subjects will be needed to address this issue

As discussed previously,  the joint optimization scheme with a 2-D non-parametric sampling pattern did not converge, possibly due to limited training data. In our future work, we will study the possibility of 2-D non-parametric sampling with more training data. In this work, we constrained the sampling pattern as the tensor product of two 1-D sampling patterns and optimized for the encoding locations. This approach reduced the number of trainable parameters, thus significantly improving the convergence of the algorithm over non-parametric strategies. We note that several alternate approaches may be used to achieve similar goals. For instance, one may search and pick a variable density pattern that yields the best performance from several randomly selected variable patterns. However, since the network has to be trained for each pattern, the anticipated training time is expected to be high.  We note that constraining the sampling pattern as the tensor product is a limitation of this work. Optimizing the parameters of a truly 2-D parametric sampling pattern such as \cite{sparkling} may improve the performance. We will explore these ideas systematically in our future work.

\bibliographystyle{IEEEtran}
\balance
\input{main.bbl}

%\bibliography{IEEEabrv,bibTexSamp}

\end{document}

%% file: fig_sampling.tex
  \centering
  \subfloat[sampling 1-D]{
  \includegraphics[width=.7\linewidth]{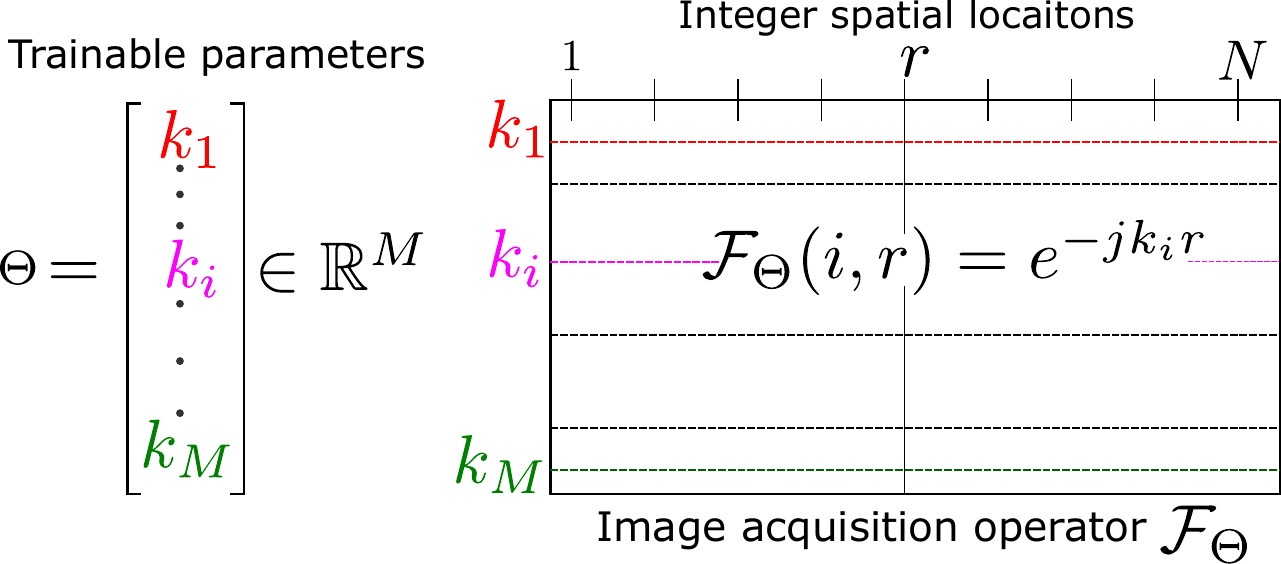}
}

  \subfloat[sampling 2-D]{
    \includegraphics[width=.9\linewidth]{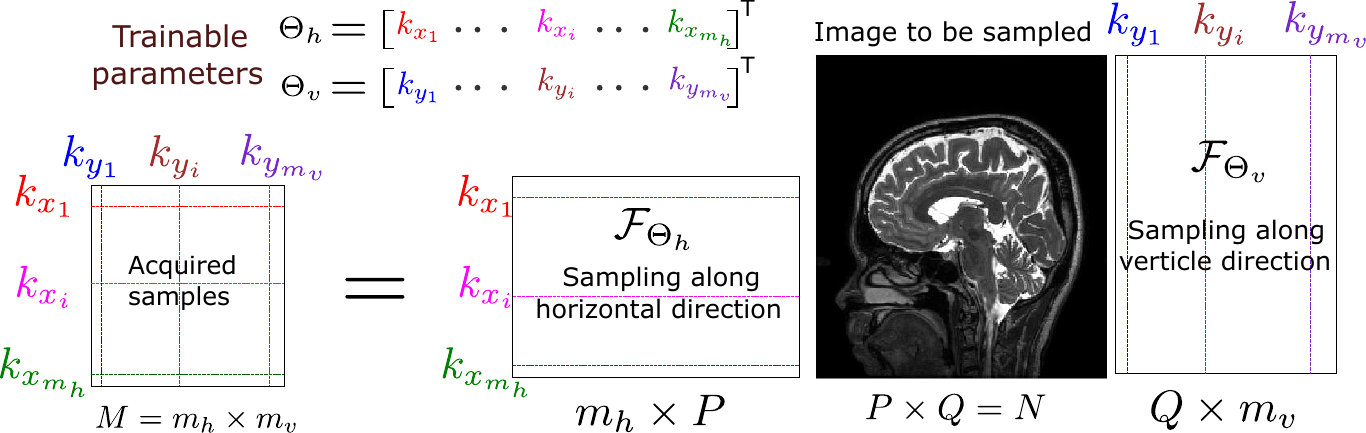}    
  }

%%% Local Variables: 
%%% mode: latex
%%% TeX-master: "hk"
%%% End: 

%% file: tab_1d_sc.tex
\begin{tabular}{l@{\hskip 0.05cm}|c@{\hskip 0.1cm}|c@{\hskip 0.1cm}|c@{\hskip 0.1cm}|c} \toprule
      & \multicolumn{2}{c|}{PSNR} & \multicolumn{2}{c}{SSIM} \\ \midrule
 Optimize        & UNET        & MoDL       & UNET        & MoDL       \\ \midrule
$\Phi$-alone        & $28.65 \pm 1.14$        & $30.65 \pm 1.43$      & $0.80 \pm 0.03$        & $0.82 \pm 0.04$       \\ 
 $\Theta$-alone      & $29.02 \pm 1.03$       & $32.46 \pm 1.07$     &  $0.80 \pm 0.03$       & $0.84 \pm 0.03$       \\
 Joint  & $29.70 \pm 1.06$       & $33.78 \pm 1.13$      & $0.82 \pm 0.03$        & $0.87 \pm 0.03$ \\

  \bottomrule      
\end{tabular}

% \begin{tabular}{l|cc|cc} \toprule
%       & \multicolumn{2}{c|}{PSNR} & \multicolumn{2}{c}{SSIM} \\ \midrule
%  Optimize        & UNET        & MoDL       & UNET        & MoDL       \\ \midrule
% $\Phi $ alone        & 30.00       & 33.42      & 0.84        & 0.85       \\ 
%  $\Theta$ alone      & 25.40       & 35.03      & .071        & 0.89       \\
%    Greedy             & --          & 36.23      & --          & 0.87 \\
% $\Theta,\Phi$ Joint  & 30.61       & 35.69      & 0.87        & 0.90 \\

%   \bottomrule      
% \end{tabular}

%%% Local Variables:
%%% mode: latex
%%% TeX-master: "hk"
%%% End:

%% file: fig_1d_single_channel.tex
{  \captionsetup[subfigure]{justification=centering}
  \centering
  \newcommand{\sz}{.24\linewidth} 
  \newcommand{\spt}{(0.15,-0.1)} 
  \newcommand{\pt}{(0,-2.2)} 
  \newcommand{\mg}{6}
\pgfdeclarelayer{background}
\pgfdeclarelayer{foreground}
\pgfsetlayers{background,main,foreground} 
\subfloat[Original image]{%
 \begin{tikzpicture}[spy using outlines={rectangle,red,magnification=\mg,size=\sz}]
    \node {
        \includegraphics[ width=\sz]{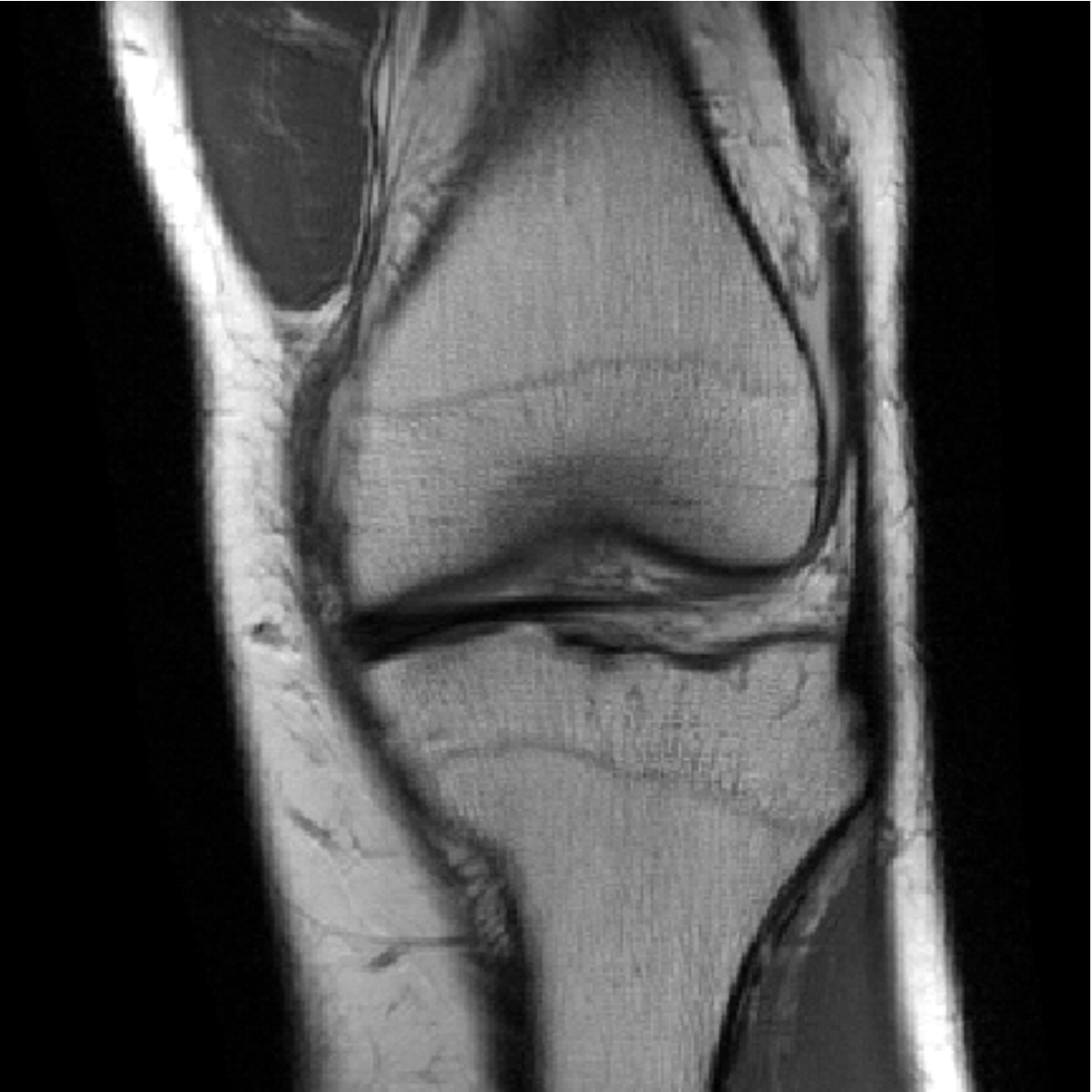}
      };
      \spy on \spt in node at \pt;
      \begin{pgfonlayer}{foreground}
   \draw [red,-triangle 45] (-.3,-1.5) -- (-.3,-2.05);
      \end{pgfonlayer}{foreground}
    \end{tikzpicture}%
  } \hspace{-.28cm}
  \subfloat[MoDL, 30.75 dB]{%
    \begin{tikzpicture}[spy using outlines={rectangle,red,magnification=\mg,size=\sz}]
      \node {
        \includegraphics[width=\sz]{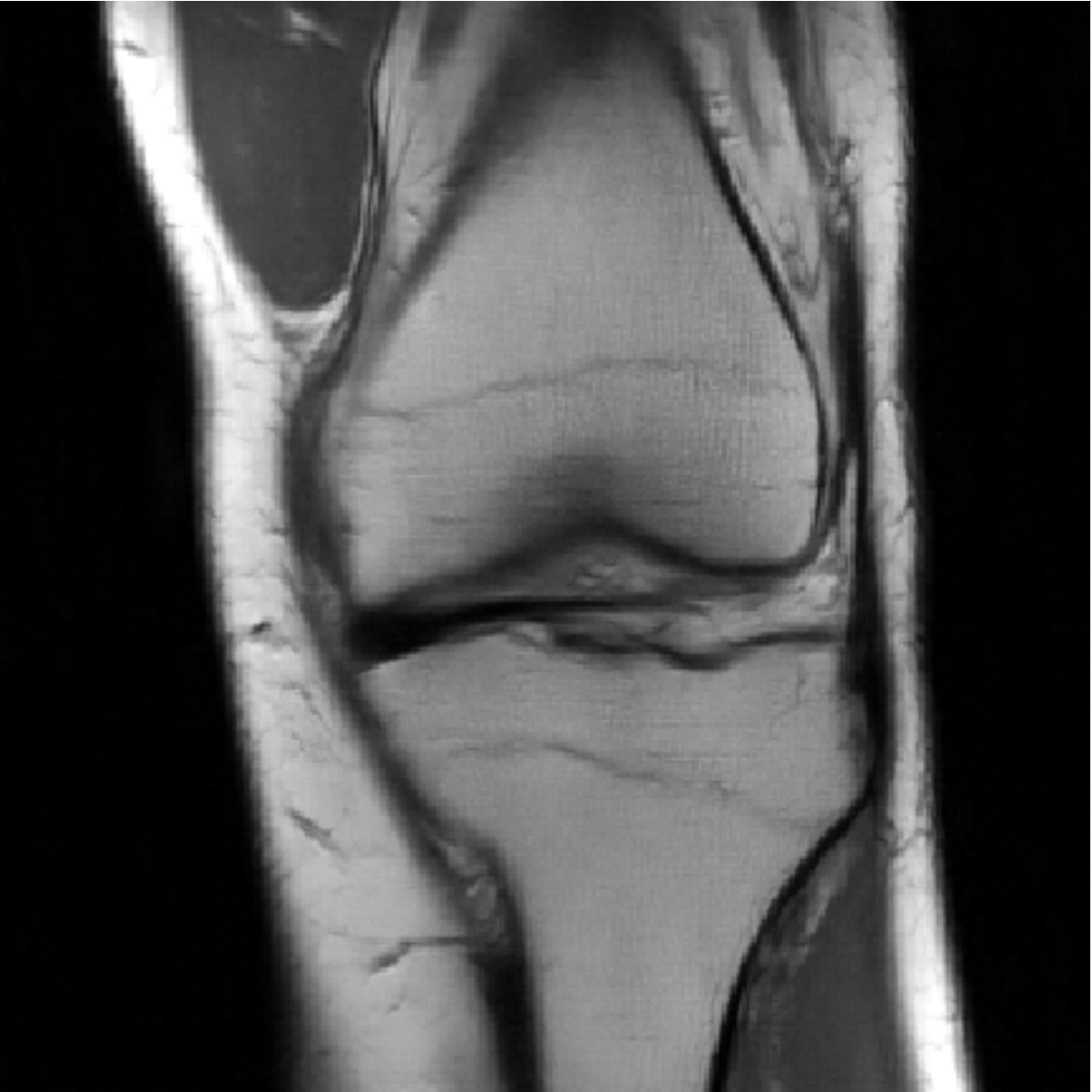}%
      };
      \spy on \spt in node at \pt;
      \begin{pgfonlayer}{foreground}
   \draw [red,-triangle 45] (-.3,-1.5) -- (-.3,-2.05);
      \end{pgfonlayer}{foreground}
    \end{tikzpicture}%
  }\hspace{-.28cm}
  \subfloat[J-UNET, 29.69 dB]{%
    \begin{tikzpicture}[spy using outlines={rectangle,red,magnification=\mg,size=\sz}]
      \node {
        \includegraphics[width=\sz]{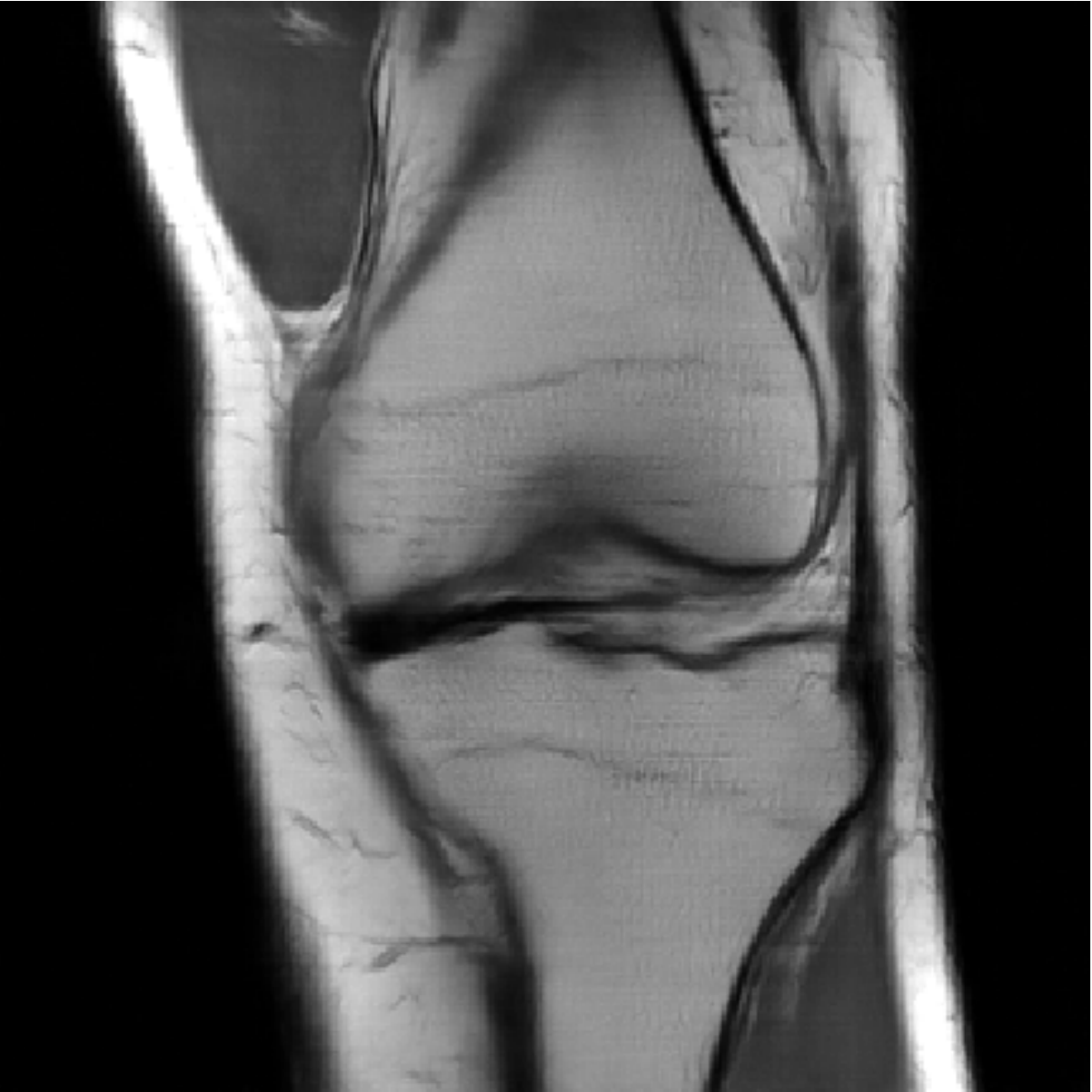}%
      };
      \spy on \spt in node at \pt;
      \begin{pgfonlayer}{foreground}
   \draw [red,-triangle 45] (-.3,-1.5) -- (-.3,-2.05);
      \end{pgfonlayer}{foreground}
    \end{tikzpicture}%
  }\hspace{-.28cm}
  \subfloat[J-MoDL, 34.19 dB]{%
    \begin{tikzpicture}[spy using outlines={rectangle,red,magnification=\mg,size=\sz}]
      \node {
        \includegraphics[width=\sz]{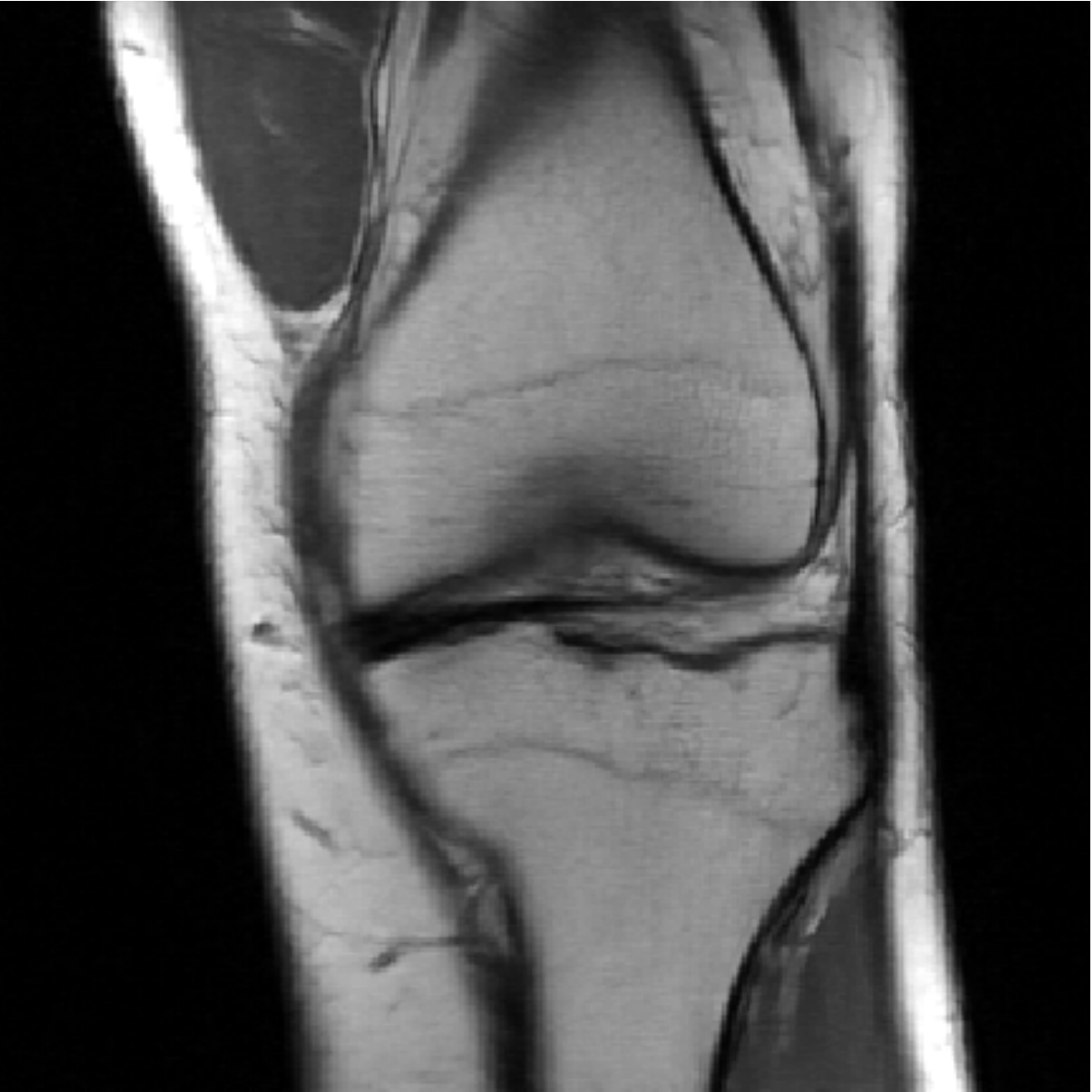}
      };
      \spy on \spt in node at \pt;
\begin{pgfonlayer}{foreground}
   \draw [red,-triangle 45] (-.3,-1.5) -- (-.3,-2.05);
      \end{pgfonlayer}{foreground}
    \end{tikzpicture}%
}%
}

%%% Local Variables:
%%% mode: latex
%%% TeX-master: "hk"
%%% End:

%% file: tab_1d_mc.tex
\begin{tabular}{l@{\hskip 0.05cm}|c@{\hskip 0.05cm}|c|c@{\hskip 0.05cm}|c} \toprule
  & \multicolumn{2}{c|}{PSNR} & \multicolumn{2}{c}{SSIM} \\ \midrule
       & UNET  & MoDL  & UNET & MoDL \\ \midrule
  Optimize & \multicolumn{4}{c}{4x acceleration} \\ \midrule
 $\Phi $-alone& $29.95 \pm 3.76 $ & $34.21 \pm 3.14$& $0.83 \pm 0.13$ & $0.91 \pm 0.04$\\ 
                    $\Theta$-alone& $28.85 \pm 3.94 $ & $37.66 \pm 3.30$& $0.86 \pm 0.04$ & $0.96 \pm 0.03$\\ 
  Joint & $34.02 \pm 3.31 $ & $41.28 \pm 3.07$& $0.93 \pm 0.04$ & $0.96 \pm 0.02$ \\ \midrule
  & \multicolumn{4}{c}{6x acceleration} \\ \midrule
 $\Phi $-alone& $29.24 \pm 3.94 $ & $32.40 \pm 3.00$& $0.82 \pm 0.13$ & $0.89 \pm 0.04$ \\ 
                   $\Theta$-alone & $24.45 \pm 3.65 $ & $33.31 \pm 3.17$& $0.78 \pm 0.09$ & $0.93\pm 0.03$ \\ 
              Joint & $29.62 \pm 2.54 $ & $35.93 \pm 2.74$& $0.89 \pm 0.05$ & $0.93\pm 0.03$ \\
  \bottomrule      
\end{tabular}

%%% Local Variables:
%%% mode: latex
%%% TeX-master: "hk"
%%% End:

%% file: fig_1d4x_multi_channel.tex
{  \captionsetup[subfigure]{justification=centering}
  \centering
  \newcommand{\sz}{.24\linewidth} 
  \newcommand{\spt}{(0.1,-0.22)} 
  \newcommand{\pt}{(0,-2.2)} 
  \newcommand{\mg}{7}
\pgfdeclarelayer{background}
\pgfdeclarelayer{foreground}
\pgfsetlayers{background,main,foreground} 
\subfloat[Original image]{%
 \begin{tikzpicture}[spy using outlines={rectangle,red,magnification=\mg,size=\sz}]
    \node {
        \includegraphics[ width=\sz]{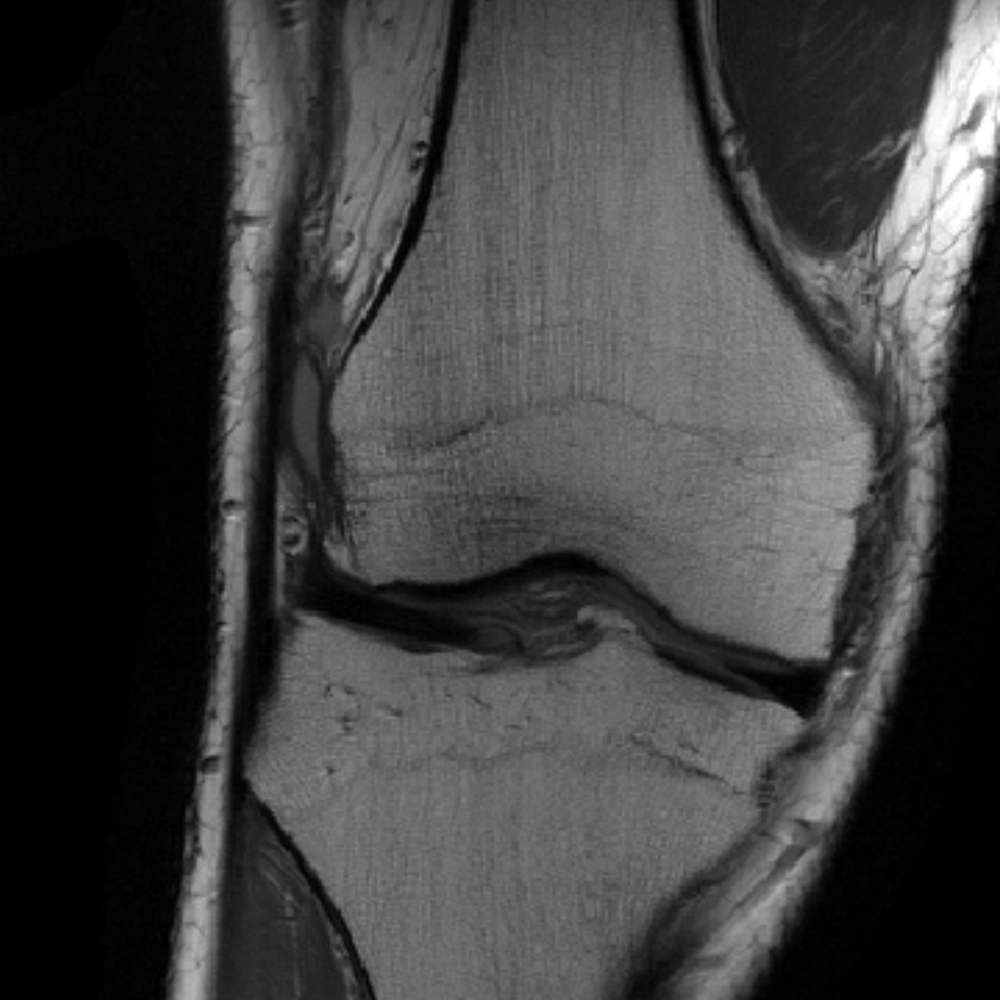}
      };
      \spy on \spt in node at \pt;
    \end{tikzpicture}%
  } \hspace{-.28cm}
  \subfloat[MoDL, 32.77 dB]{%
    \begin{tikzpicture}[spy using outlines={rectangle,red,magnification=\mg,size=\sz}]
      \node {
        \includegraphics[width=\sz]{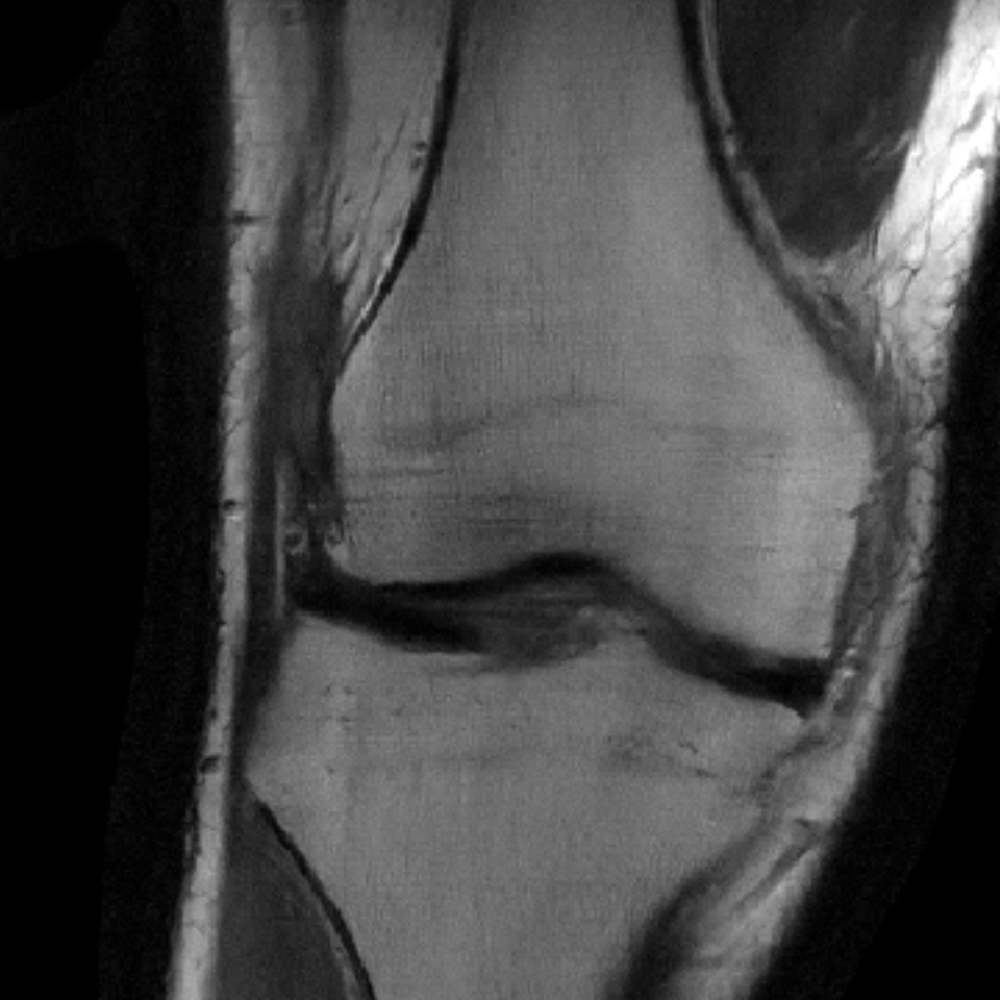}%
      };
      \spy on \spt in node at \pt;
    \end{tikzpicture}%
  }\hspace{-.28cm}
  \subfloat[J-UNET, 34.98 dB]{%
    \begin{tikzpicture}[spy using outlines={rectangle,red,magnification=\mg,size=\sz}]
      \node {
        \includegraphics[width=\sz]{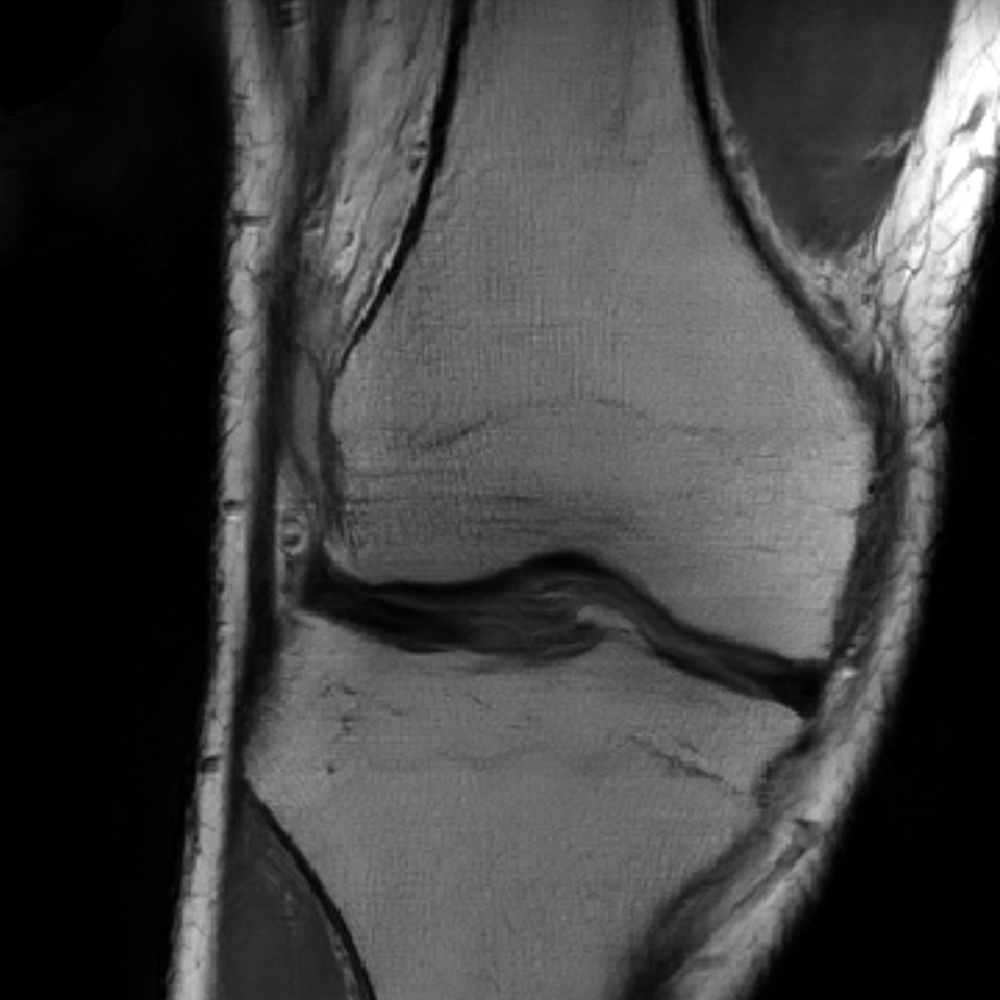}%
      };
      \spy on \spt in node at \pt;
    \end{tikzpicture}%
  }\hspace{-.28cm}
  \subfloat[J-MoDL, 40.76 dB]{%
    \begin{tikzpicture}[spy using outlines={rectangle,red,magnification=\mg,size=\sz}]
      \node {
        \includegraphics[width=\sz]{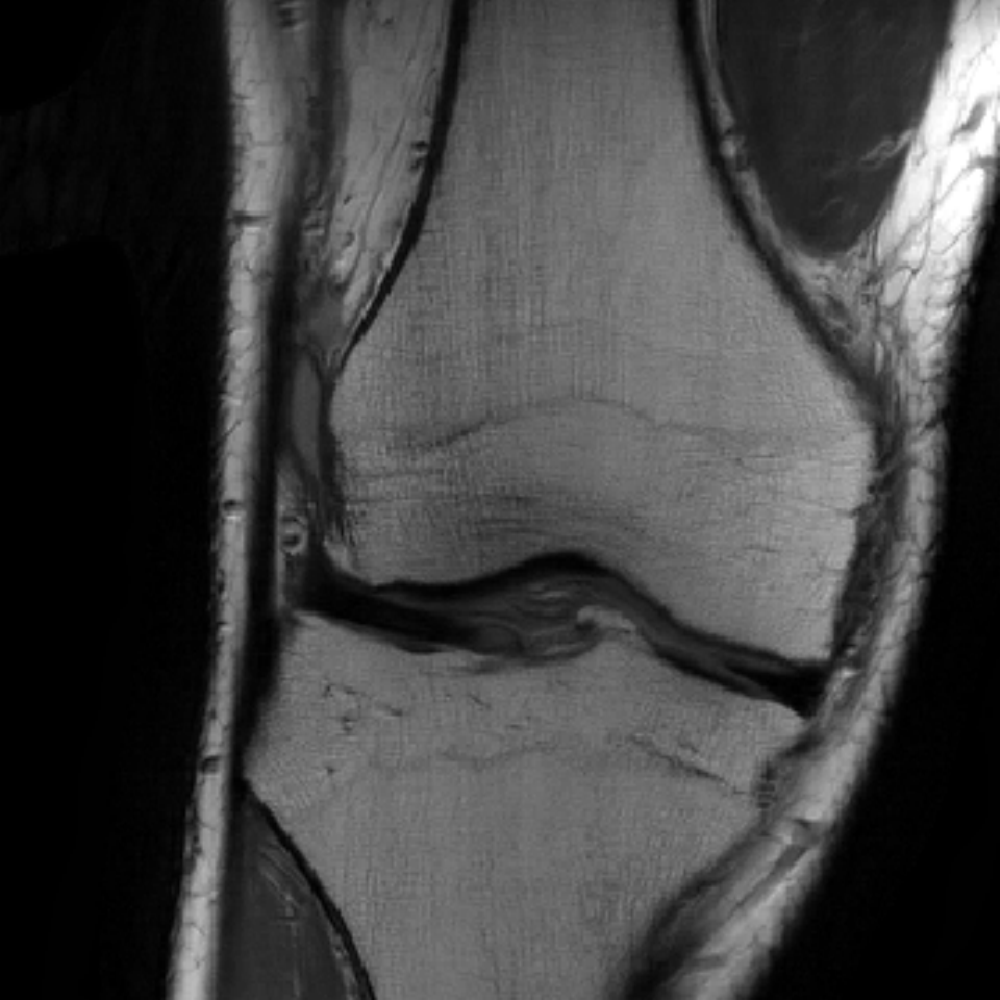}
      };
      \spy on \spt in node at \pt;
    \end{tikzpicture}%
}%
}

%%% Local Variables:
%%% mode: latex
%%% TeX-master: "hk"
%%% End:

%% file: fig_landscape.tex
\centering
  \includegraphics[width=.95\linewidth]{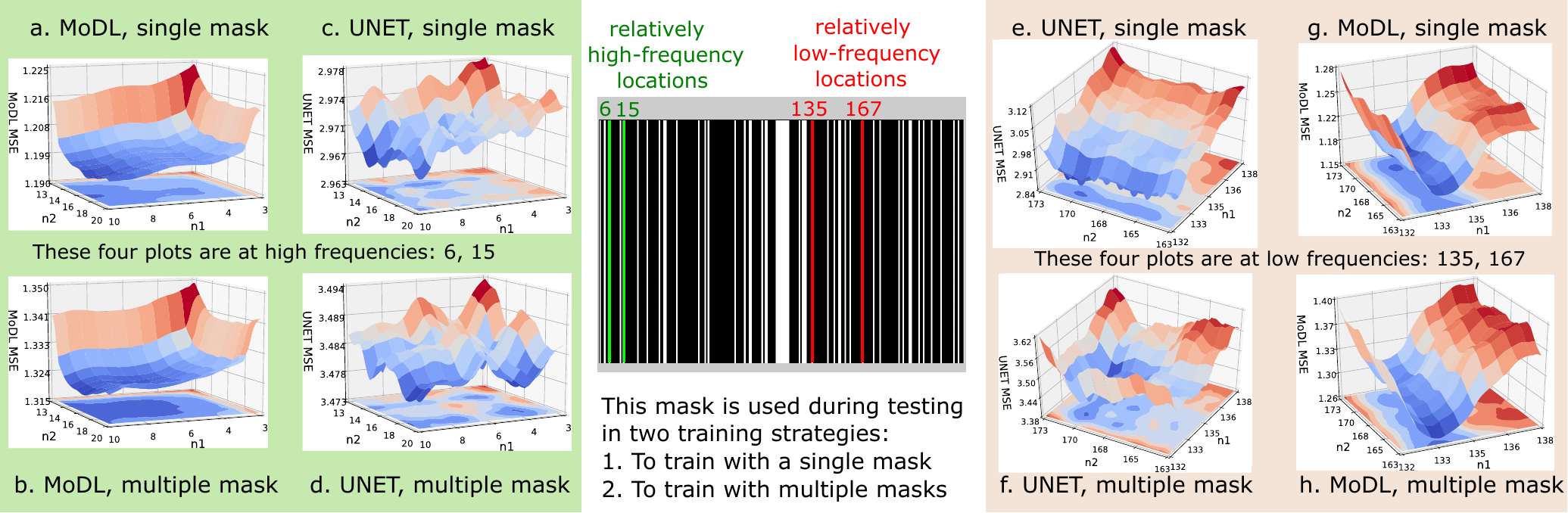}
% \subfloat[MoDL high-freq]{
%   \includegraphics[width=.19\linewidth]{modl_high}
% }
%   \subfloat[UNet high-freq]{
%     \includegraphics[width=.19\linewidth]{unet_high}    
%   }
% \subfloat[Mask]{
%   \includegraphics[width=.19\linewidth]{mskLandscape}
% }
%   \subfloat[MoDL low-freq]{
%   \includegraphics[width=.19\linewidth]{modl_low}
% }
%   \subfloat[UNet low-freq]{
%     \includegraphics[width=.19\linewidth]{unet_low}    
%   }

%%% Local Variables: 
%%% mode: latex
%%% TeX-master: "cleanLatest"
%%% End: 

%% file: tab_2d_mc.tex
%\begin{tabular}{l|cc|cc} \toprule
\begin{tabular}{c@{\hskip 0.05cm}|c@{\hskip 0.1cm}|c@{\hskip 0.1cm}|c@{\hskip 0.1cm}|c} \toprule

  & \multicolumn{2}{c|}{PSNR} & \multicolumn{2}{c}{SSIM} \\ \midrule
  Optimize     & UNET  & MoDL  & UNET & MoDL \\ \midrule
 $\Phi $-alone& $27.34 \pm 1.14$ & $34.19 \pm 1.03$ & $0.82 \pm 0.02$ & $0.94 \pm 0.01$ \\ 
$\Theta$-alone& $28.56 \pm 0.93$ & $37.47 \pm 0.57$ & $0.85 \pm 0.02$ & $0.94 \pm 0.01$ \\ 
 Joint & $34.31 \pm 0.81$ & $37.60 \pm 0.56$ & $0.94 \pm 0.01$ & $0.96 \pm 0.01$ \\   \bottomrule      
\end{tabular}

%%% Local Variables:
%%% mode: latex
%%% TeX-master: "hk"
%%% End:

%% file: fig_2d6x_mc.tex
{  \captionsetup[subfigure]{justification=centering,labelformat=empty}
  \centering
  \newcommand{\sz}{.13\linewidth}
  \newcommand{\sm}{-.24cm}
  \newcommand{\spt}{(0.47,-0.34)} 
  \newcommand{\pt}{(0,-2.3)} 
  \newcommand{\mg}{4}
\pgfdeclarelayer{background}
\pgfdeclarelayer{foreground}
\pgfsetlayers{background,main,foreground} 
\subfloat[]{%
 \begin{tikzpicture}[spy using outlines={rectangle,red,magnification=\mg,size=\sz}]
    \node { 
        \includegraphics[ width=\sz]{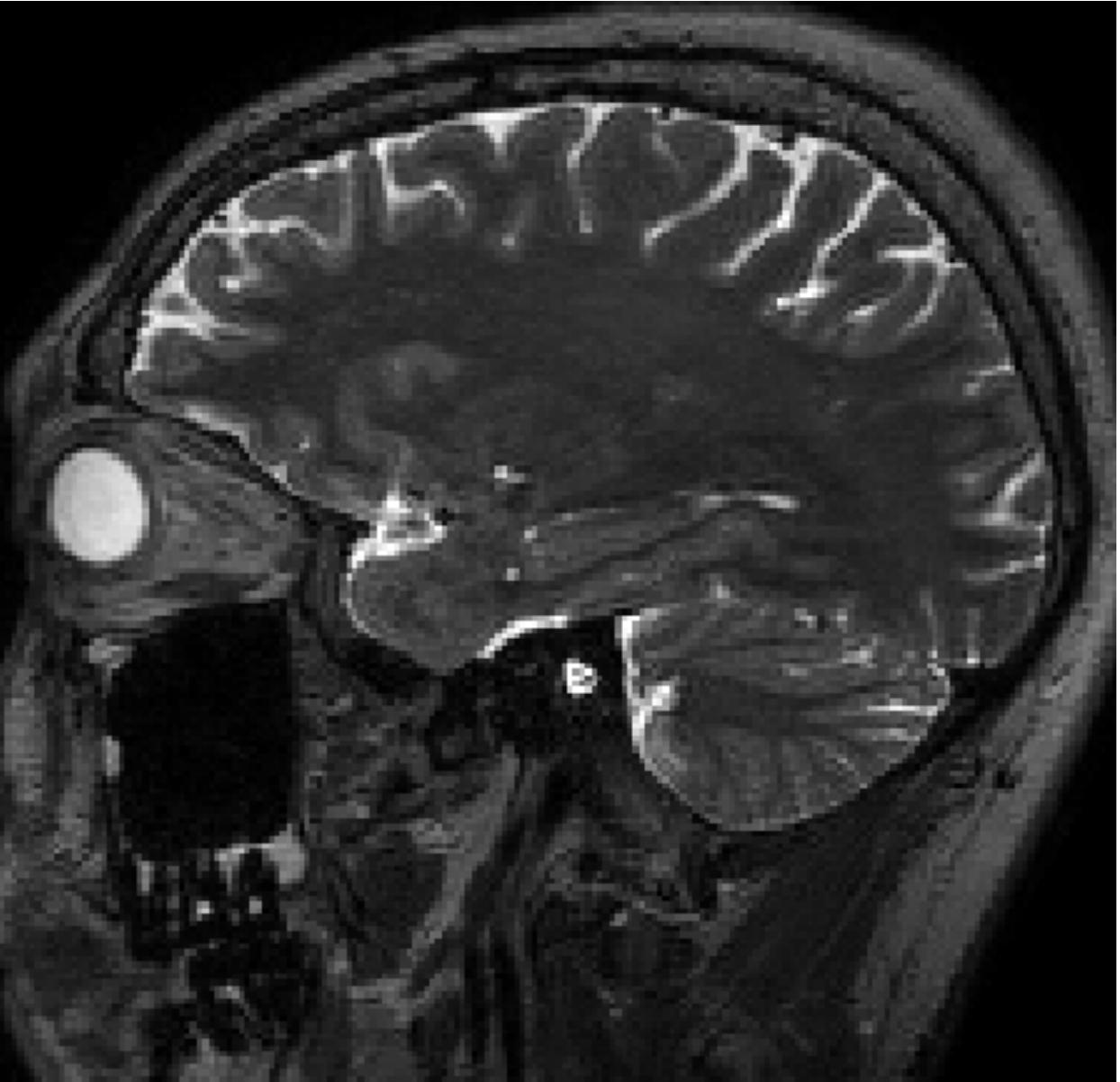}
      };
      \spy on \spt in node at \pt;
  \begin{pgfonlayer}{foreground}
%    \draw [red,-triangle 45] (-.2,-2.4) -- (-.2,-1.9);
    \draw [red,-triangle 45] (-.2,-2.1) -- (-.42,-2.7);
      \end{pgfonlayer}{foreground}%
    \end{tikzpicture}%
  }\hspace{\sm}
  \subfloat[]{%
    \begin{tikzpicture}[spy using outlines={rectangle,red,magnification=\mg,size=\sz}]%
      \node {%
        \includegraphics[width=\sz]{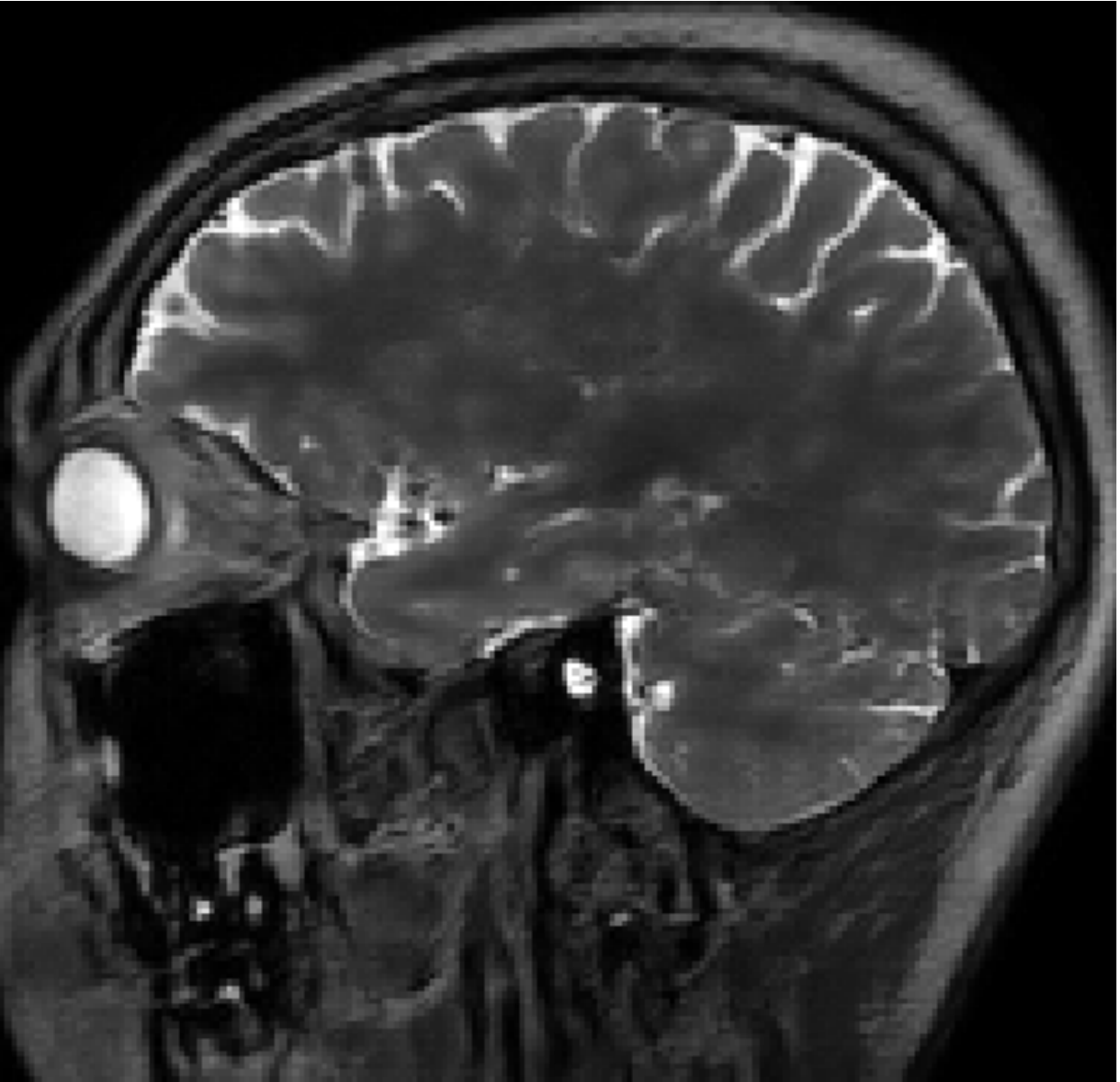}%
      };%
      \spy on \spt in node at \pt;%
\begin{pgfonlayer}{foreground}%
%  \draw [red,-triangle 45] (-.2,-2.4) -- (-.2,-1.9);%
    \draw [red,-triangle 45] (-.2,-2.1) -- (-.42,-2.7);
      \end{pgfonlayer}{foreground}%
    \end{tikzpicture}%
  } \hspace{\sm}
  \subfloat[]{%
    \begin{tikzpicture}[spy using outlines={rectangle,red,magnification=\mg,size=\sz}]
      \node {
        \includegraphics[width=\sz]{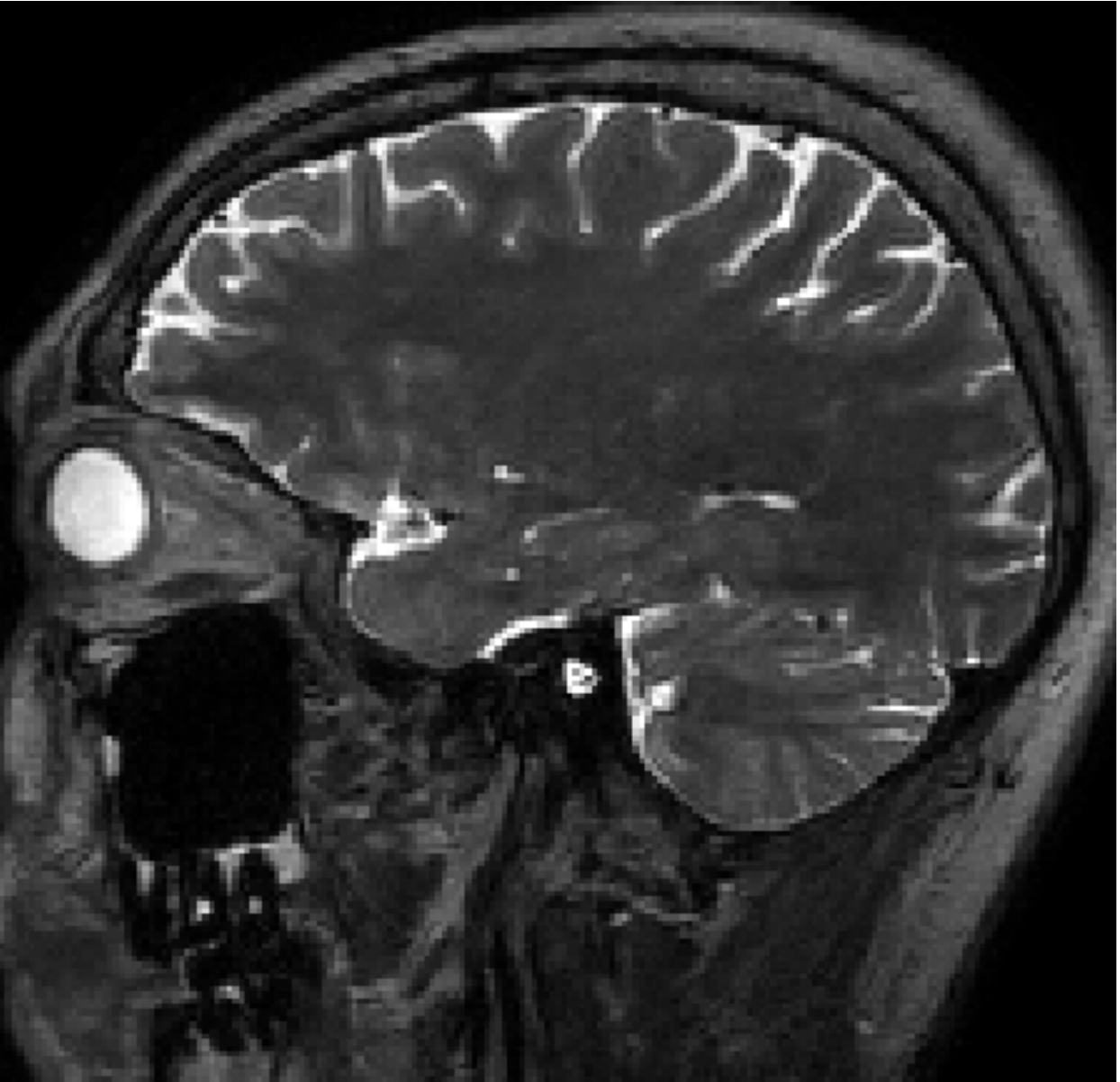}%
      };
      \spy on \spt in node at \pt;
  \begin{pgfonlayer}{foreground}
%    \draw [red,-triangle 45] (-.2,-2.4) -- (-.2,-1.9);
        \draw [red,-triangle 45] (-.2,-2.1) -- (-.42,-2.7);
      \end{pgfonlayer}{foreground}
  \end{tikzpicture}%
  }\hspace{\sm}
  \subfloat[]{%
    \begin{tikzpicture}[spy using outlines={rectangle,red,magnification=\mg,size=\sz}]
      \node {
        \includegraphics[width=\sz]{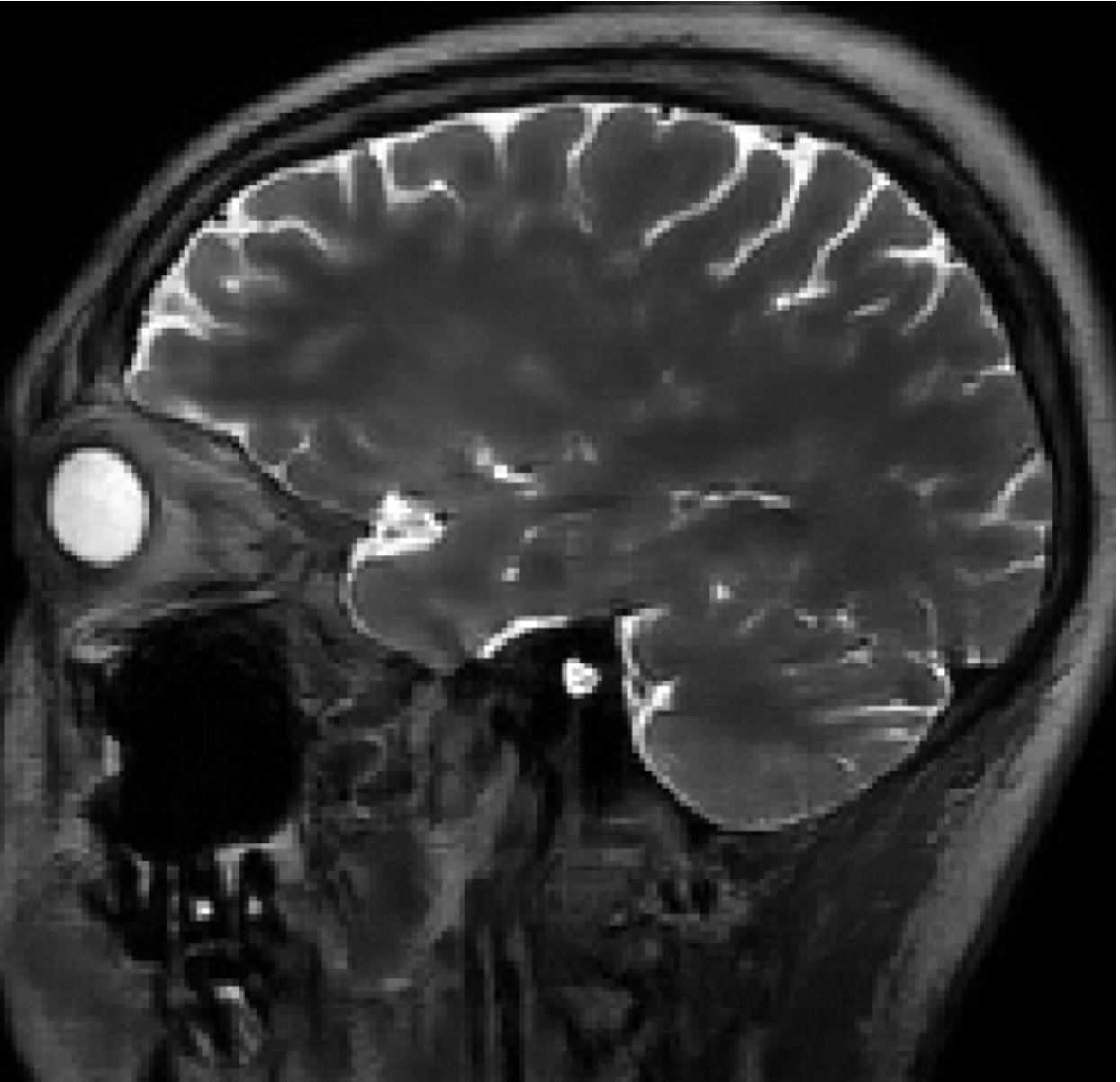}%
      };
      \spy on \spt in node at \pt;
  \begin{pgfonlayer}{foreground}
%    \draw [red,-triangle 45] (-.2,-2.4) -- (-.2,-1.9);
        \draw [red,-triangle 45] (-.2,-2.1) -- (-.42,-2.7);
      \end{pgfonlayer}{foreground}
  \end{tikzpicture}%
  }\hspace{\sm}
  \subfloat[]{%
    \begin{tikzpicture}[spy using outlines={rectangle,red,magnification=\mg,size=\sz}]
      \node {
        \includegraphics[width=\sz]{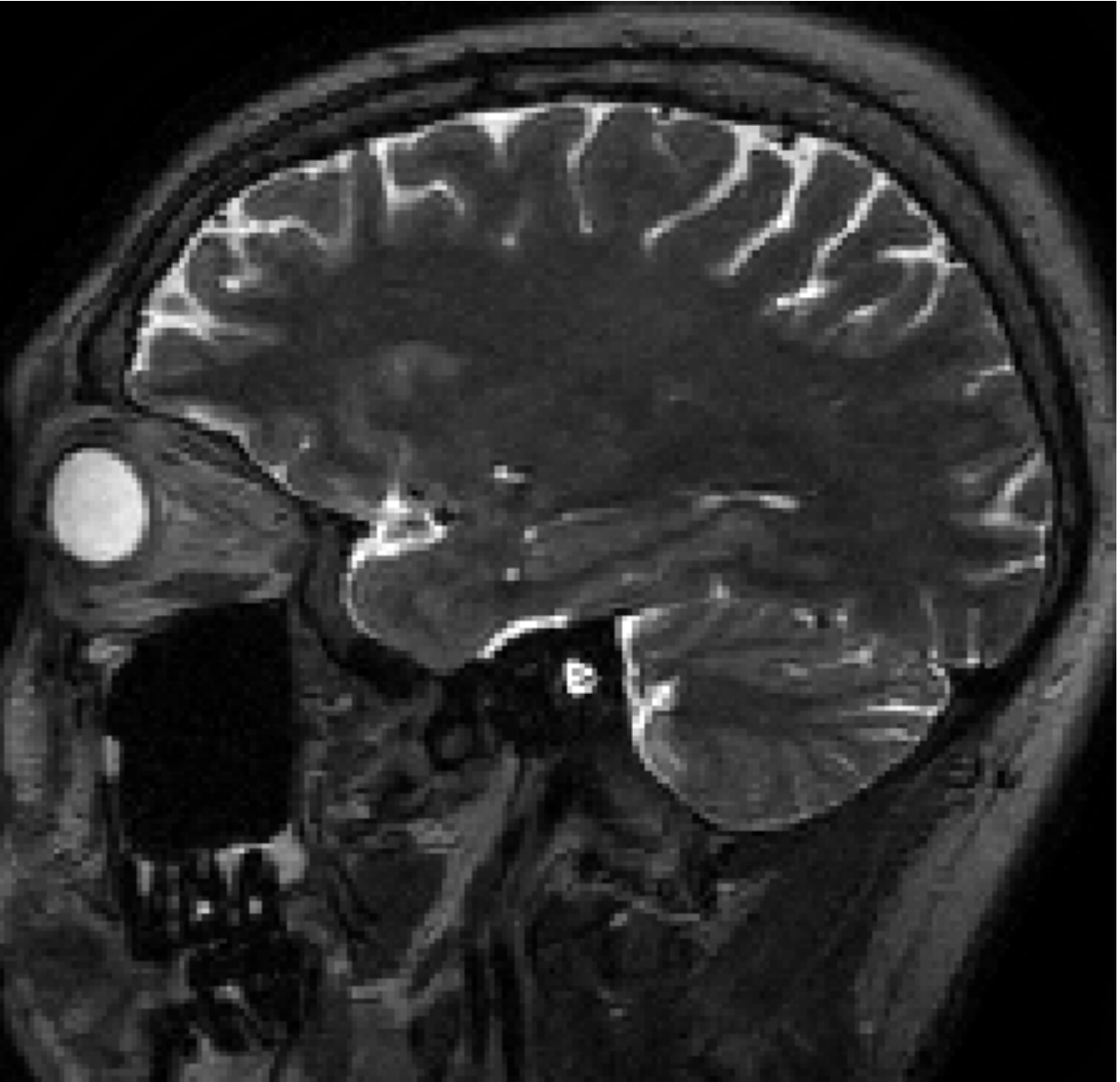}%
      };
      \spy on \spt in node at \pt;
  \begin{pgfonlayer}{foreground}
%    \draw [red,-triangle 45] (-.2,-2.4) -- (-.2,-1.9);
        \draw [red,-triangle 45] (-.2,-2.1) -- (-.42,-2.7);
      \end{pgfonlayer}{foreground}
  \end{tikzpicture}%
  }\hspace{\sm}
  \subfloat[]{%
    \begin{tikzpicture}[spy using outlines={rectangle,red,magnification=\mg,size=\sz}]
      \node {
        \includegraphics[width=\sz]{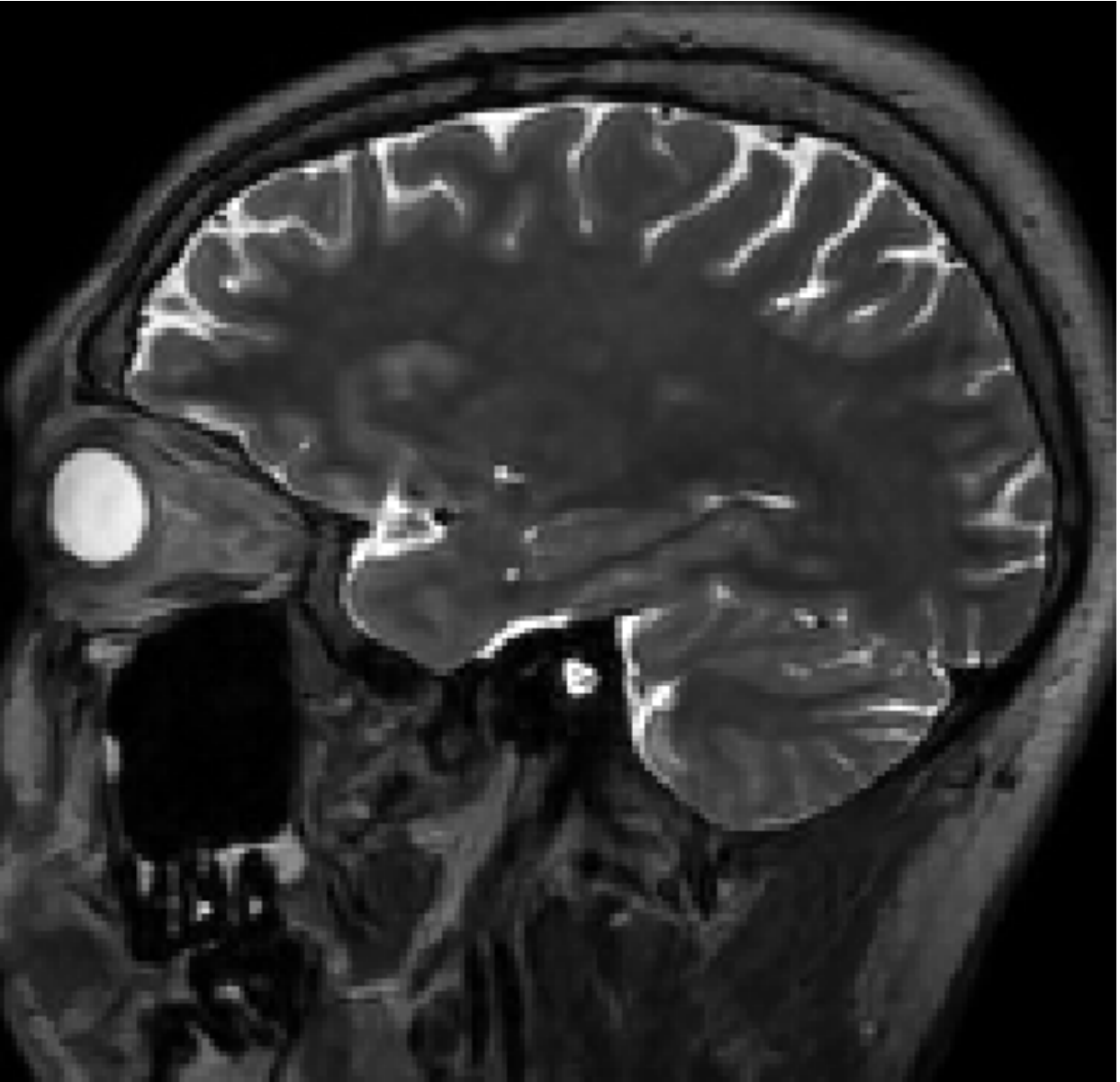}%
      };
      \spy on \spt in node at \pt;
  \begin{pgfonlayer}{foreground}
%    \draw [red,-triangle 45] (-.2,-2.4) -- (-.2,-1.9);
    \draw [red,-triangle 45] (-.2,-2.1) -- (-.42,-2.7);
      \end{pgfonlayer}{foreground}
  \end{tikzpicture}%
  }\hspace{\sm}
  \subfloat[]{%
    \begin{tikzpicture}[spy using outlines={rectangle,red,magnification=\mg,size=\sz}]
      \node {
        \includegraphics[width=\sz]{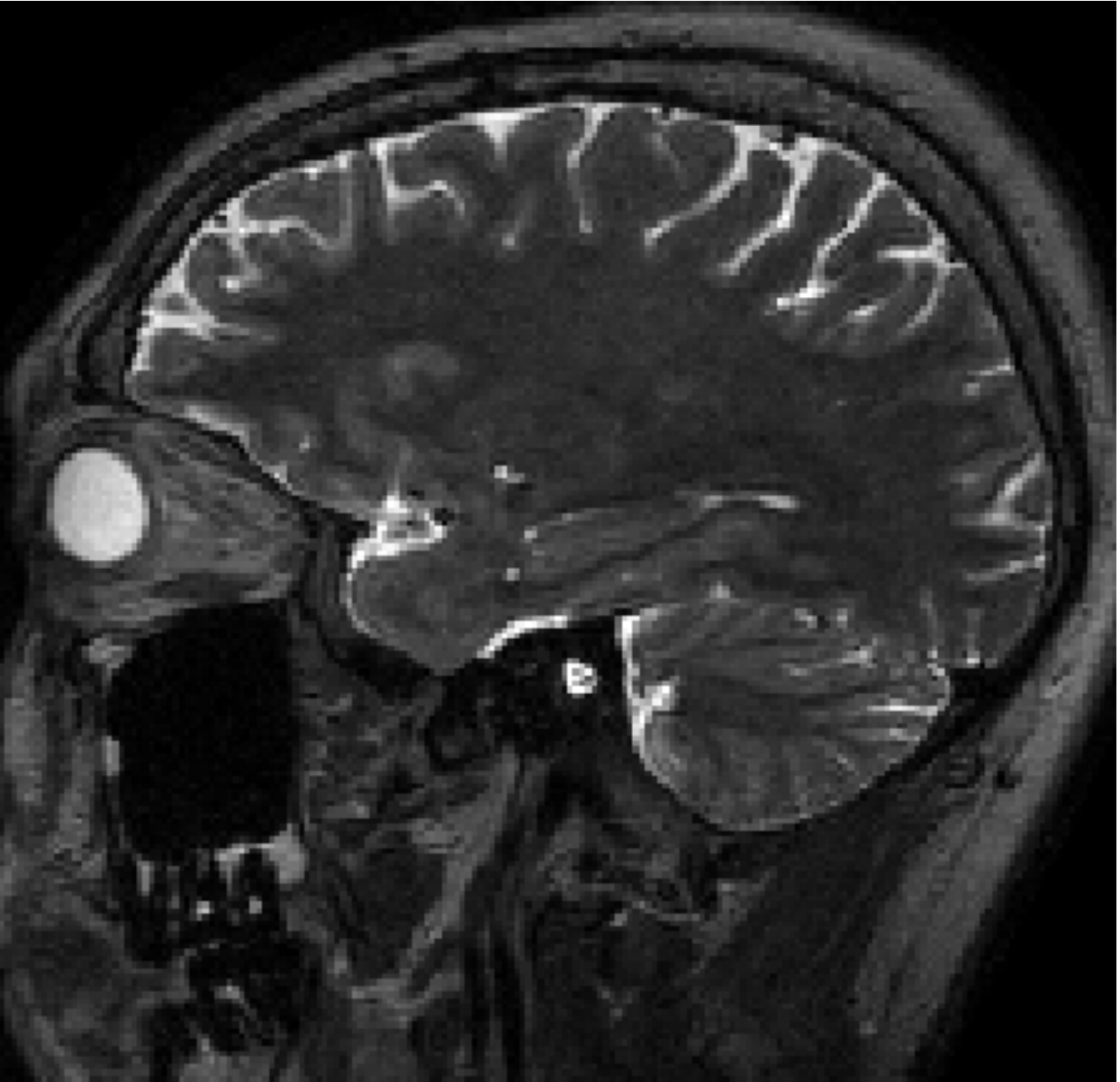}
      };
      \spy on \spt in node at \pt;
\begin{pgfonlayer}{foreground}
%  \draw [red,-triangle 45] (-.2,-2.4) -- (-.2,-1.9);
    \draw [red,-triangle 45] (-.2,-2.1) -- (-.42,-2.7);
      \end{pgfonlayer}{foreground}
    \end{tikzpicture}%
}%

\vspace{-.8cm}
\subfloat[a. Original]{
  \includegraphics[width=\sz]{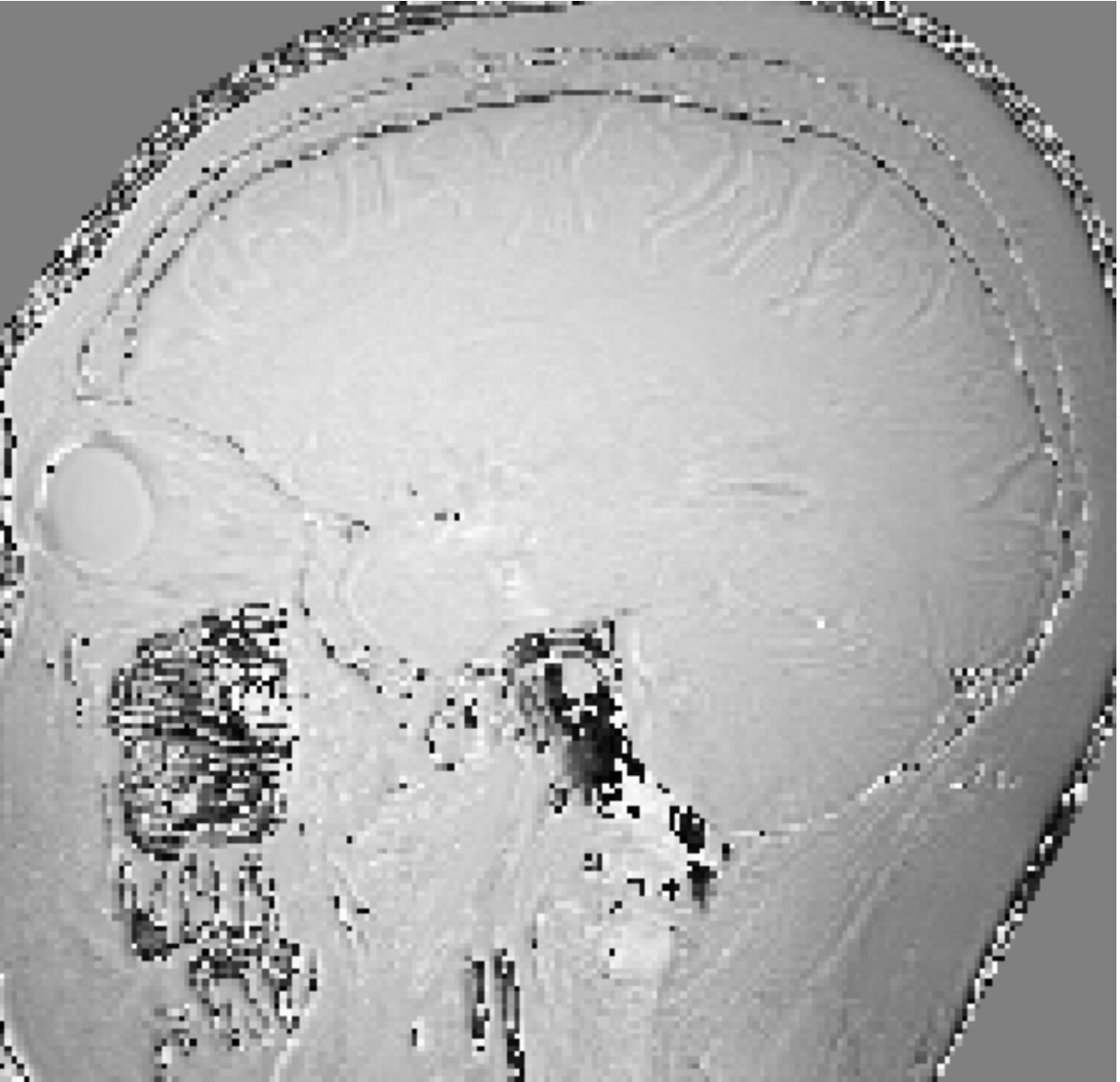}
}
\subfloat[]{%
  \includegraphics[width=\sz]{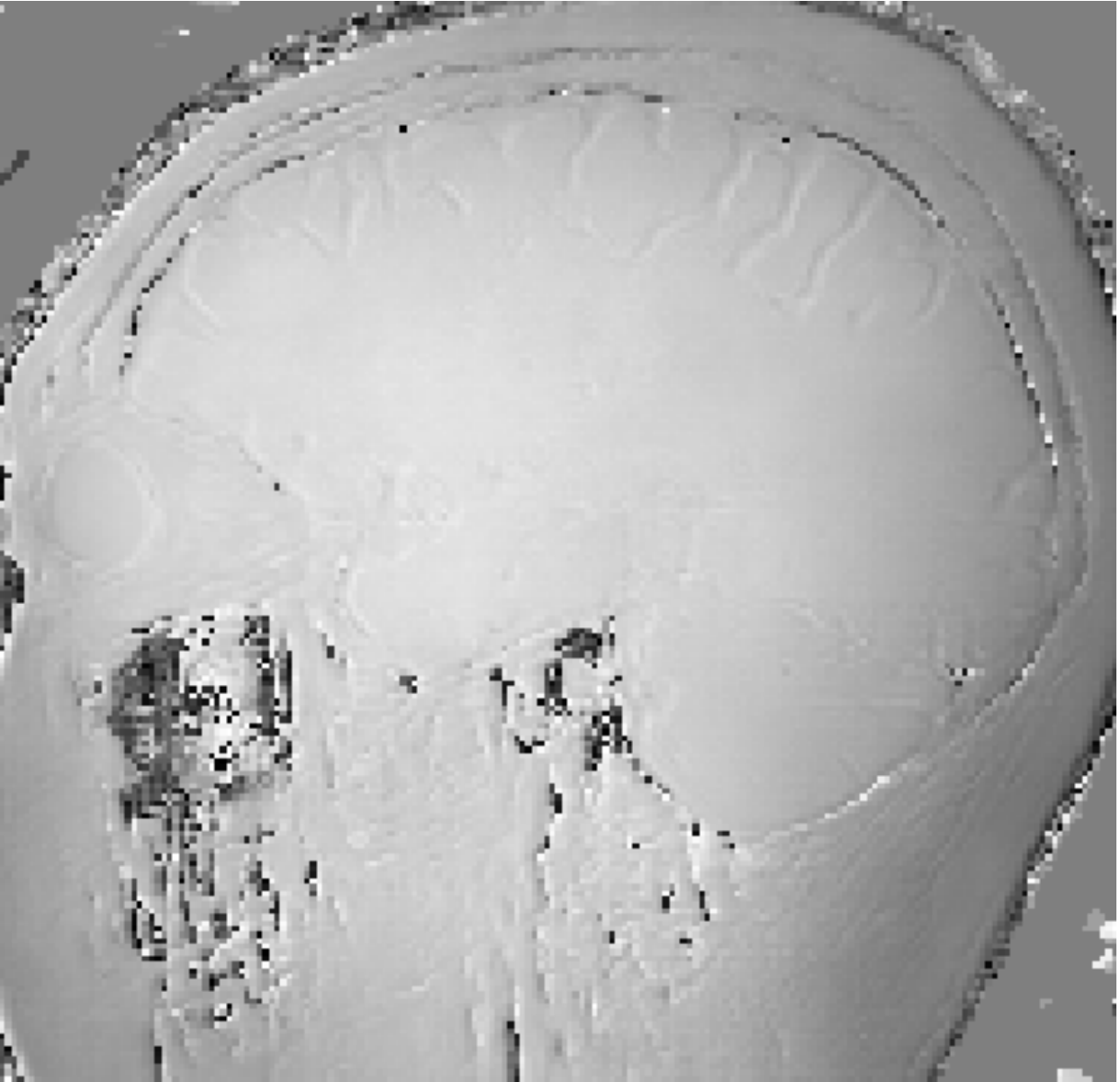}
}%
\subfloat[]{%
  \includegraphics[width=\sz]{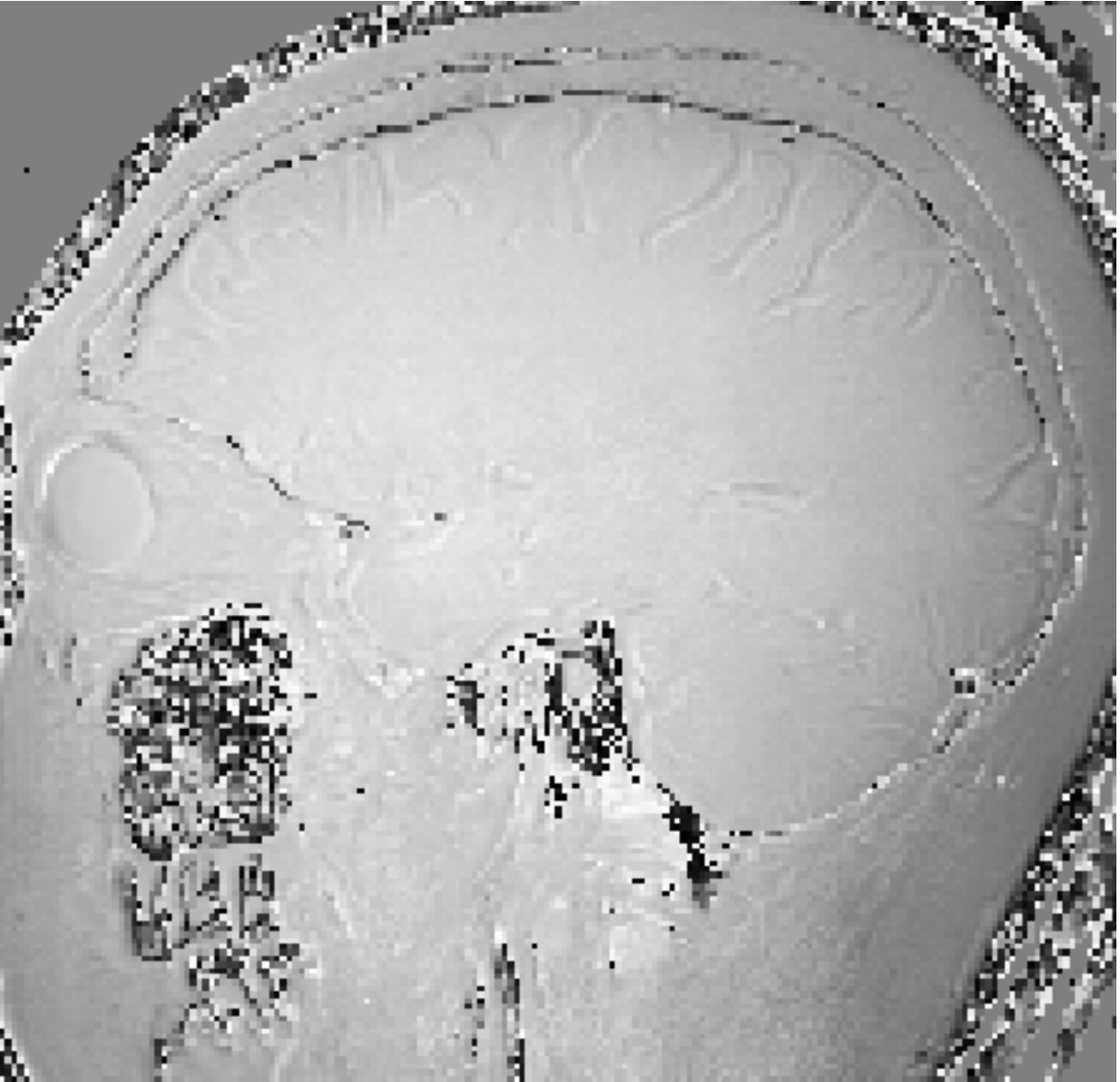}
}%
\subfloat[]{%
  \includegraphics[width=\sz]{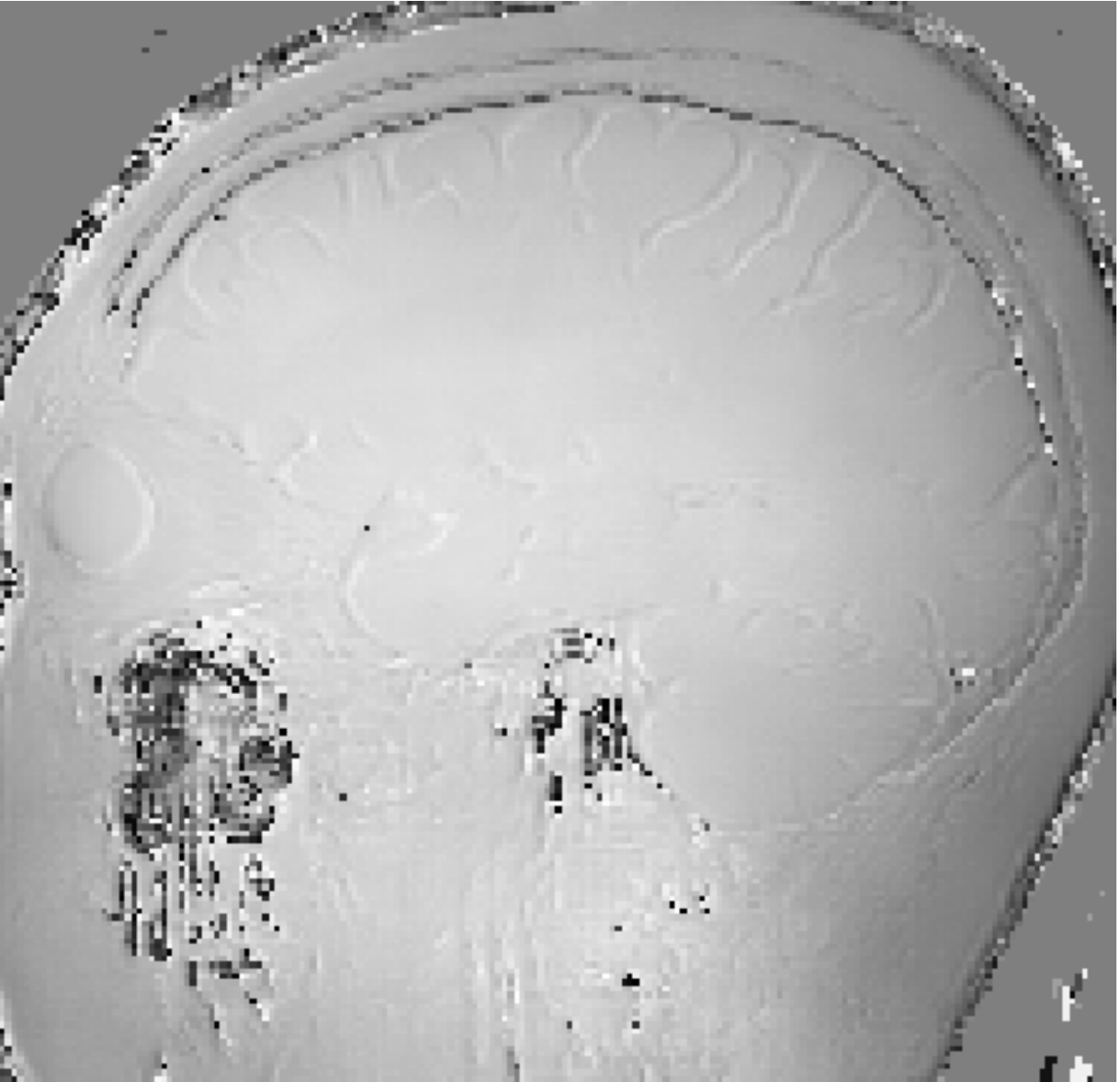}
}%
\subfloat[]{%
  \includegraphics[width=\sz]{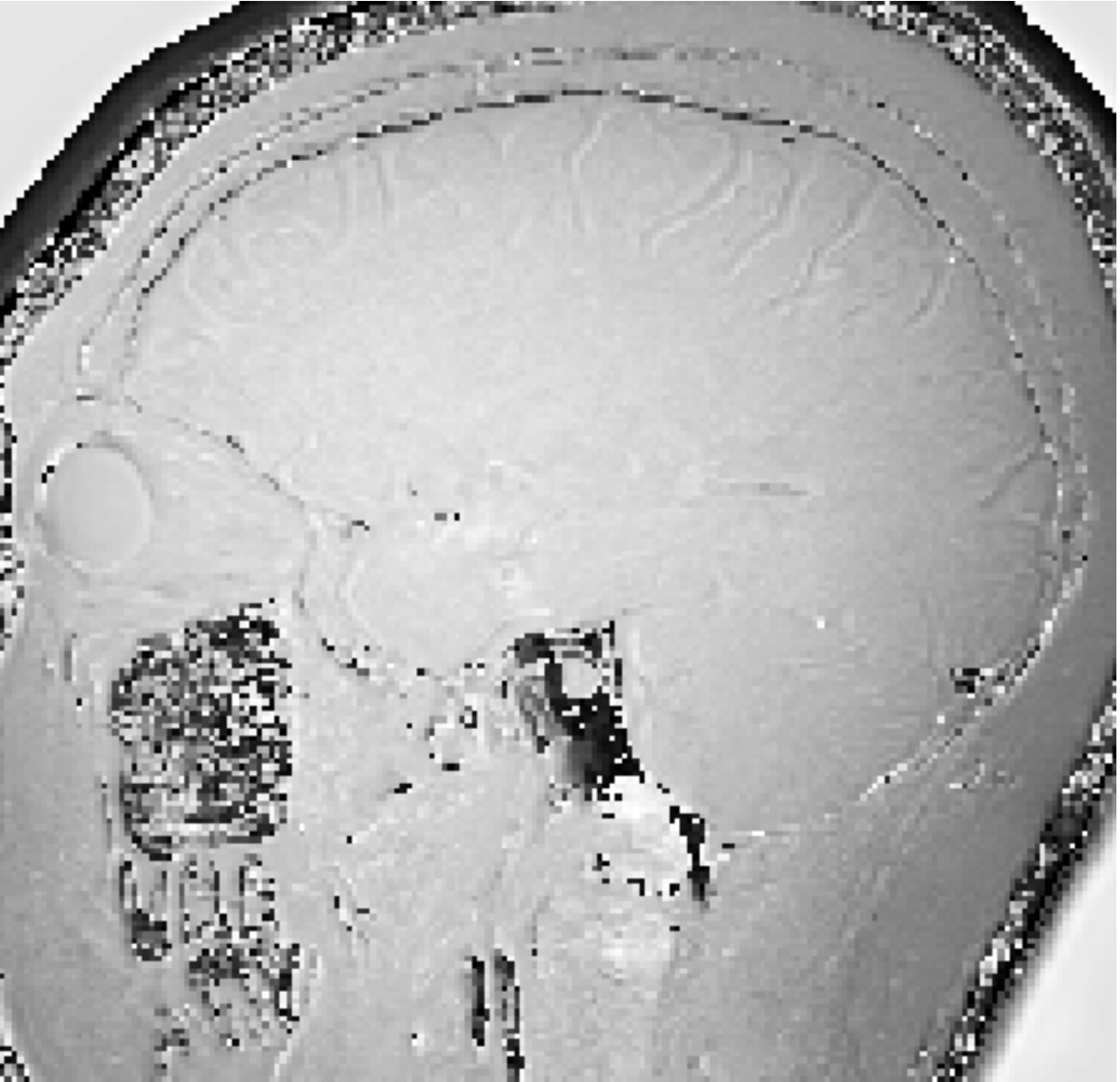}
}%
\subfloat[]{%
  \includegraphics[width=\sz]{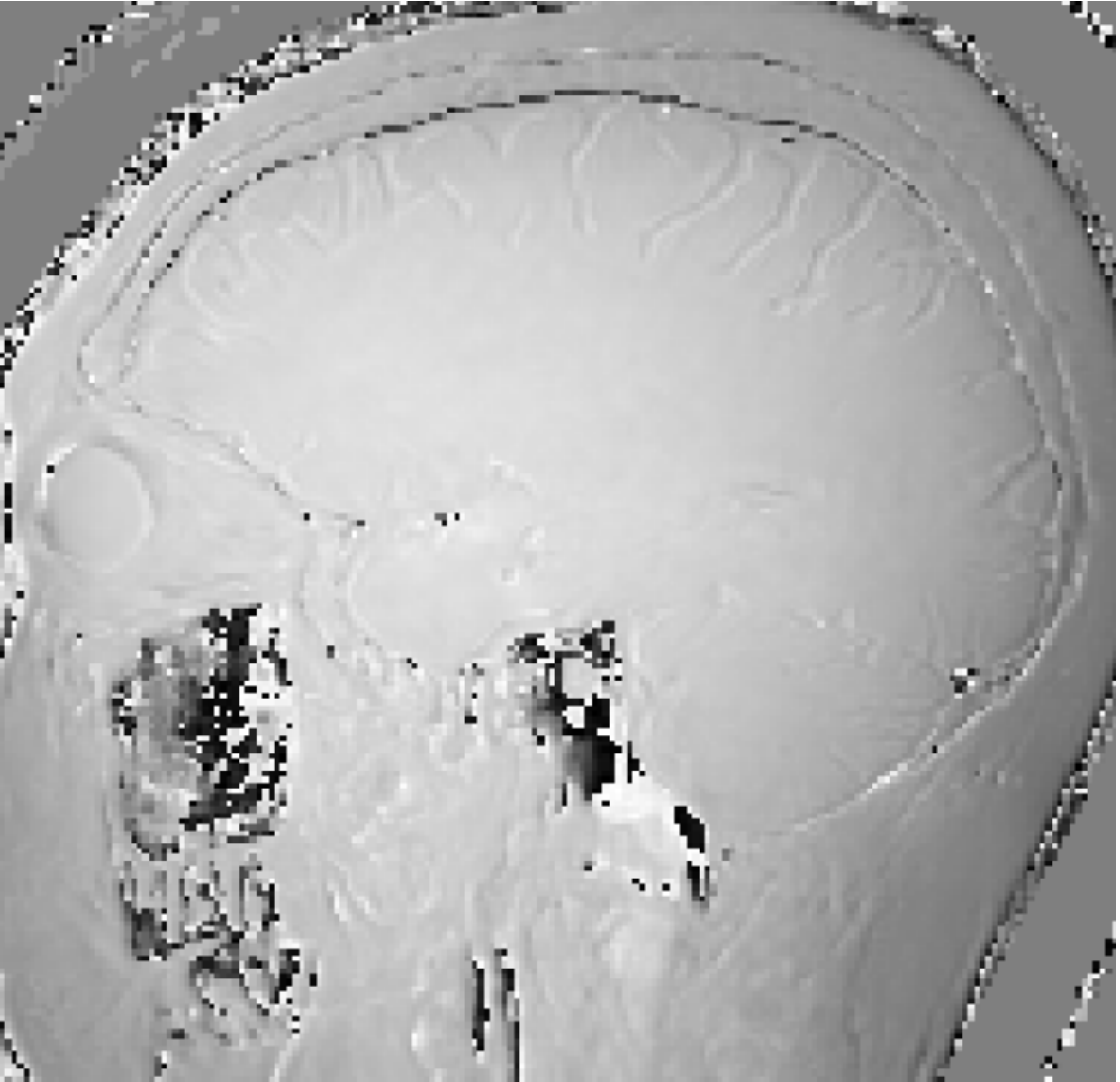}
}%
\subfloat[]{%
  \includegraphics[width=\sz]{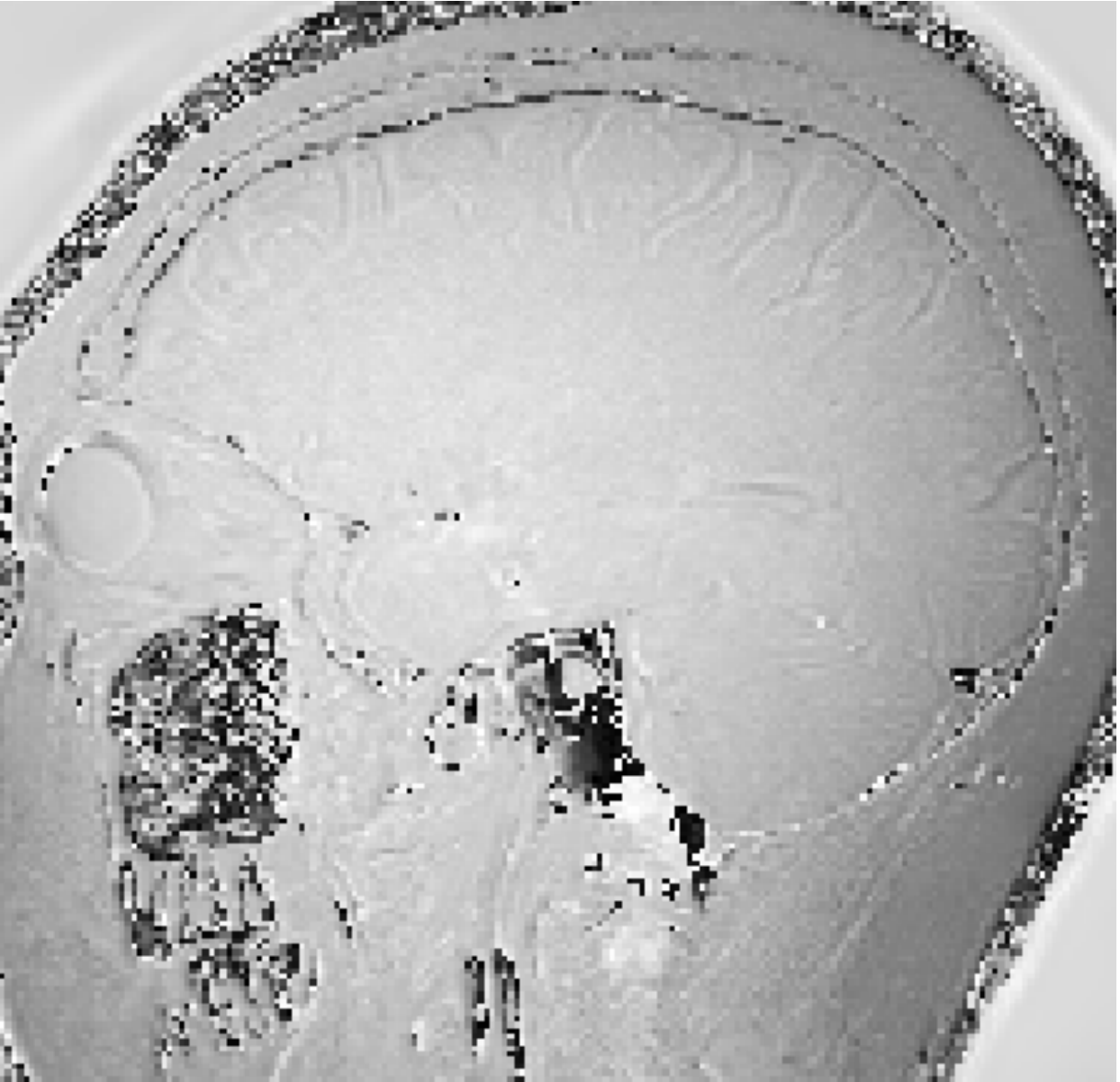}
}%

\vspace{-.8cm}
\crule[white]{\sz}{1cm} \hspace{.01cm}
\subfloat[b. $\Phi$-UNet 0.82]{%
  \includegraphics[width=\sz]{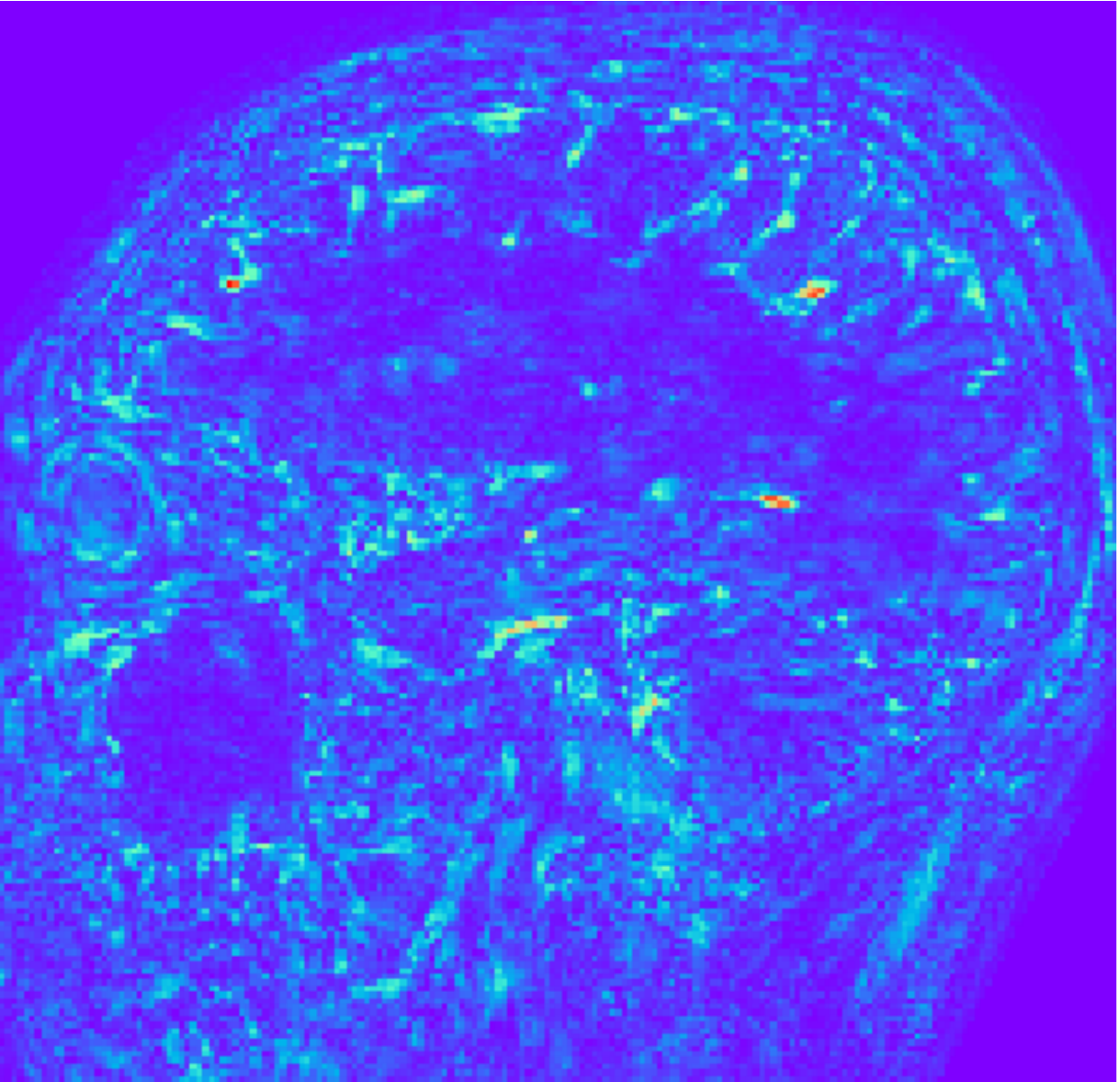}
}%
\subfloat[c. $\Phi$-MoDL 0.91]{%
  \includegraphics[width=\sz]{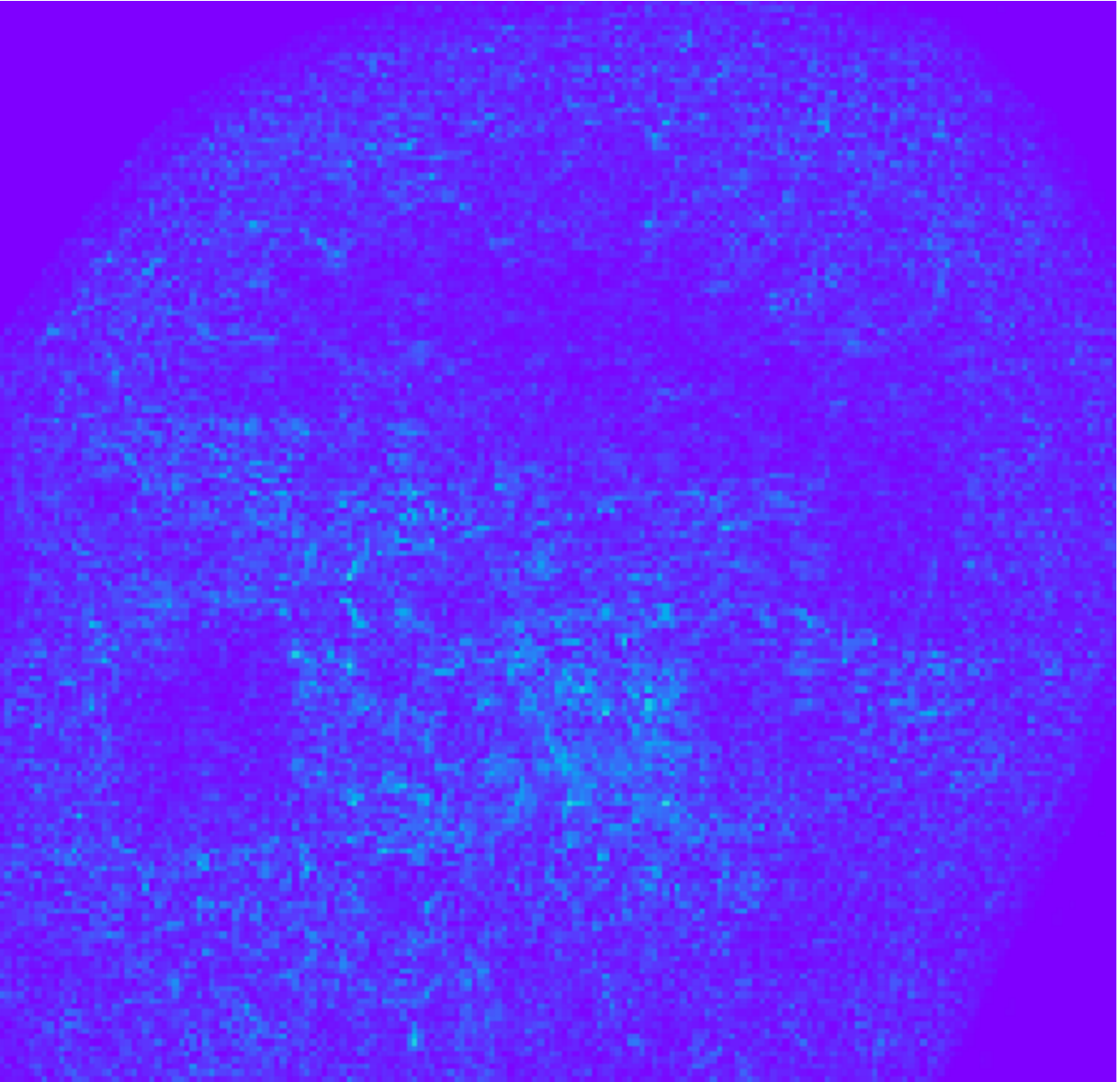}
}%
\subfloat[d. $\Theta$-UNet 0.84]{%
  \includegraphics[width=\sz]{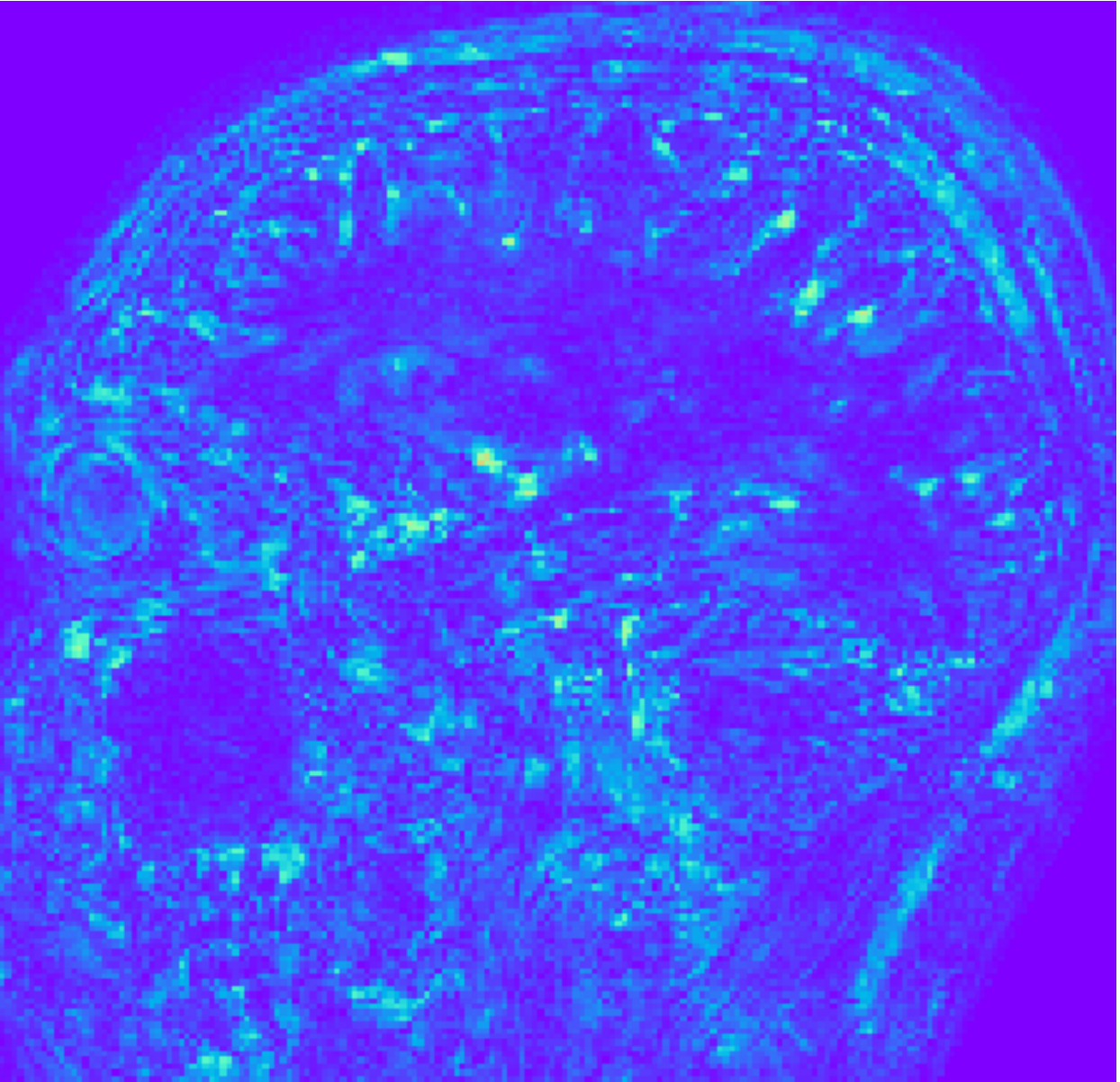}
}%
\subfloat[e. $\Theta$-MoDL 0.93]{%
  \includegraphics[width=\sz]{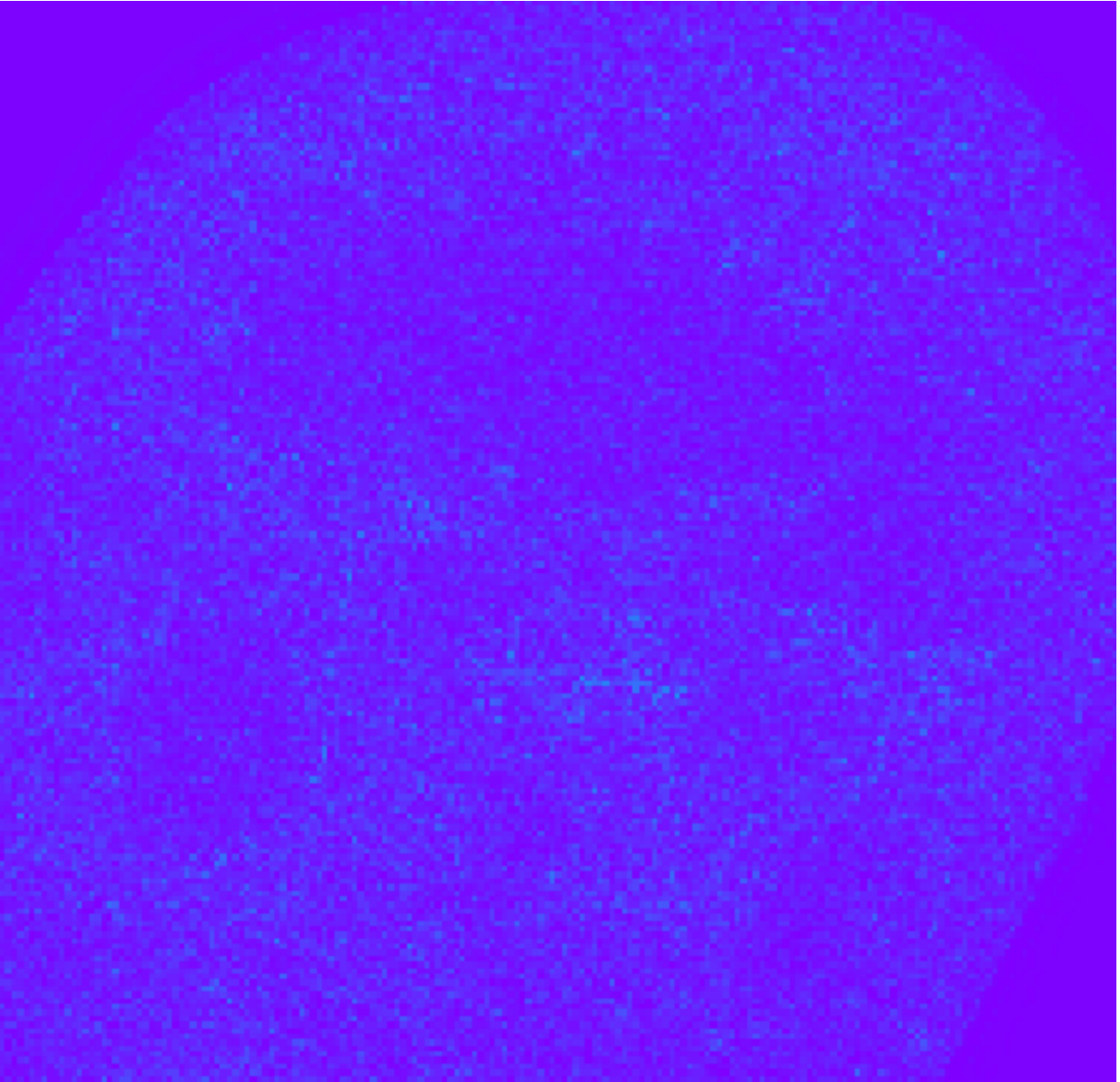}
}%
\subfloat[f. J-UNet 0.94]{%
  \includegraphics[width=\sz]{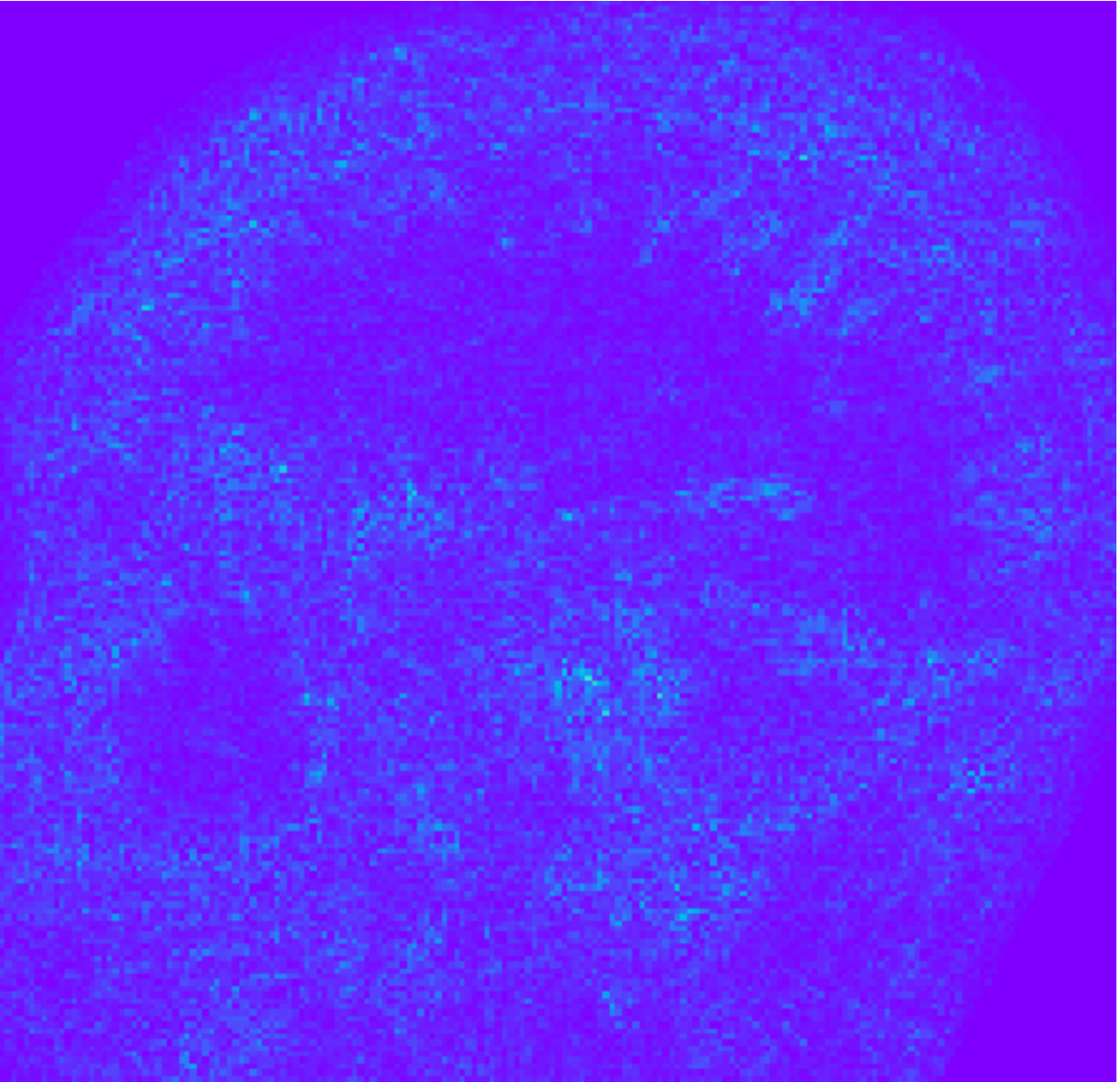}
}%
\subfloat[g. J-MoDL 0.96]{%
  \includegraphics[width=\sz]{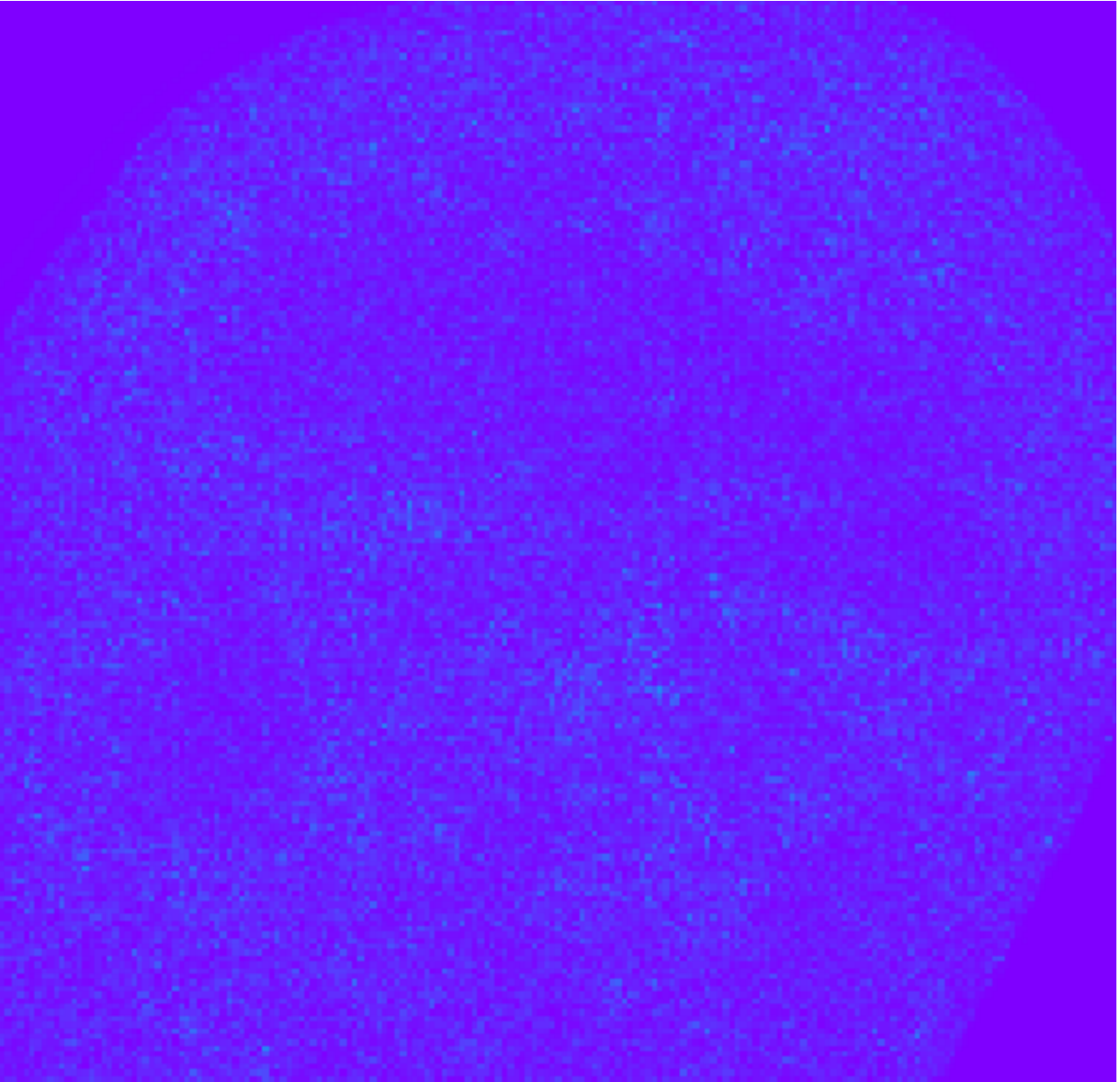}
}%

}

%%% Local Variables:
%%% mode: latex
%%% TeX-master: "hk"
%%% End:

%% file: fig_2d10x_vd_rand_learn.tex
{  \captionsetup[subfigure]{justification=centering}
  \centering
  \newcommand{\sz}{.3\linewidth} % width of each includegraphics  subfigure
  
  \subfloat[random]{
    \includegraphics[ width=\sz]{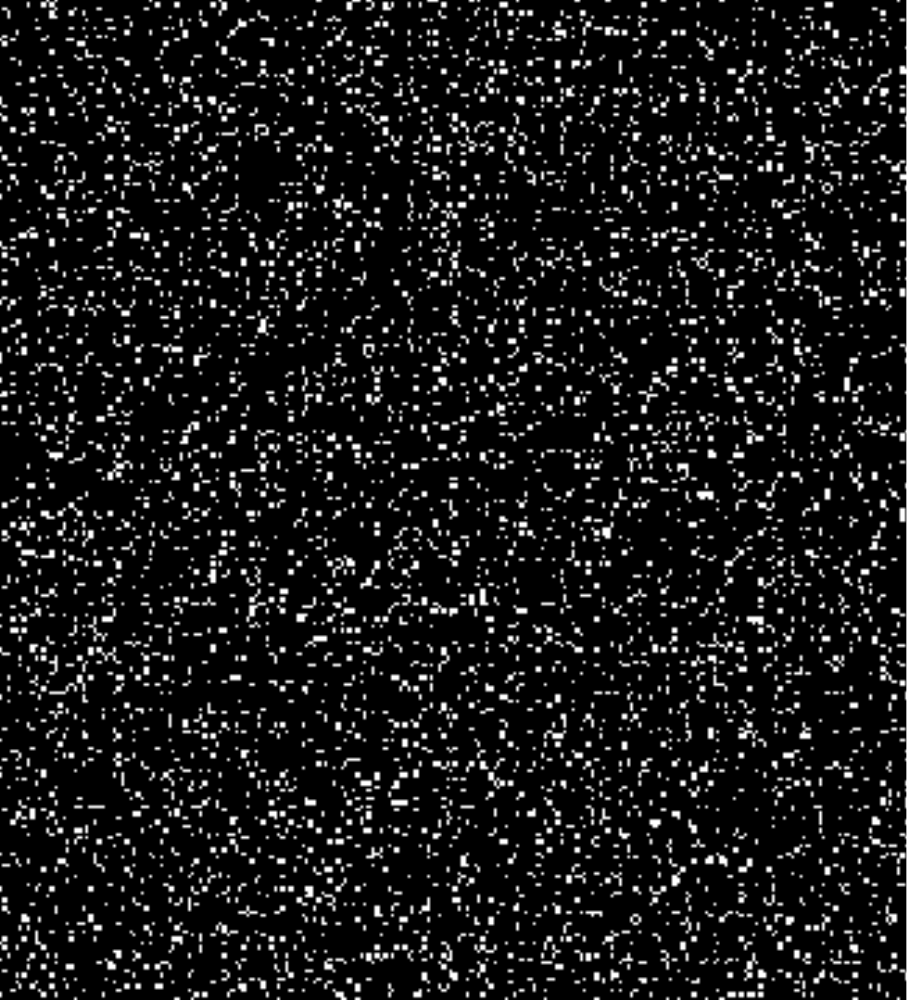}
   }
  \subfloat[VD]{
    \includegraphics[width=\sz]{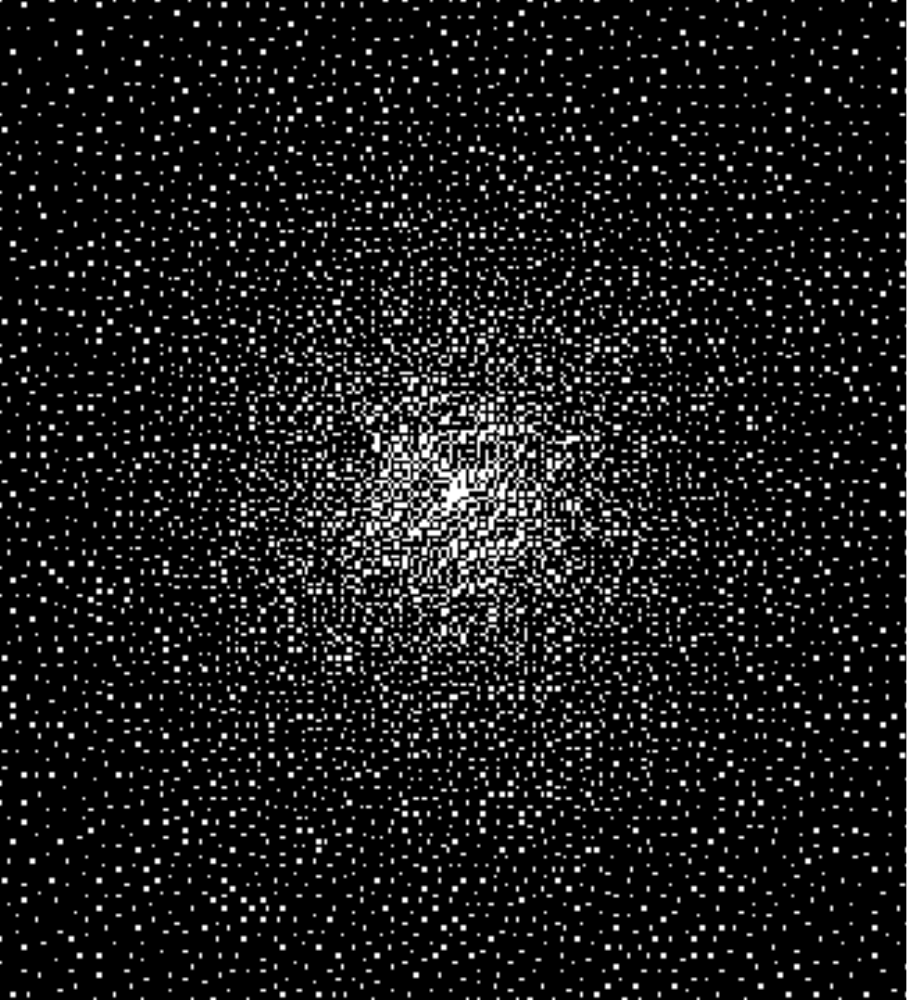}
   }
  \subfloat[learned]{
    \includegraphics[width=\sz]{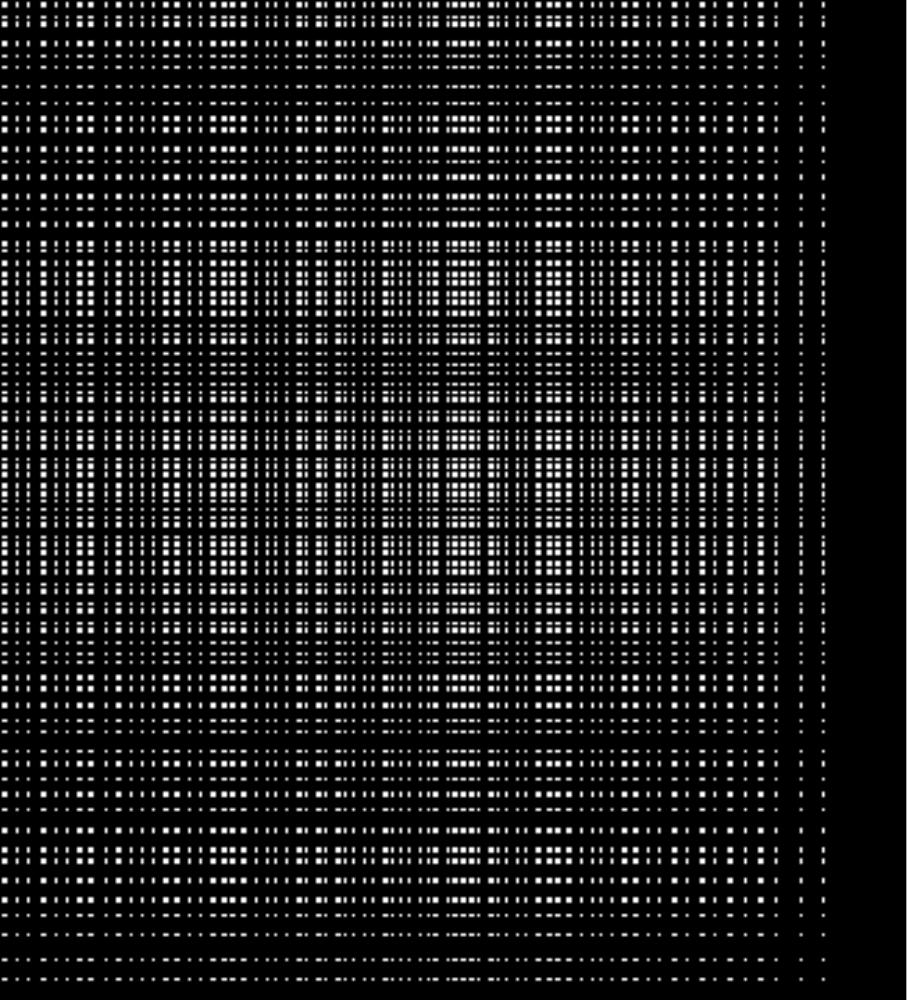}
  }

  \subfloat[random, 18.57]{
    \includegraphics[ width=\sz]{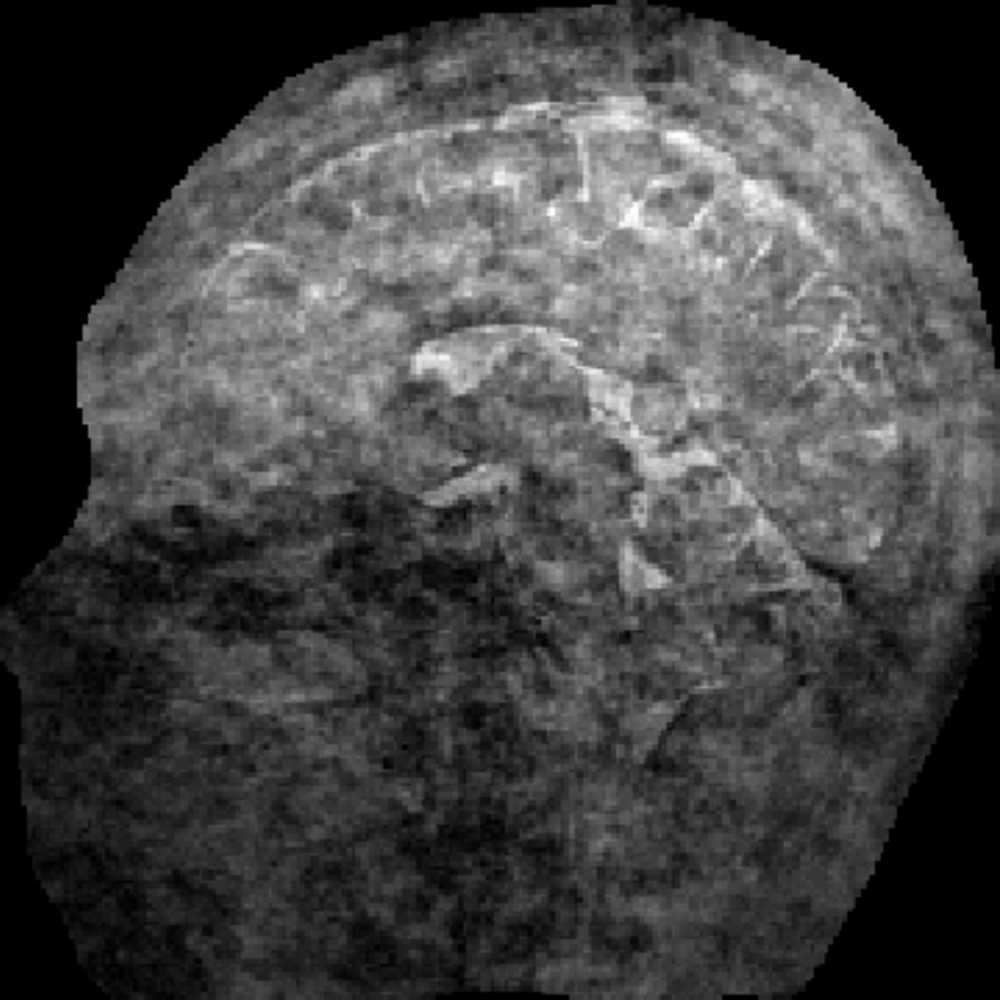}
   }
  \subfloat[VD, 21.28]{
    \includegraphics[width=\sz]{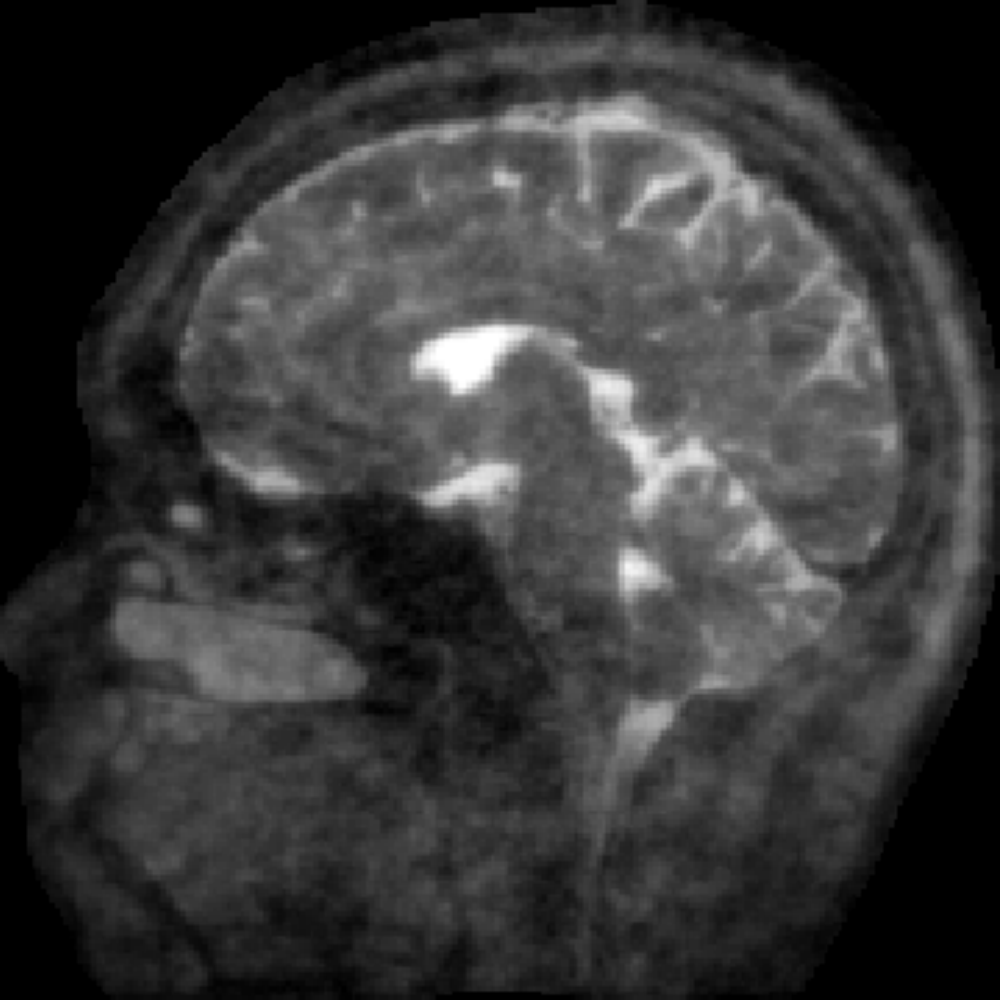}
   }
  \subfloat[learned, 21.68]{
    \includegraphics[width=\sz]{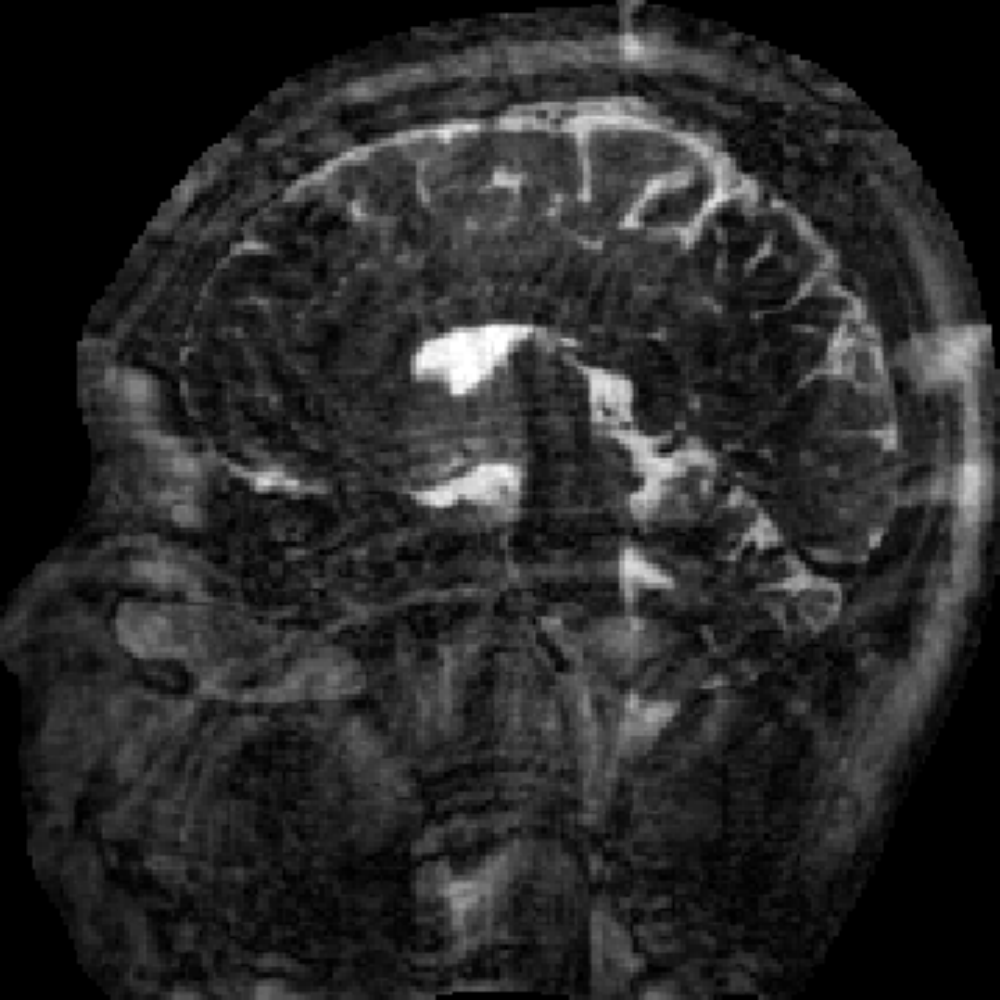}
  }

  \subfloat[random, 28.56]{
    \includegraphics[ width=\sz]{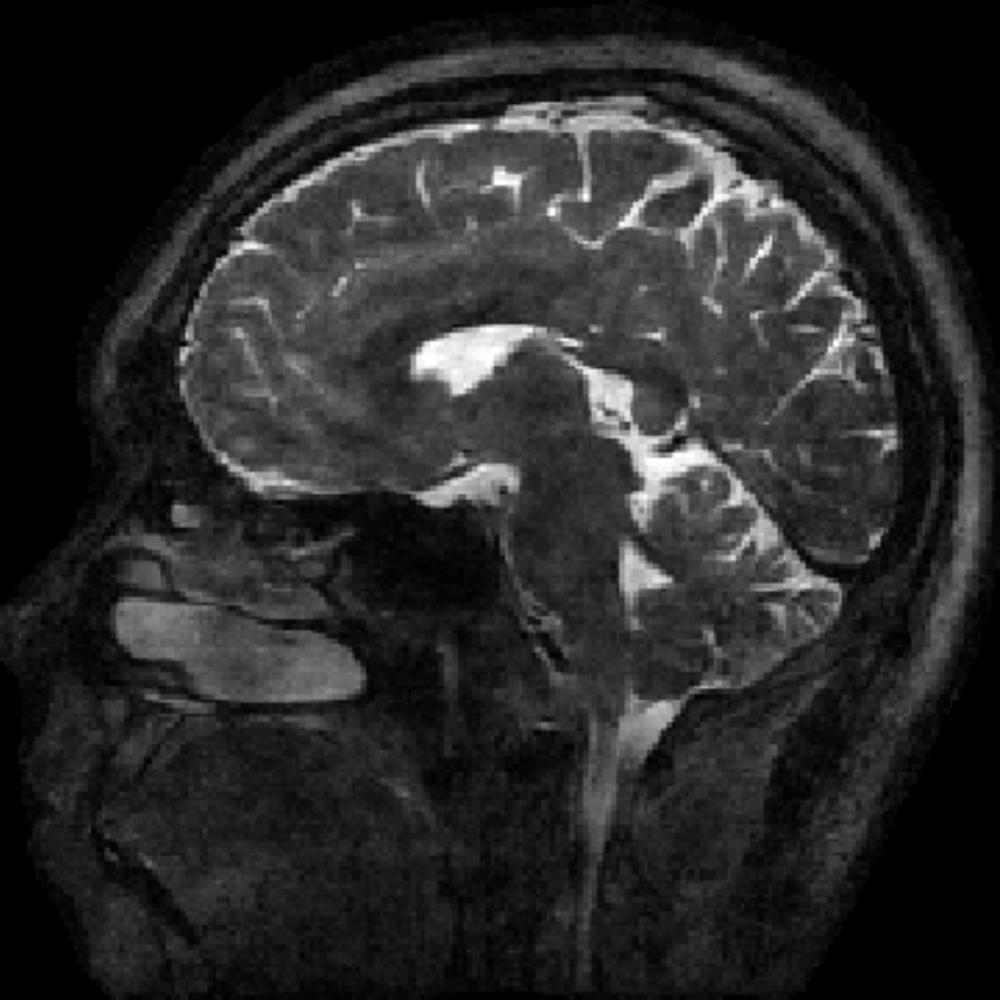}
   }
  \subfloat[VD, 34.37]{
    \includegraphics[width=\sz]{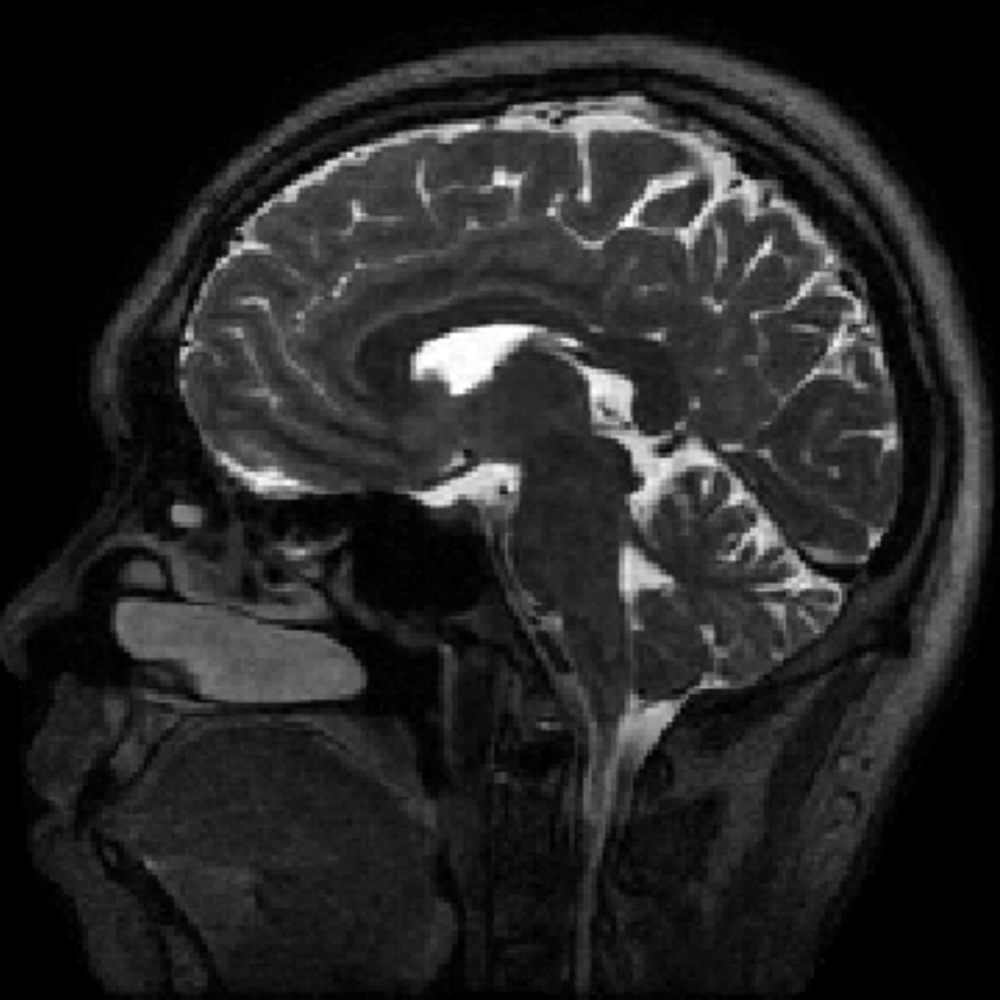}
   }
  \subfloat[learned, 38.67]{
    \includegraphics[width=\sz]{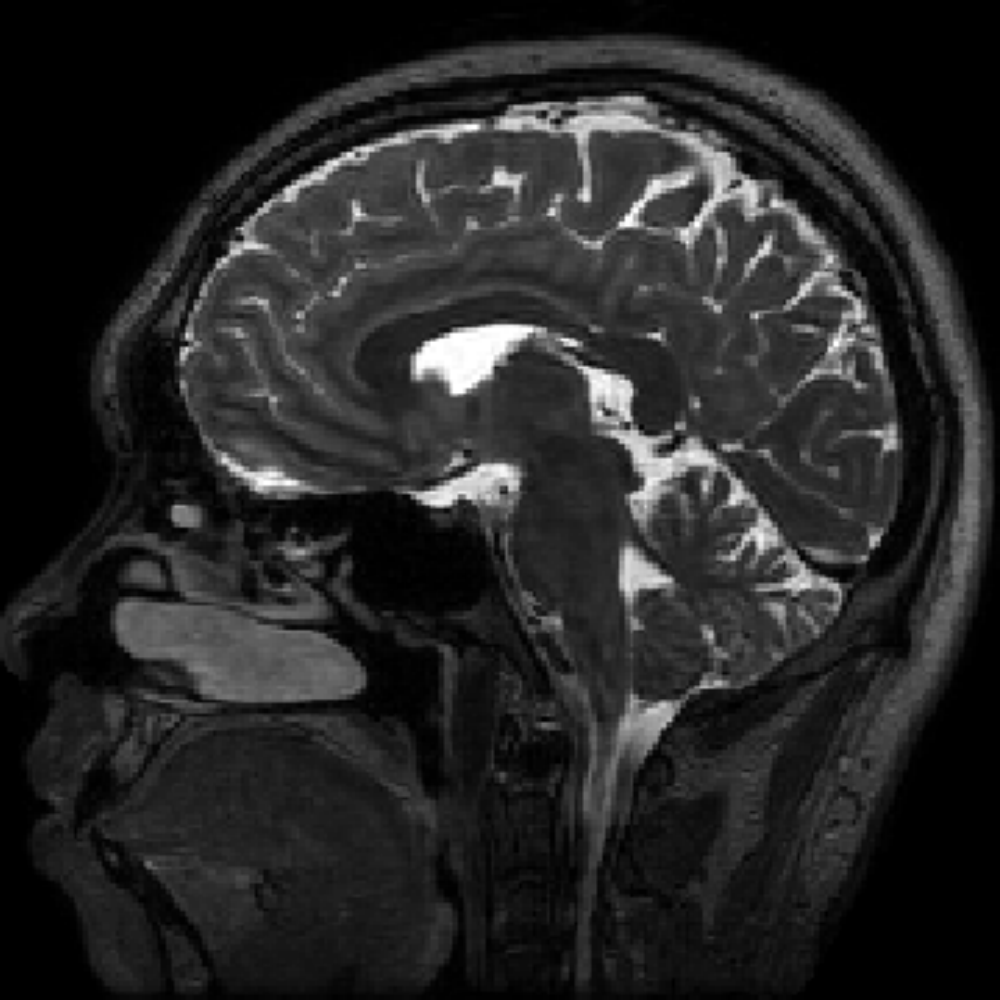}
  }
}

%%% Local Variables:
%%% mode: latex
%%% TeX-master: "hk"
%%% End:

%% file: fig_2d8x_high_noise.tex
{  \captionsetup[subfigure]{justification=centering}
  \centering
  \newcommand{\sz}{.32\linewidth} % width of each includegraphics  subfigure

  \newcommand{\spt}{(.4,0.5)} % at which location to magnify
  \newcommand{\pt}{(0,-2.3)} %location of magnified point
  \newcommand{\mg}{3} %magnification
  \pgfdeclarelayer{background}
  \pgfdeclarelayer{foreground}
  \pgfsetlayers{background,main,foreground} 

  \subfloat[mask, $\sigma=0.05$]{%
 \begin{tikzpicture}[spy using  outlines={rectangle,red,magnification=\mg,size=\sz}]%
    \node {%      
      \includegraphics[ width=\sz]{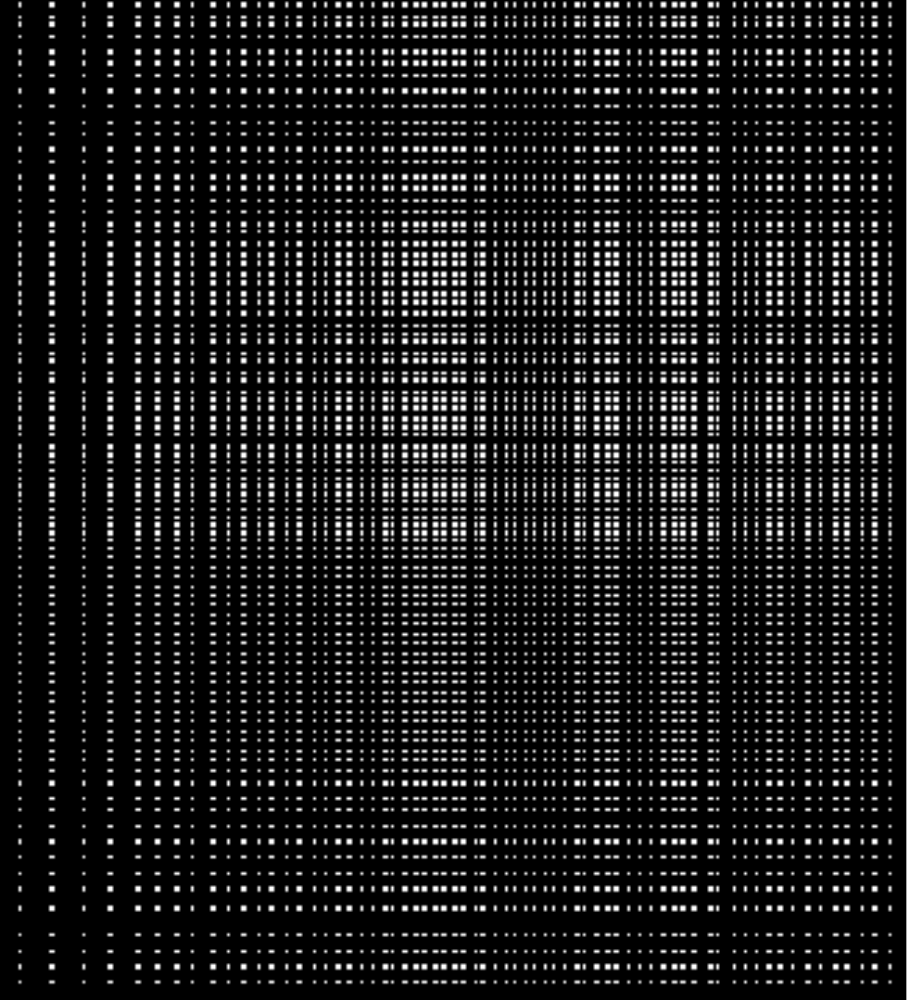}    %  
      };%
     \end{tikzpicture}%
   }%
  \subfloat[mask, $\sigma=3.0$]{%
 \begin{tikzpicture}[spy using  outlines={rectangle,red,magnification=\mg,size=\sz}]%
    \node {%      
      \includegraphics[width=\sz]{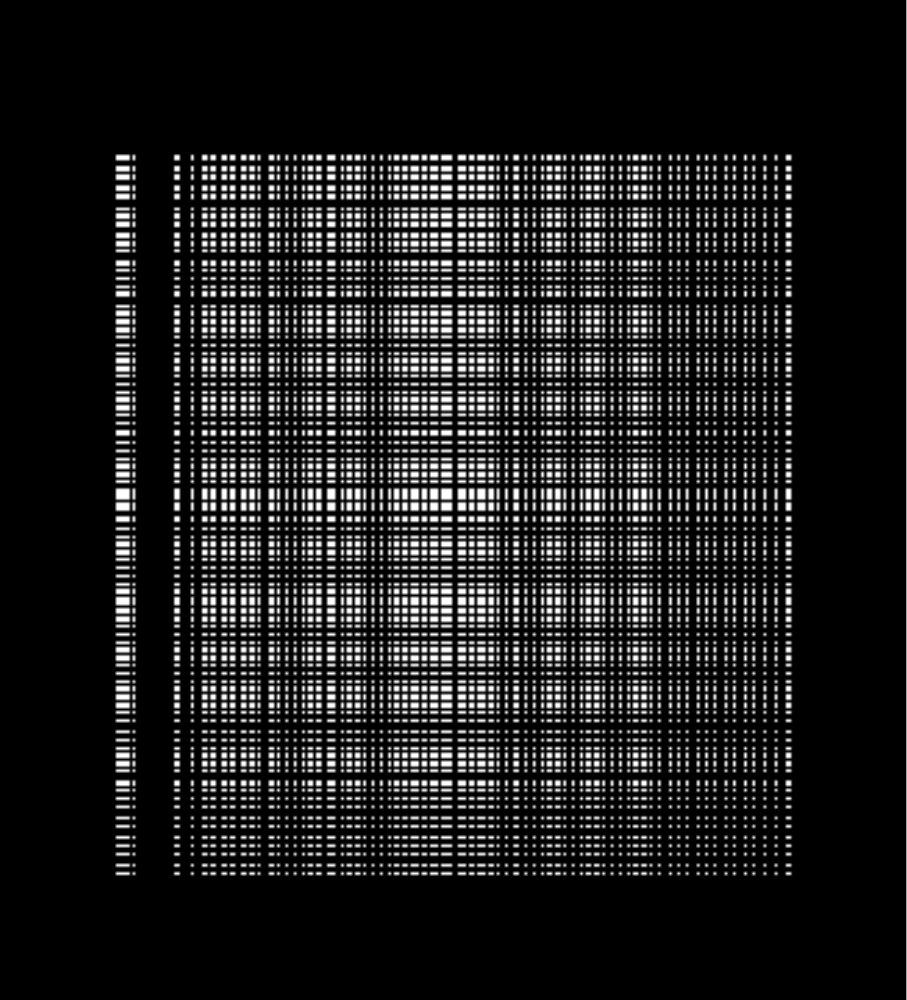}      
      };%
     \end{tikzpicture}%
   }%
  \subfloat[mask, $\sigma=4.0$]{%
 \begin{tikzpicture}[spy using  outlines={rectangle,red,magnification=\mg,size=\sz}]%
    \node {%      
      \includegraphics[width=\sz]{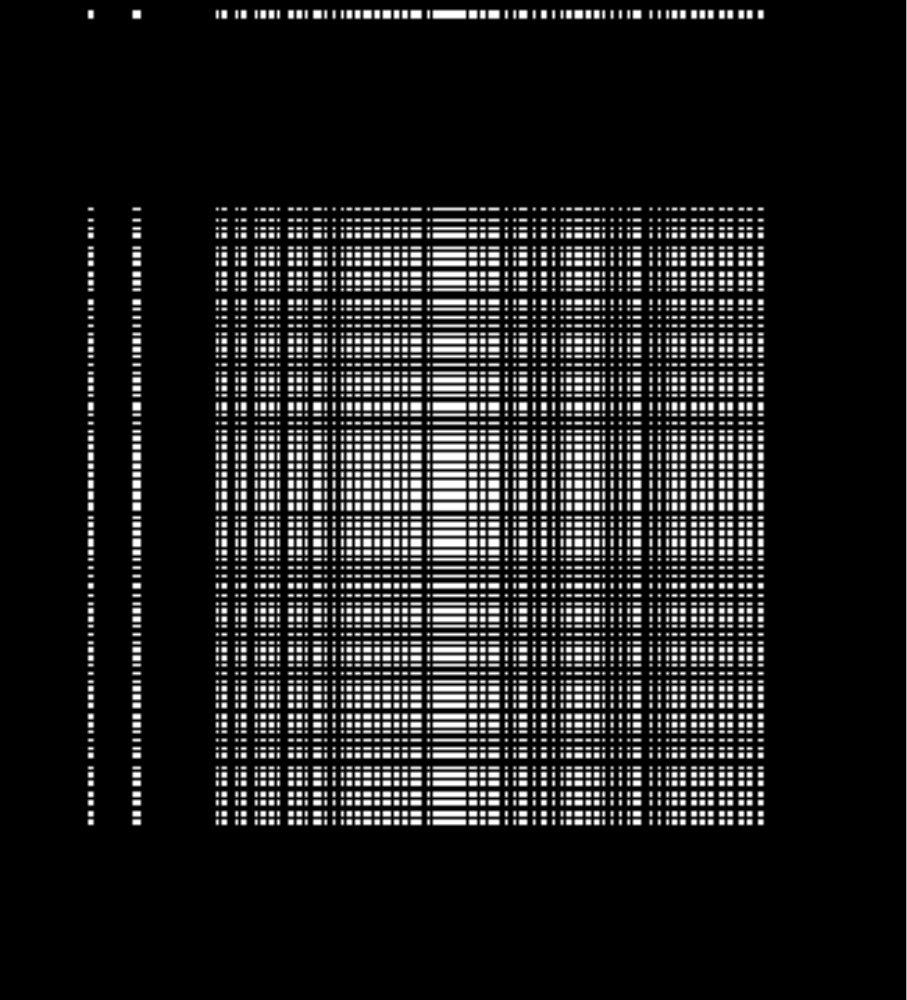}%
       };%
     \end{tikzpicture}%
  }%

  % \subfloat[$\mathcal A^Hb$, 23.33 dB]{
  %   \includegraphics[ width=\sz]{sl113atb058f}
  %  }
  % \subfloat[$\mathcal A^Hb$, 20.44 dB]{
  %   \includegraphics[width=\sz]{sl113atb38f}
  %  }
  % \subfloat[$\mathcal A^Hb$, 16.37 dB]{
  %   \includegraphics[width=\sz]{sl113atb48f}
  % }

  \subfloat[Recon., 42.25]{%
 \begin{tikzpicture}[spy using  outlines={rectangle,red,magnification=\mg,size=\sz}]
    \node {%      
      \includegraphics[ width=\sz]{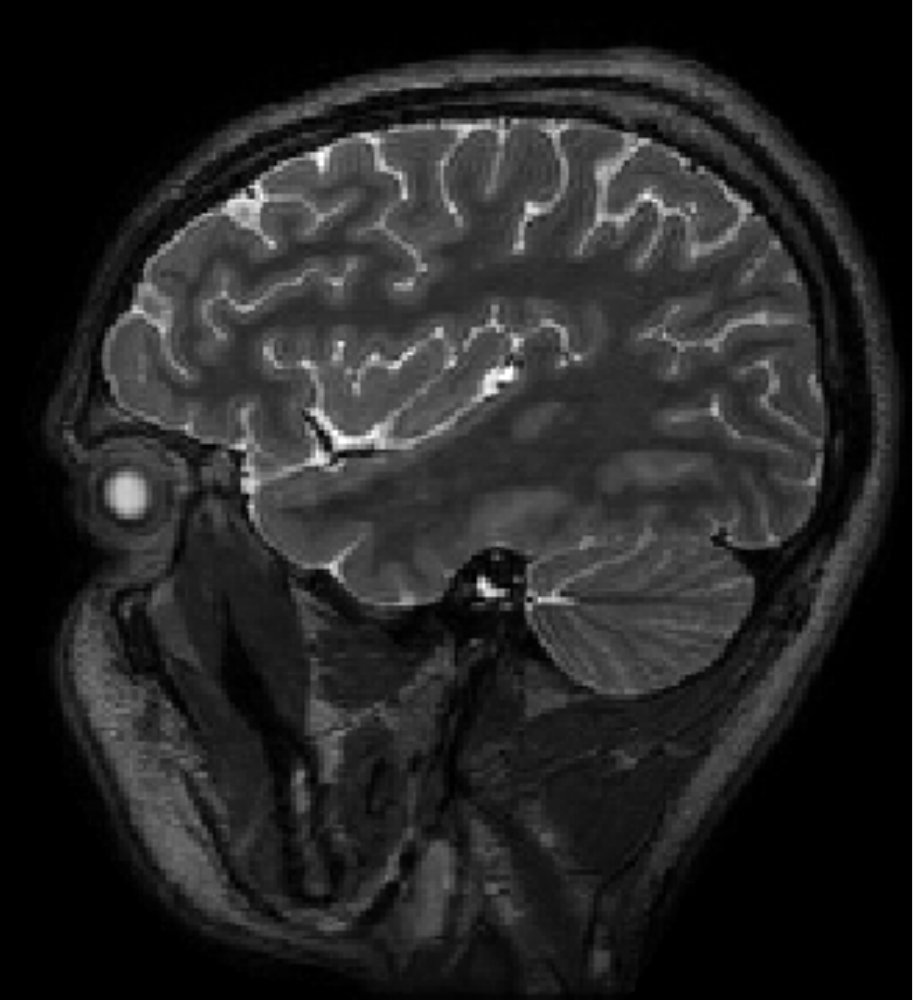}%
      };%
      \spy on \spt in node at \pt;%
    \end{tikzpicture}%
  }%
  \subfloat[Recon., 33.76]{%
    \begin{tikzpicture}[spy using outlines={rectangle,red,magnification=\mg,size=\sz}]
      \node {%
          \includegraphics[width=\sz]{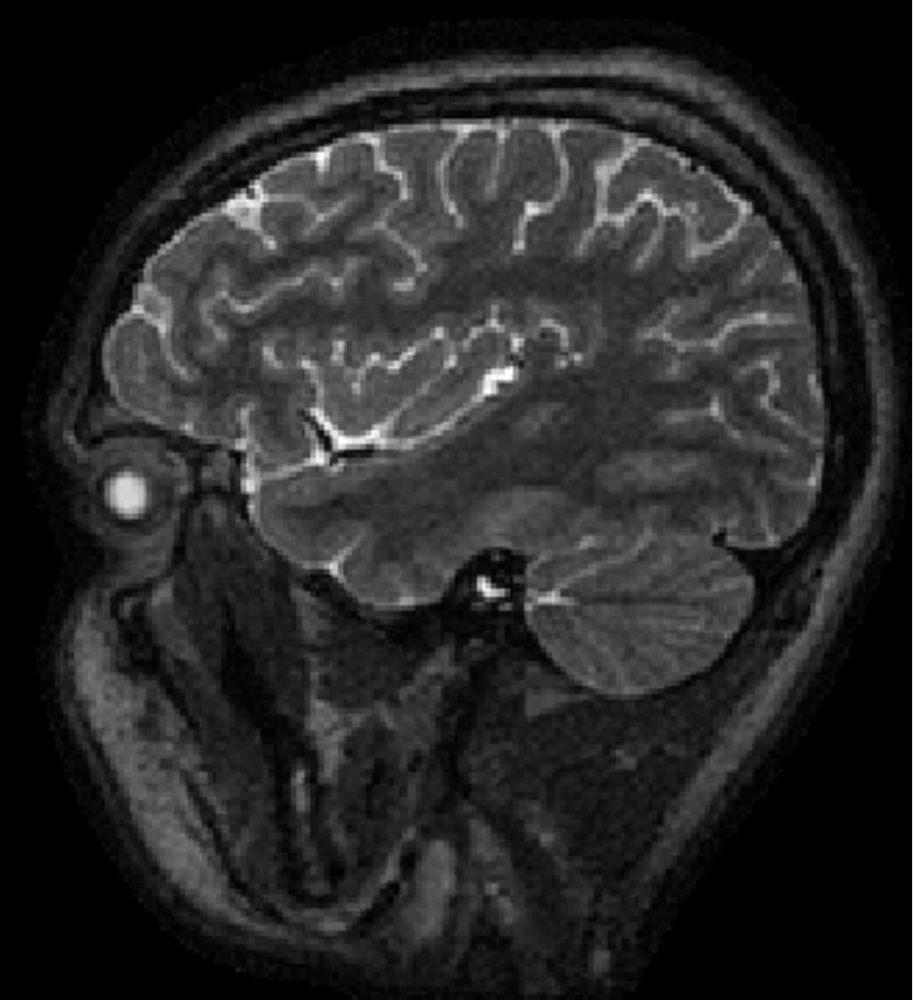}%
      };%
      \spy on \spt in node at \pt;%
    \end{tikzpicture}%
  }%
  \subfloat[Recon., 32.52]{%
    \begin{tikzpicture}[spy using outlines={rectangle,red,magnification=\mg,size=\sz}]%
      \node {%
       \includegraphics[width=\sz]{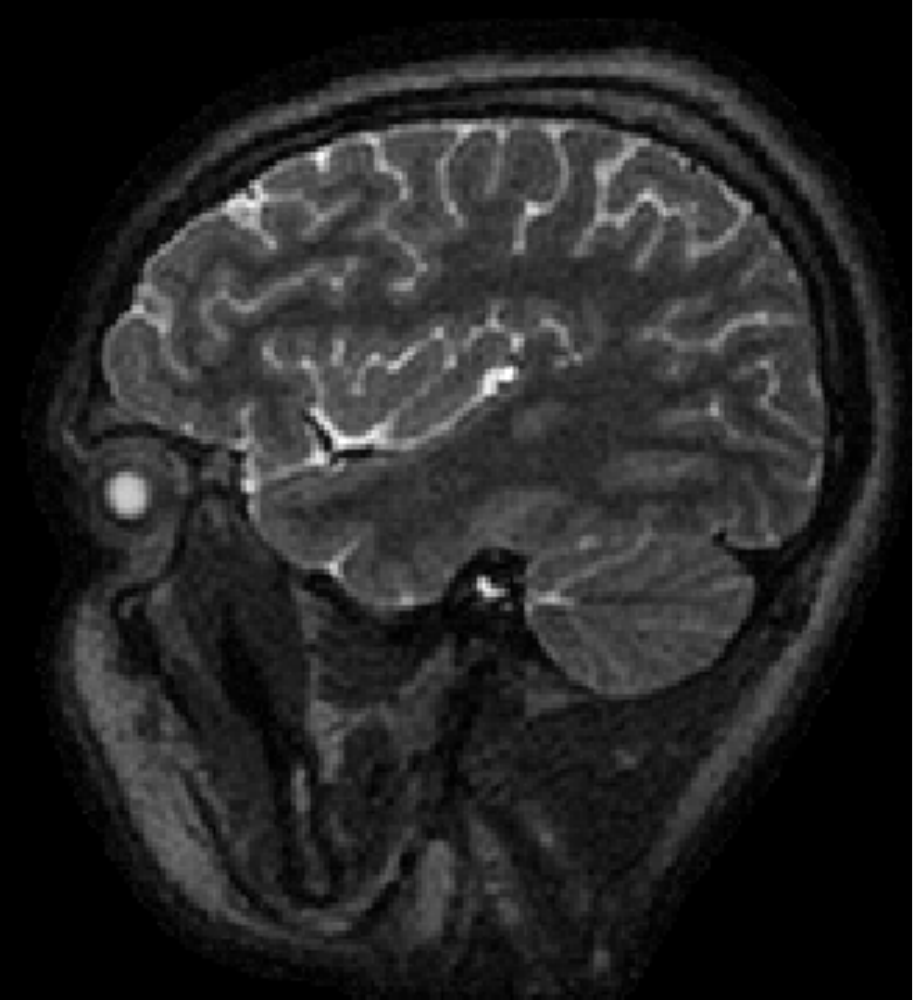}%
      };%
      \spy on \spt in node at \pt;%
    \end{tikzpicture}%
  }%
}

%%% Local Variables:
%%% mode: latex
%%% TeX-master: "hk"
%%% End:

%% file: fig_converge.tex
\centering
\subfloat[Proposed model-based framework: J-MoDL ]{
  \begin{tikzpicture}
 \begin{axis}[
   width=.49\linewidth,
   height=2.2in,
   xlabel=Number of training epochs,
   ylabel=Training loss,
   grid=both,   
   grid style={line width=.1pt, draw=gray!10},
   major grid style={line width=.2pt,draw=gray!50},
   minor tick num=1,
   legend style={at={(0.01,0.06)},anchor=west,legend columns=-1,draw=none, nodes={scale=0.7, transform shape}}
   ]
   \addplot[red,line  width=1]  table [x=a, y=b, col sep=comma] {data.csv};
   \addlegendentry{Exp.~1};
   \addplot[mark=.,blue,line   width=1]  table [x=a, y=c, col sep=comma] {data.csv};
   \addlegendentry{Exp.~2};
   \addplot[mark=.,green,line width=1]  table [x=a, y=d, col sep=comma] {data.csv};
   \addlegendentry{Exp.~3};

    \node [above right] at (rel axis cs:0.27,0.2) {\includegraphics[width=5cm]{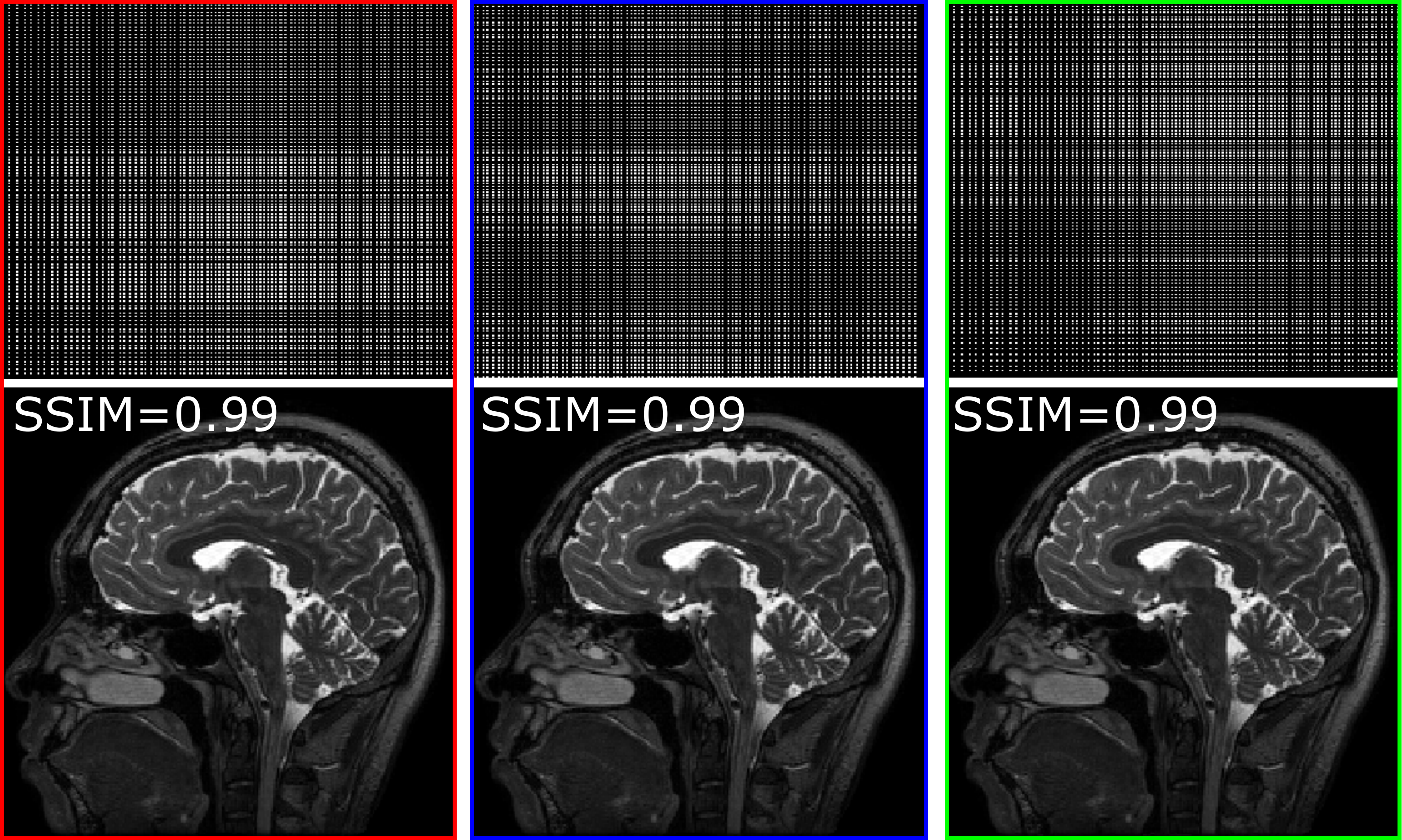}};
 \end{axis}

\end{tikzpicture}
}
\subfloat[Direct-inversion based framework: J-UNET]{
  \begin{tikzpicture}
 \begin{axis}[
   width=.49\linewidth,
   height=2.2in,
   xlabel=Number of training epochs,
   %ylabel=Training loss,
   grid=both,   
   grid style={line width=.1pt, draw=gray!10},
   major grid style={line width=.2pt,draw=gray!50},
   minor tick num=1,
   legend style={at={(0.01,0.06)},anchor=west,legend columns=-1, draw=none, nodes={scale=0.7, transform shape}}
   ]
    \node [above right] at (rel axis cs:0.27,0.2) {\includegraphics[width=5cm]{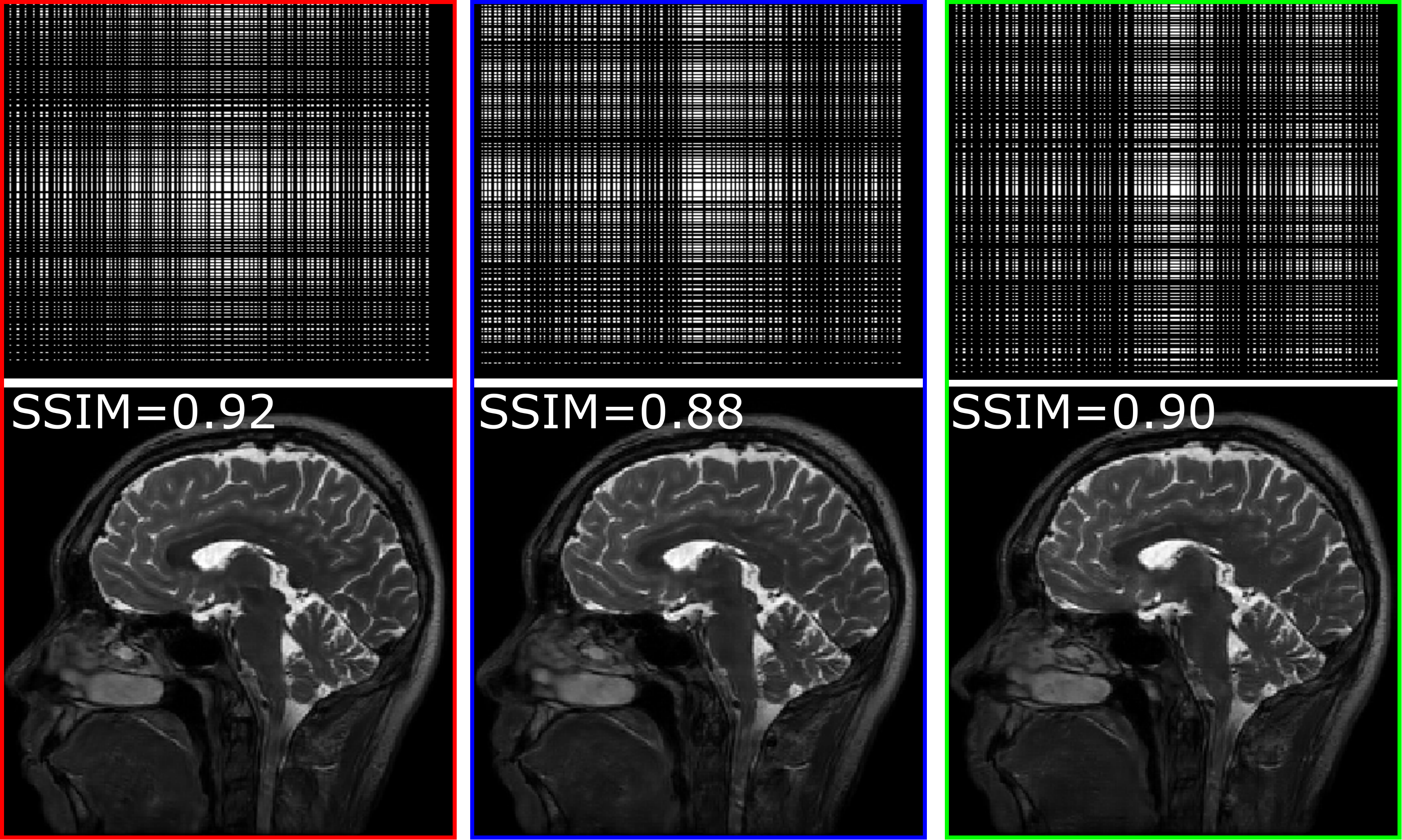}};

   \addplot[red,line  width=1]  table [x=a, y=b, col sep=comma] {dataunet.csv};
   \addlegendentry{Exp.~1};
   \addplot[mark=.,blue,line   width=1]  table [x=a, y=c, col sep=comma] {dataunet.csv};
   \addlegendentry{Exp.~2};
   \addplot[mark=.,green,line width=1]  table [x=a, y=d, col sep=comma] {dataunet.csv};
   \addlegendentry{Exp.~3};

 \end{axis}

\end{tikzpicture}
  
}

%%% Local Variables: 
%%% mode: latex
%%% TeX-master: "hk"
%%% End: 

%% file: fig_2d8x_loupe.tex
{  \captionsetup[subfigure]{justification=centering}
  \centering
  \newcommand{\sz}{.16\linewidth} 
  \newcommand{\spt}{(0.5,-0.23)} 
  \newcommand{\pt}{(0,-3)} 
  \newcommand{\mg}{4}
  \pgfdeclarelayer{background}
  \pgfdeclarelayer{foreground}
  \pgfsetlayers{background,main,foreground} 
 
\subfloat[Initial mask]{%
 \begin{tikzpicture}[spy using outlines={rectangle,red,magnification=\mg,size=\sz}]
    \node {
        \includegraphics[ width=\sz]{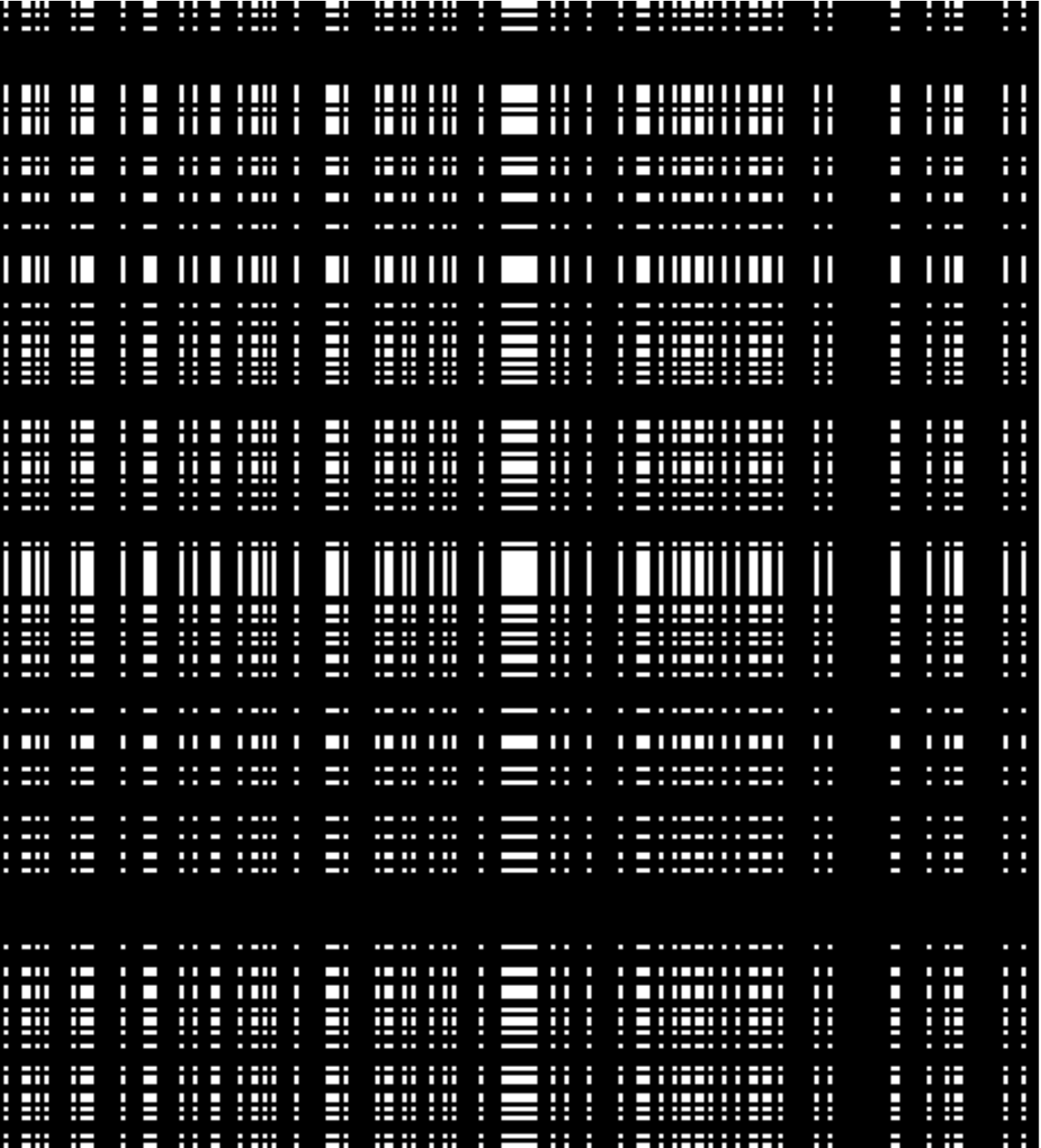}
      };
    \end{tikzpicture}%
  } \hspace{-.28cm}
  \subfloat[MC-LOUPE]{%
    \begin{tikzpicture}[spy using outlines={rectangle,red,magnification=\mg,size=\sz}]
      \node {
        \includegraphics[width=\sz]{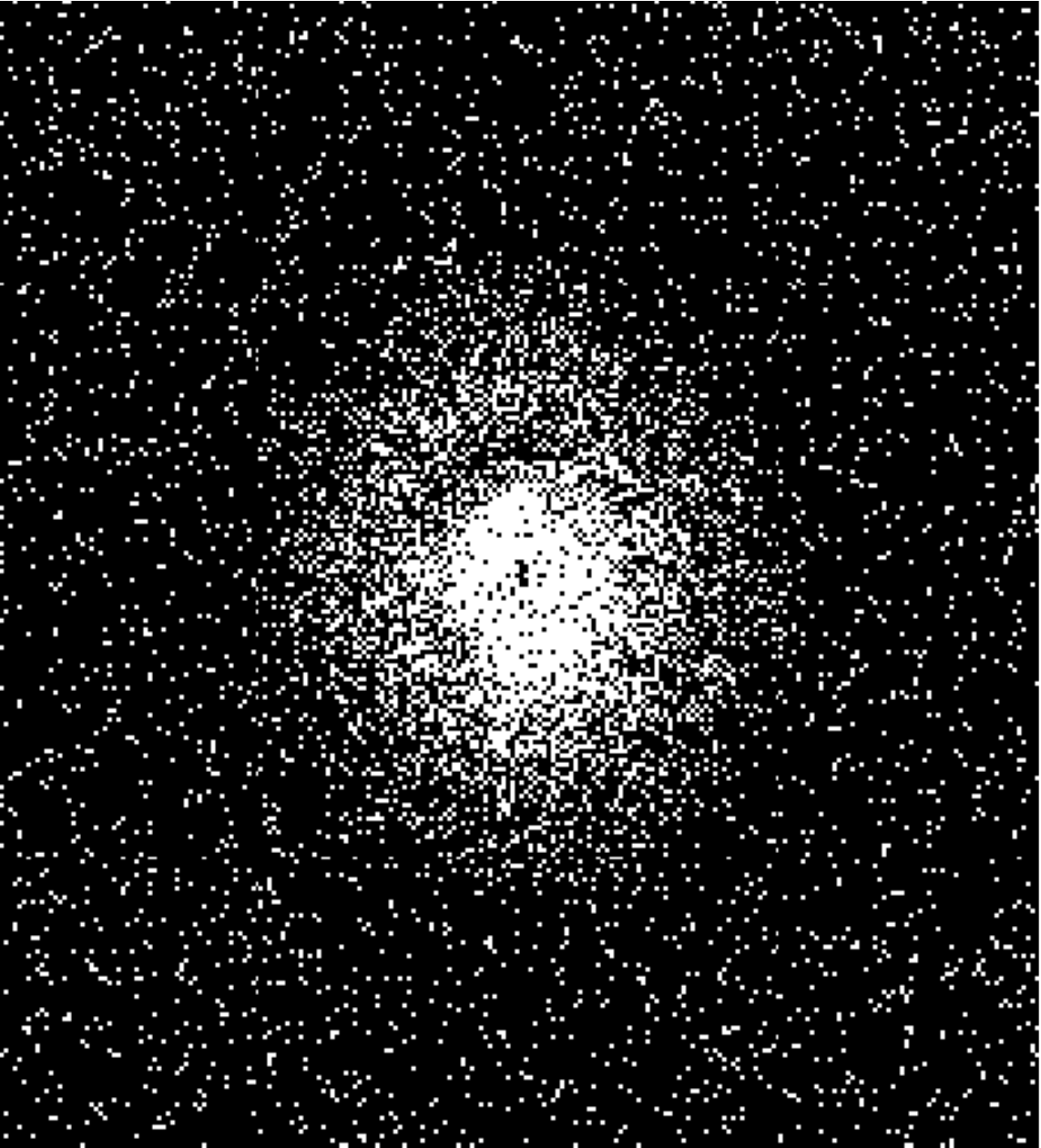}%
      };
    \end{tikzpicture}%
  }\hspace{-.28cm}
  \subfloat[J-ISTANet]{%
    \begin{tikzpicture}[spy using outlines={rectangle,red,magnification=\mg,size=\sz}]
      \node {
        \includegraphics[width=\sz]{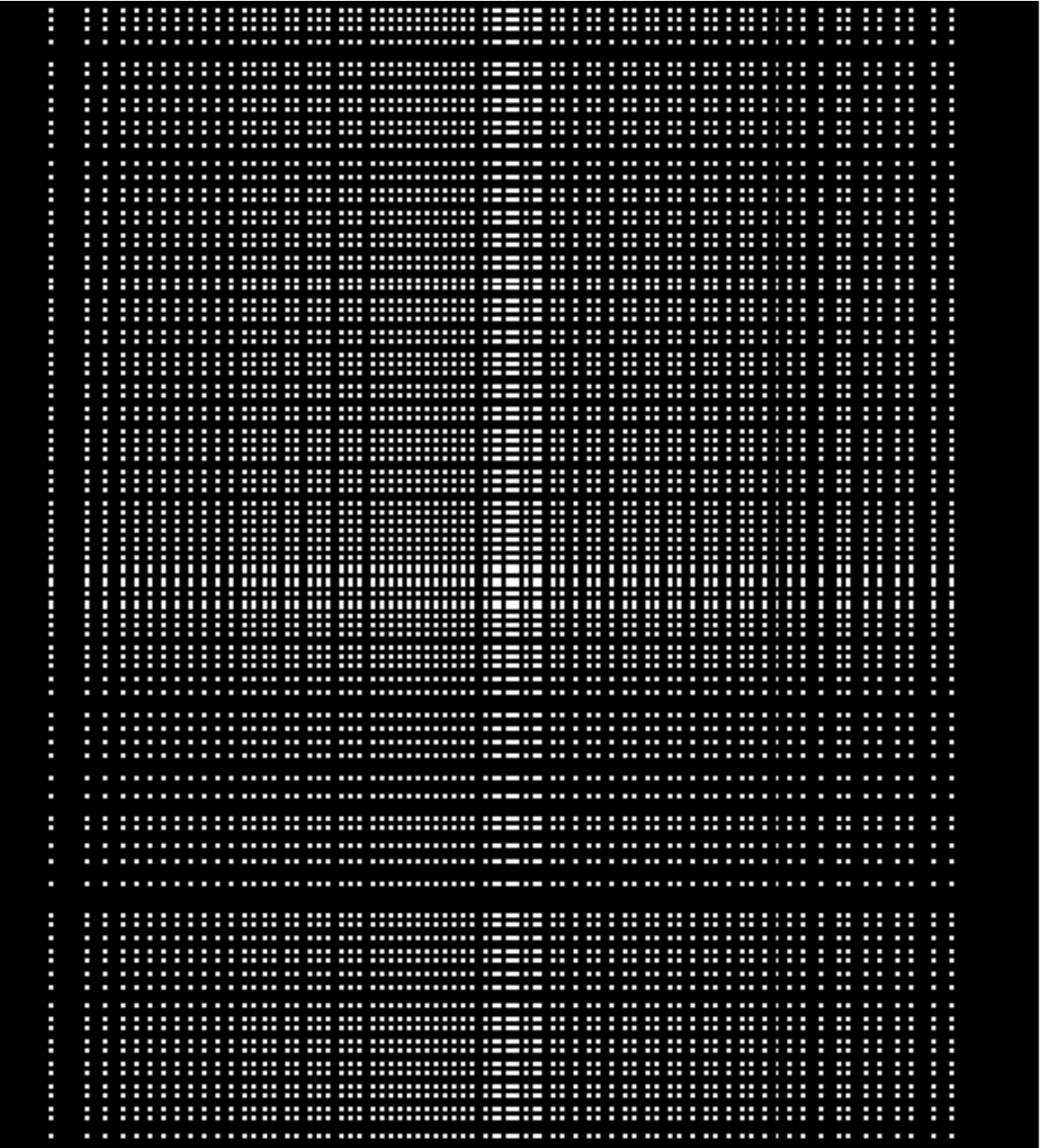}%
      };
    \end{tikzpicture}%
  }\hspace{-.28cm}
  \subfloat[J-UNET]{%
    \begin{tikzpicture}[spy using outlines={rectangle,red,magnification=\mg,size=\sz}]
      \node {
        \includegraphics[width=\sz]{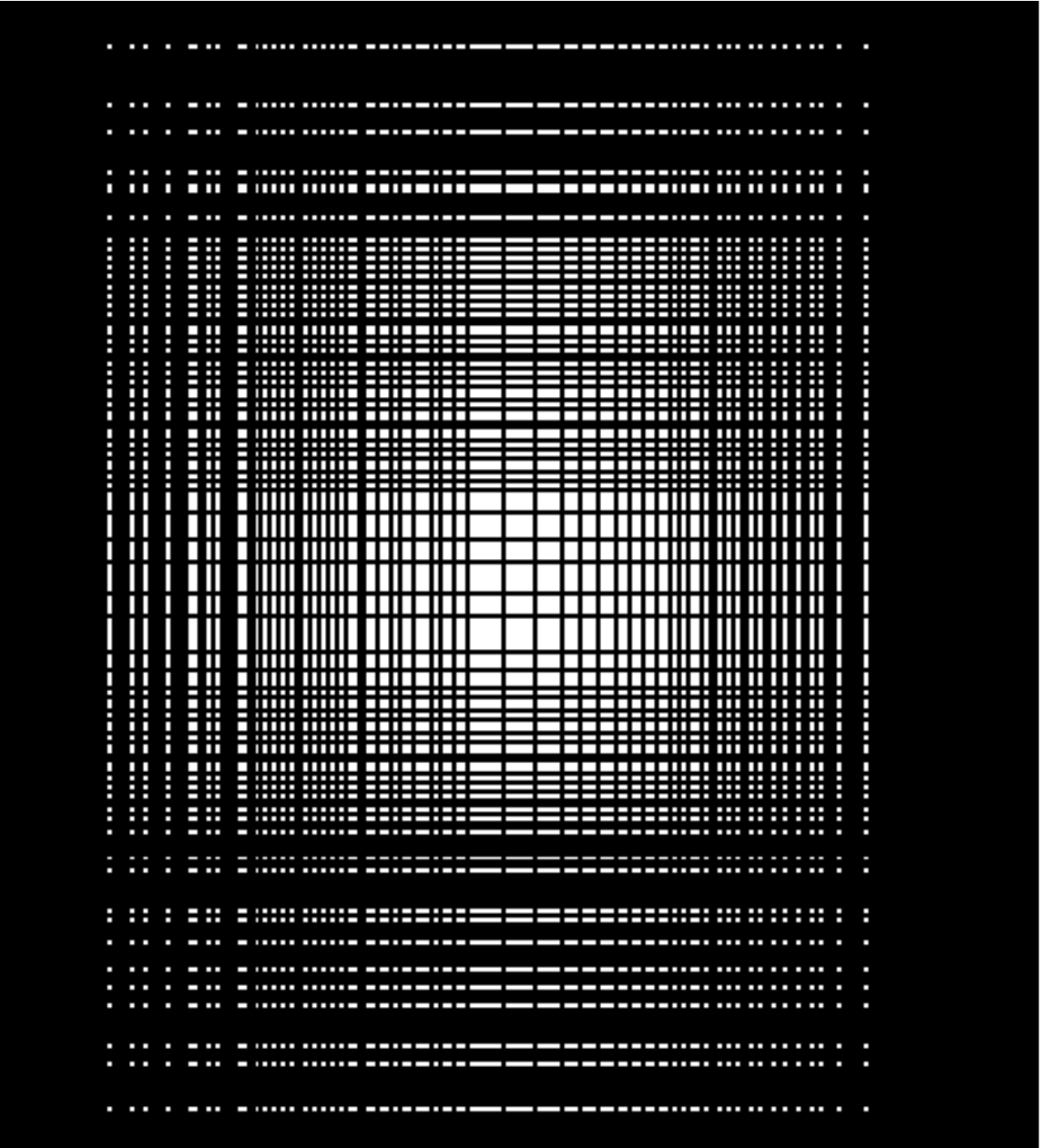}%
      };
    \end{tikzpicture}%
  }\hspace{-.28cm}
  \subfloat[J-MoDL, K=1]{%
    \begin{tikzpicture}[spy using outlines={rectangle,red,magnification=\mg,size=\sz}]
      \node {
        \includegraphics[width=\sz]{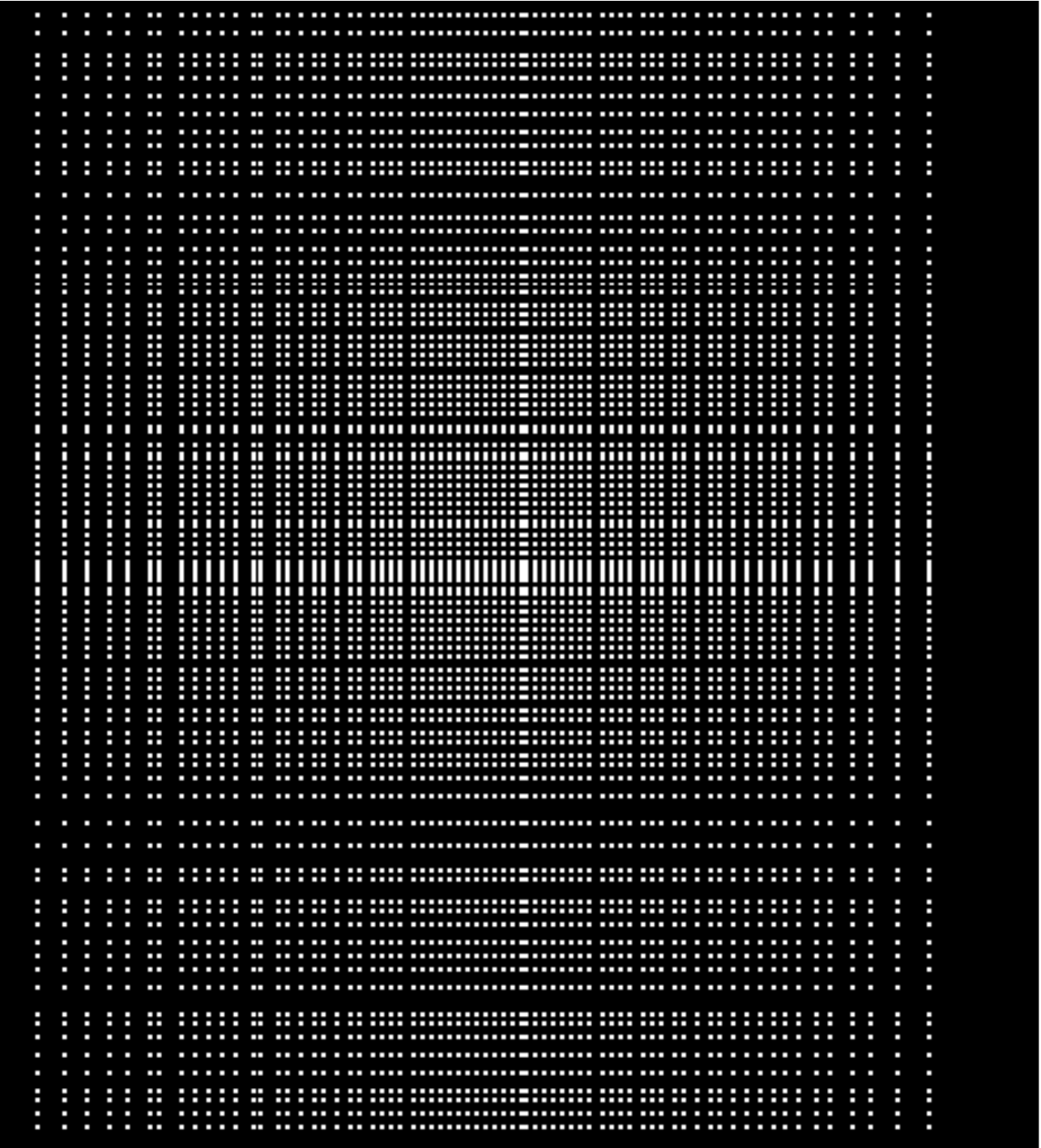}%
      };
    \end{tikzpicture}%
  }\hspace{-.28cm}
  \subfloat[J-MoDL,K=5]{%
    \begin{tikzpicture}[spy using outlines={rectangle,red,magnification=\mg,size=\sz}]
      \node {
        \includegraphics[width=\sz]{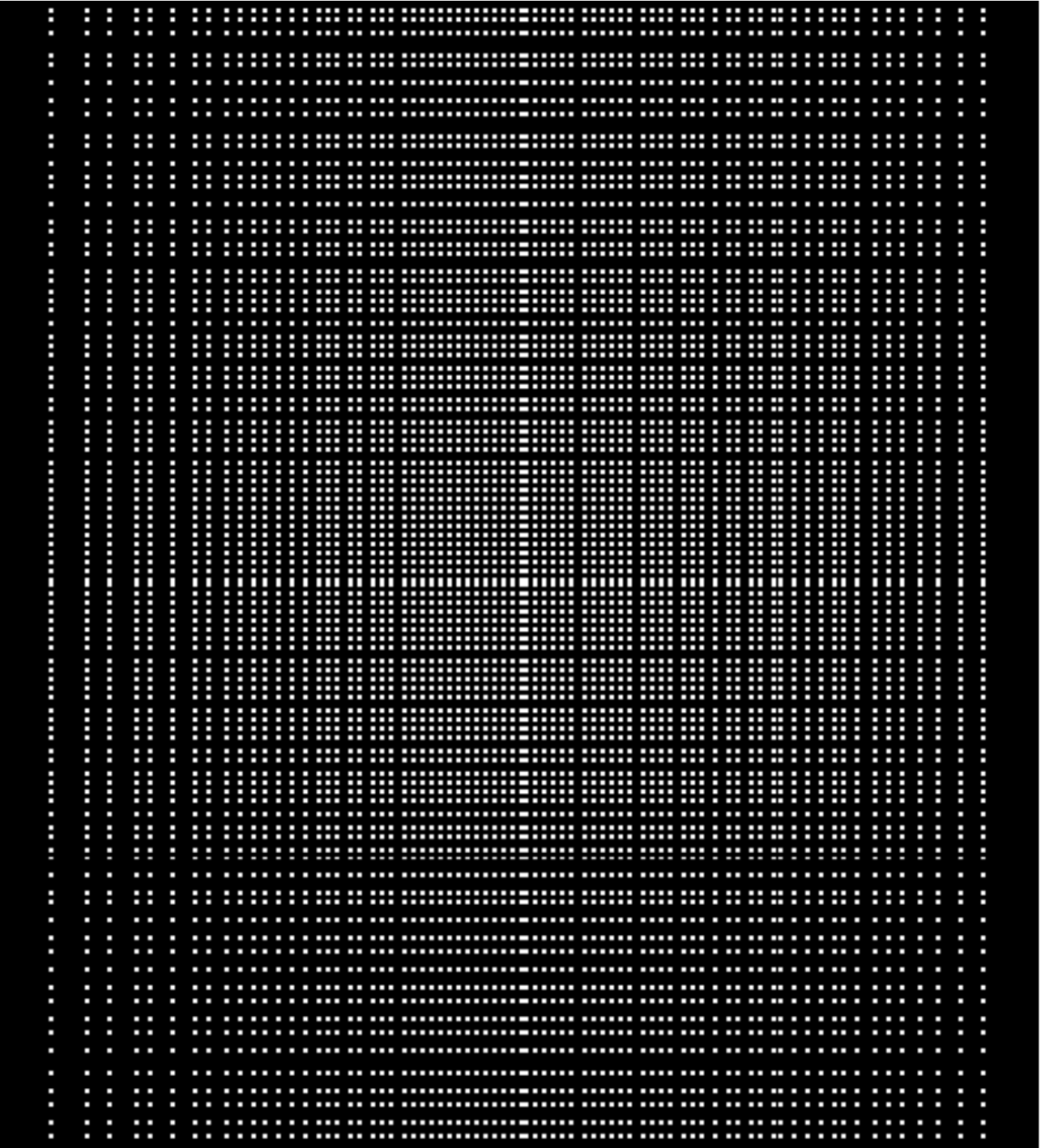}
      };
    \end{tikzpicture}%
}%
\vspace{.1cm}  
\subfloat[Original image]{%
 \begin{tikzpicture}[spy using outlines={rectangle,red,magnification=\mg,size=\sz}]
    \node {
        \includegraphics[ width=\sz]{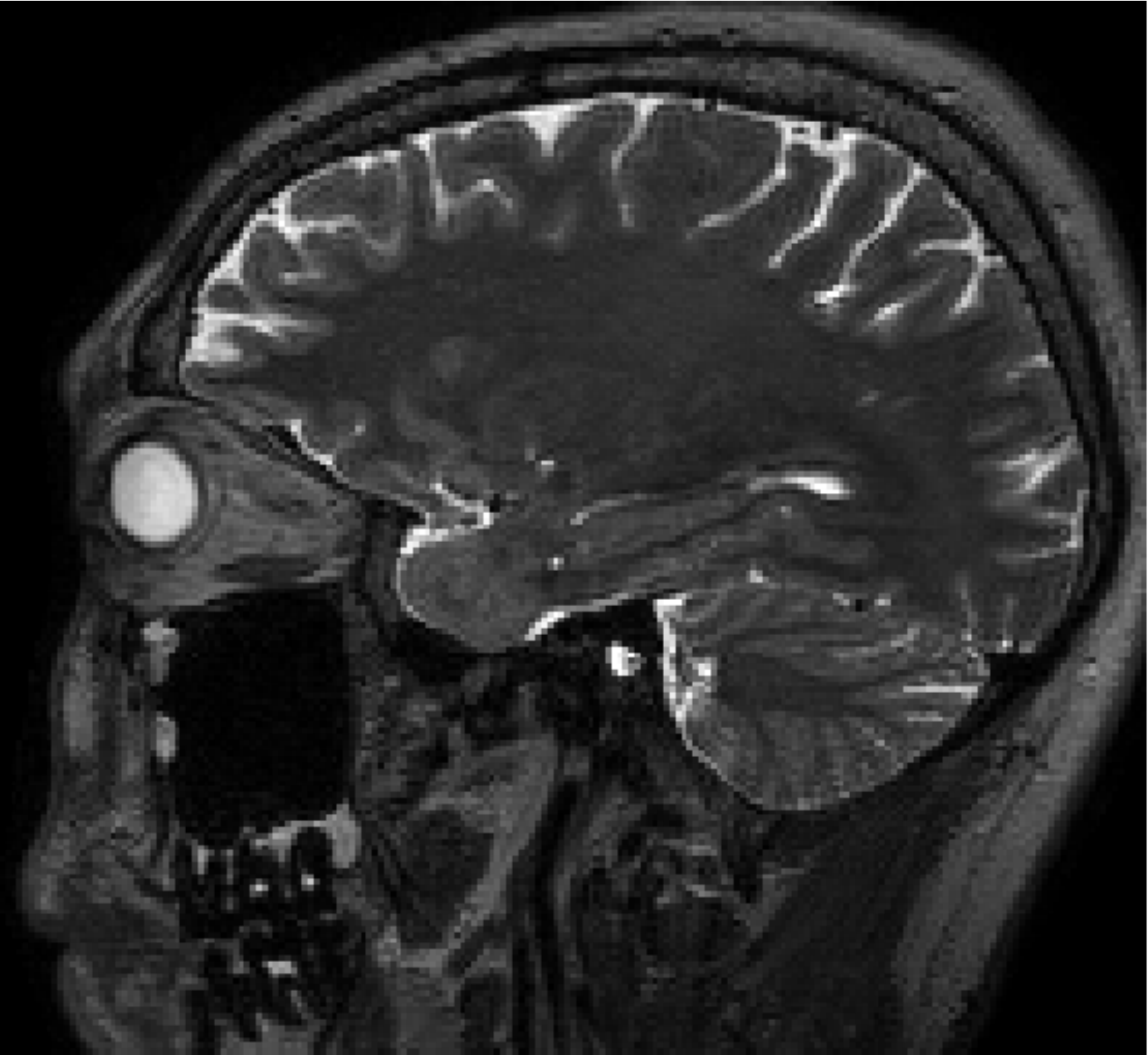}
      };
      \spy on \spt in node at \pt;
    \end{tikzpicture}%
  } \hspace{-.28cm}
  \subfloat[MC-LOUPE, 35.12 dB]{%
    \begin{tikzpicture}[spy using outlines={rectangle,red,magnification=\mg,size=\sz}]
      \node {
        \includegraphics[width=\sz]{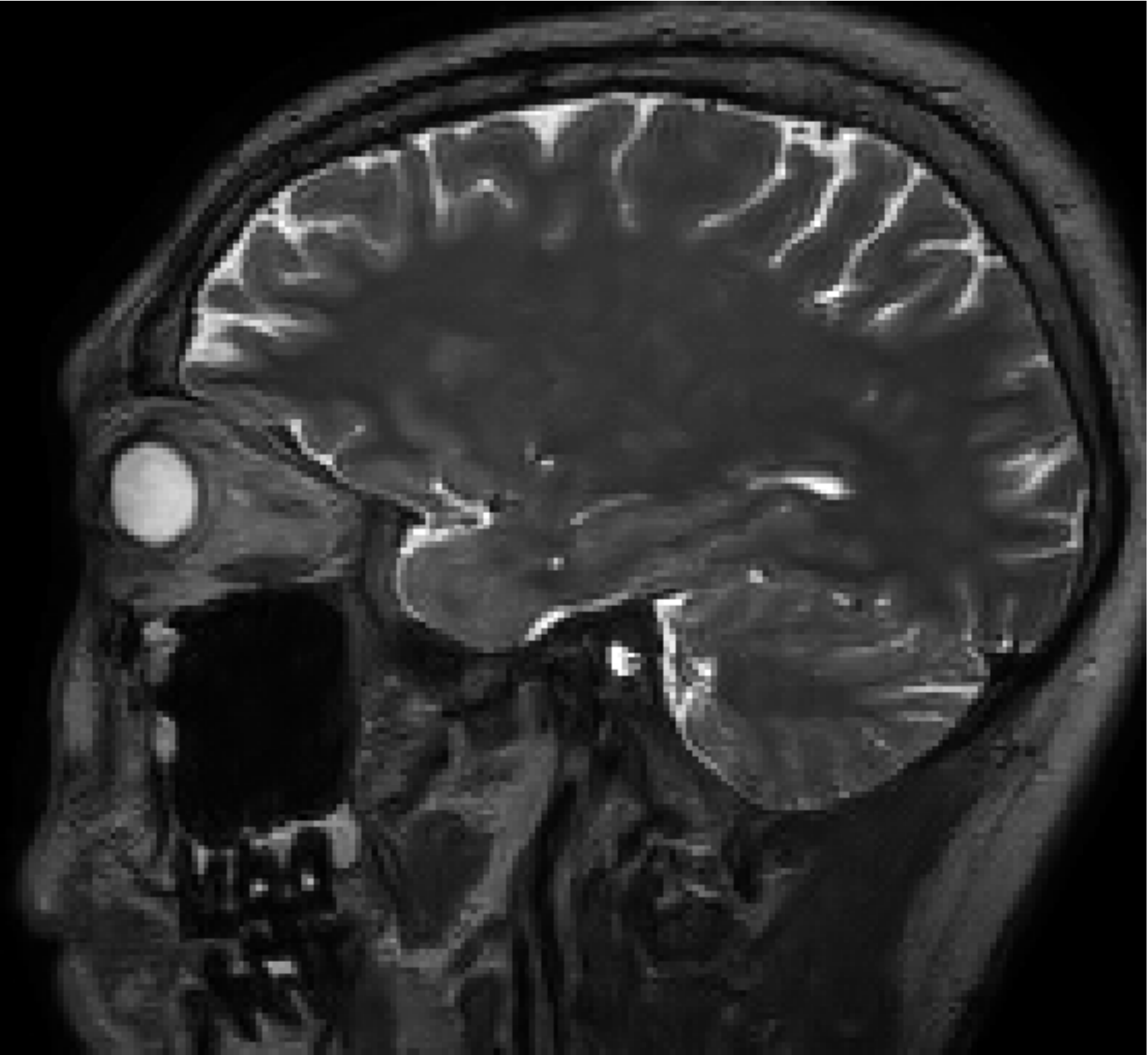}%
      };
      \spy on \spt in node at \pt;
    \end{tikzpicture}%
  }\hspace{-.28cm}
  \subfloat[J-ISTANet, K=5, \hspace{.5cm} 32.20 dB]{%
    \begin{tikzpicture}[spy using outlines={rectangle,red,magnification=\mg,size=\sz}]
      \node {
        \includegraphics[width=\sz]{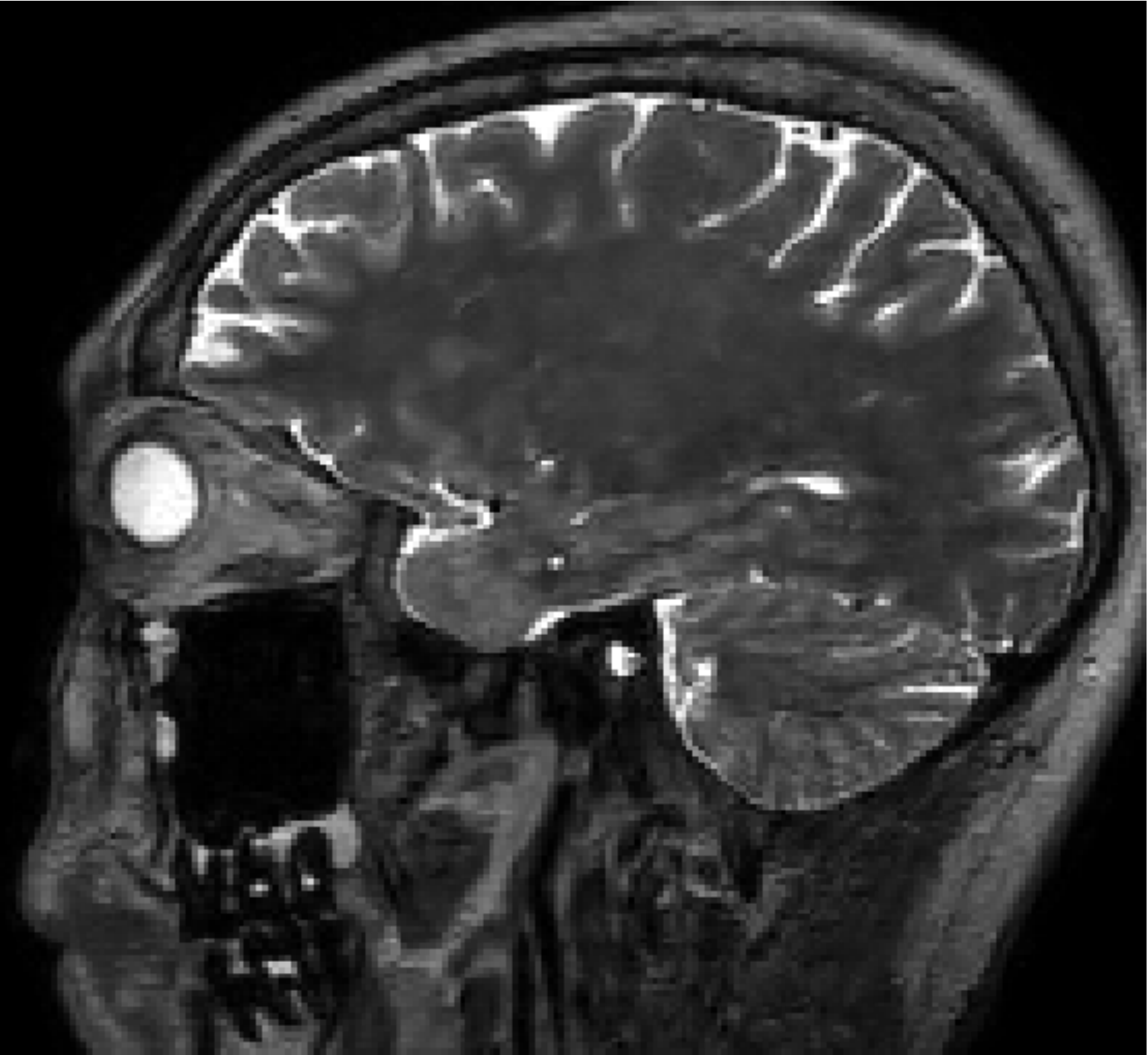}%
      };
      \spy on \spt in node at \pt;
    \end{tikzpicture}%
  }\hspace{-.28cm}
  \subfloat[J-UNET, \hspace{1cm} 31.38 dB]{%
    \begin{tikzpicture}[spy using outlines={rectangle,red,magnification=\mg,size=\sz}]
      \node {
        \includegraphics[width=\sz]{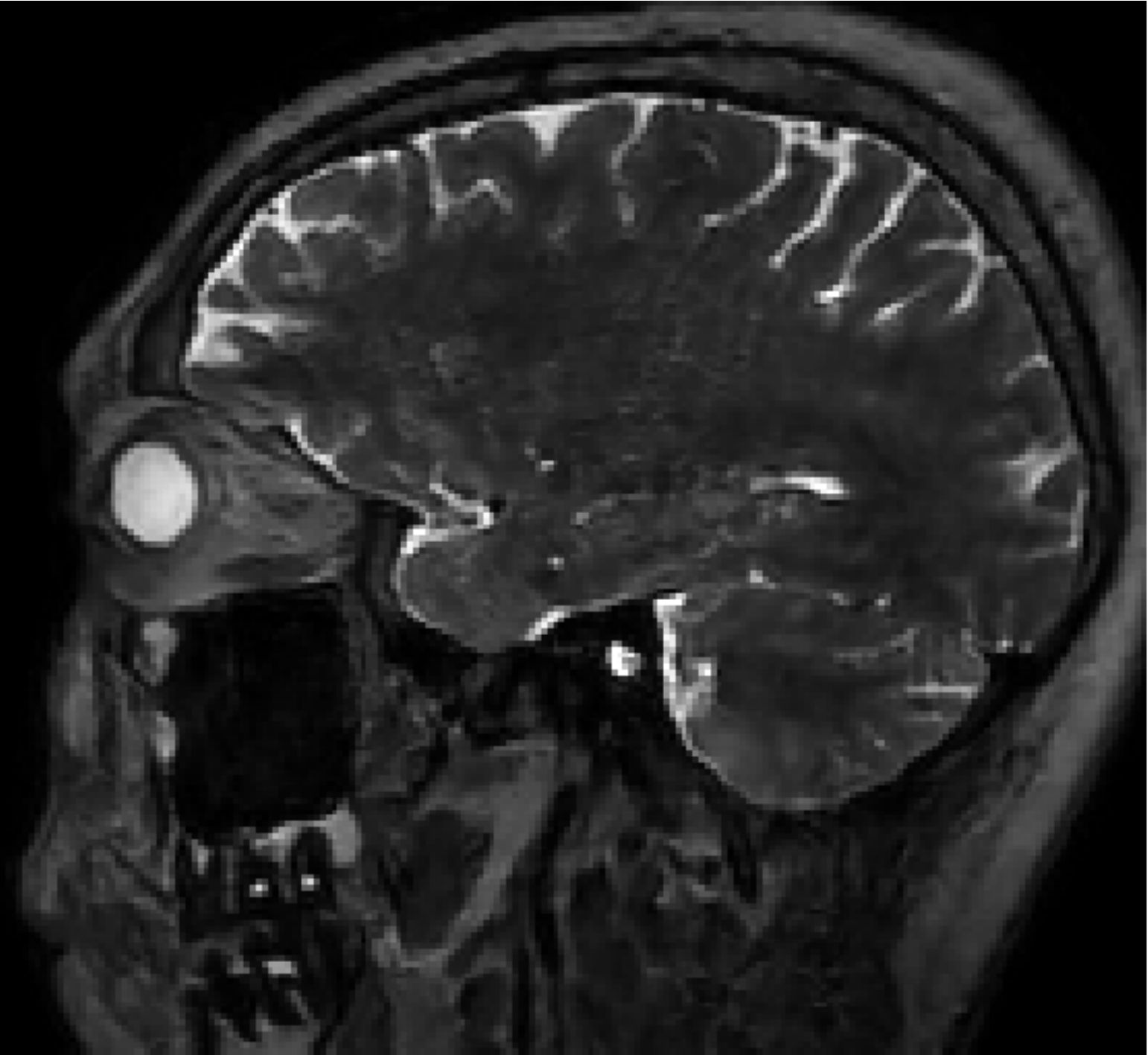}%
      };
      \spy on \spt in node at \pt;
    \end{tikzpicture}%
  }\hspace{-.28cm}
  \subfloat[J-MoDL, K=1, 31.85 dB]{%
    \begin{tikzpicture}[spy using outlines={rectangle,red,magnification=\mg,size=\sz}]
      \node {
        \includegraphics[width=\sz]{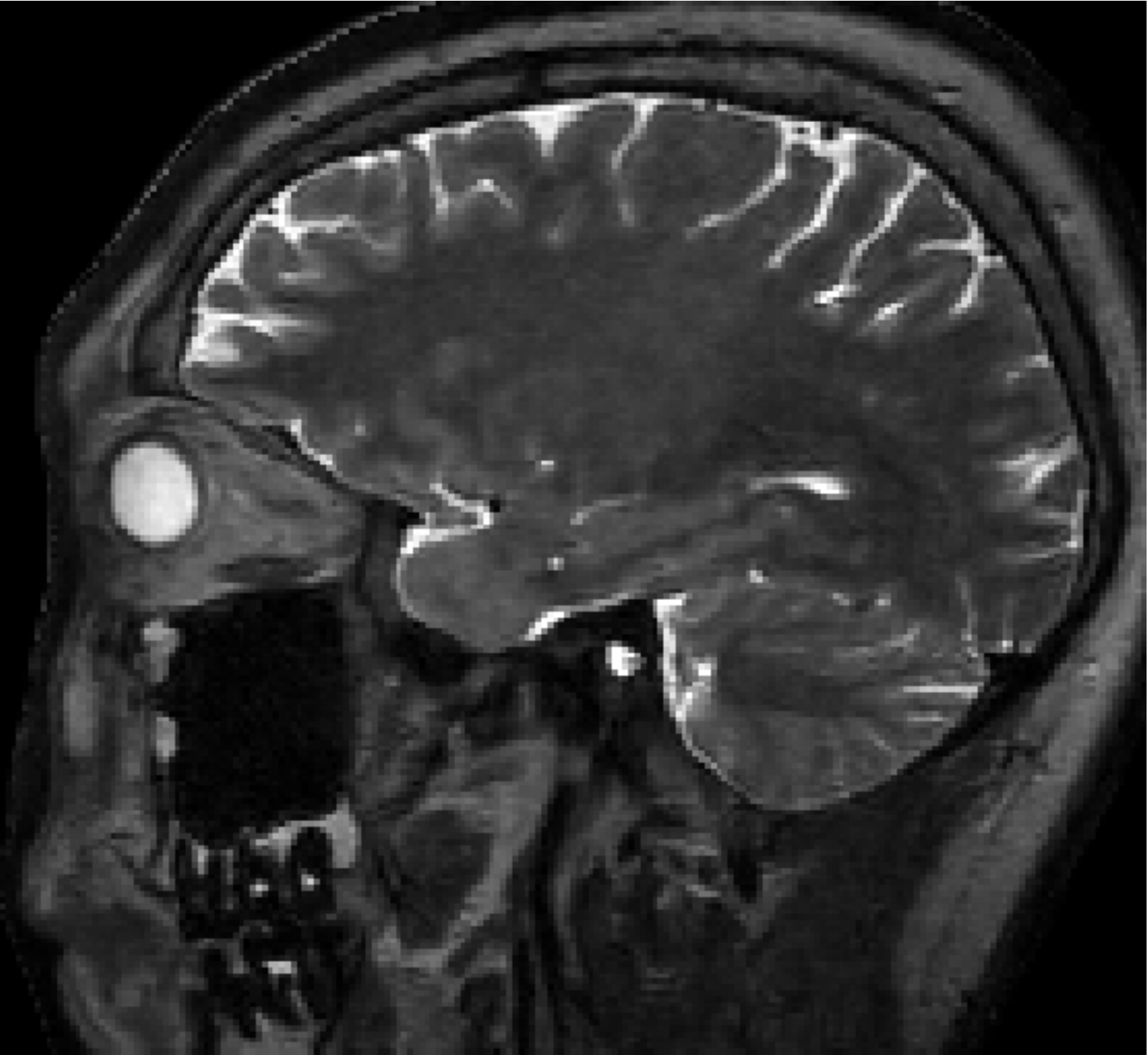}%
      };
      \spy on \spt in node at \pt;
    \end{tikzpicture}%
  }\hspace{-.28cm}
  \subfloat[J-MoDL, K=5, 37.45 dB]{%
    \begin{tikzpicture}[spy using outlines={rectangle,red,magnification=\mg,size=\sz}]
      \node {
        \includegraphics[width=\sz]{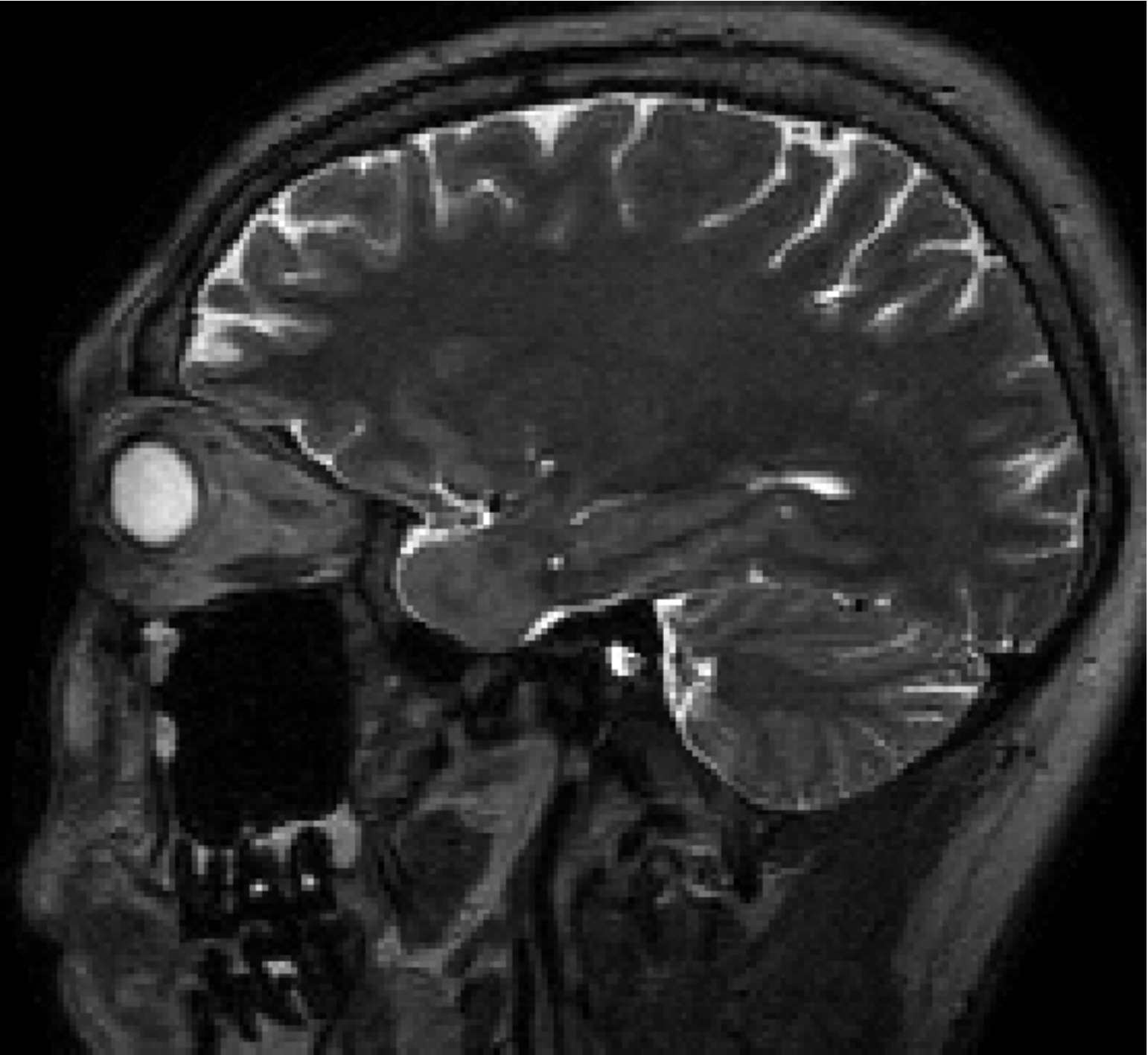}
      };
      \spy on \spt in node at \pt;
    \end{tikzpicture}%
}%
}

%%% Local Variables:
%%% mode: latex
%%% TeX-master: "hk"
%%% End:

%% file: tab_2d_loupe.tex
\begin{tabular}{l@{\hskip 0.04cm}|c|@{\hskip 0.04cm}c|@{\hskip 0.05cm}c|@{\hskip 0.05cm}c} \toprule
          & \multicolumn{2}{c|}{PNSR}          & \multicolumn{2}{c}{SSIM}   \\ \midrule
Algorithm & $\Phi$ alone   & $\Theta\; \Phi$ Joint            & $\Phi$ alone   & $\Theta\; \Phi$ Joint\\  \midrule
MC-LOUPE     & NA             & $33.68 \pm 3.23$ & NA             & $0.92 \pm 0.02$\\ 
ISTA K=5   &$28.66 \pm 1.58$& $34.38 \pm 1.38$ & $0.86 \pm 0.03$& $0.94 \pm 0.01$\\
UNET      &$26.27 \pm 1.33$& $30.63 \pm 1.22$ & $0.79 \pm 0.03$& $0.91 \pm 0.02$\\
MoDL K=1  &$30.68 \pm 1.53$& $31.85 \pm 0.84$ & $0.89 \pm 0.02$& $0.92 \pm 0.01$\\
MoDL K=5  &$32.72 \pm 1.34$& $36.38 \pm 0.54$ & $0.92 \pm 0.02$& $0.95 \pm 0.01$\\

%     Algorithm   & PSNR             & SSIM \\ \midrule
% MC-LOUPE         & $33.68 \pm 3.23$ & $0.92 \pm 0.02$ \\ 
% J-ISTANet       & $32.40 \pm 0.77$ & $0.91 \pm 0.01$ \\
% J-UNET        & $30.63 \pm 1.22$ & $0.91 \pm 0.02$ \\
% J-MoDL, K=1   & $31.85 \pm 0.84$ & $0.92 \pm 0.01$ \\
% J-MoDL, K=5   & $36.38 \pm 0.54$ & $0.95 \pm 0.01$ \\

  \bottomrule      
\end{tabular}

%%% Local Variables:
%%% mode: latex
%%% TeX-master: "hk"
%%% End:

%% file: main.bbl

%% file: ms.bbl
\begin{thebibliography}{10}
\providecommand{\url}[1]{#1}
\csname url@samestyle\endcsname
\providecommand{\newblock}{\relax}
\providecommand{\bibinfo}[2]{#2}
\providecommand{\BIBentrySTDinterwordspacing}{\spaceskip=0pt\relax}
\providecommand{\BIBentryALTinterwordstretchfactor}{4}
\providecommand{\BIBentryALTinterwordspacing}{\spaceskip=\fontdimen2\font plus
\BIBentryALTinterwordstretchfactor\fontdimen3\font minus
  \fontdimen4\font\relax}
\providecommand{\BIBforeignlanguage}[2]{{%
\expandafter\ifx\csname l@#1\endcsname\relax
\typeout{** WARNING: IEEEtran.bst: No hyphenation pattern has been}%
\typeout{** loaded for the language `#1'. Using the pattern for}%
\typeout{** the default language instead.}%
\else
\language=\csname l@#1\endcsname
\fi
#2}}
\providecommand{\BIBdecl}{\relax}
\BIBdecl

\bibitem{wangCTtmi2017}
H.~Chen, Y.~Zhang \emph{et~al.}, ``{Low-Dose {CT} with a Residual
  Encoder-Decoder Convolutional Neural Network},'' \emph{{IEEE} Trans. Med.
  Imag.}, vol.~36, no.~12, pp. 2524--2535, 2017.

\bibitem{jong2019kspace}
Y.~Han, L.~Sunwoo, and J.~C. Ye, ``k-space deep learning for accelerated
  {MRI},'' \emph{{IEEE} Trans. Med. Imag.}, 2019.

\bibitem{dagan}
G.~Yang, S.~Yu \emph{et~al.}, ``{DAGAN}: Deep de-aliasing generative
  adversarial networks for fast compressed sensing {MRI} reconstruction,''
  \emph{{IEEE} Trans. Med. Imag.}, vol.~37, no.~6, pp. 1310--1321, 2017.

\bibitem{gan_cyclic}
T.~M. Quan, T.~Nguyen-Duc, and W.-K. Jeong, ``Compressed sensing {MRI}
  reconstruction using a generative adversarial network with a cyclic loss,''
  \emph{{IEEE} Trans. Med. Imag.}, vol.~37, no.~6, pp. 1488--1497, 2018.

\bibitem{sigmanet}
K.~Hammernik, J.~Schlemper \emph{et~al.}, ``{Sigma-Net}: Systematic evaluation
  of iterative deep neural networks for fast parallel {MR} image
  reconstruction,'' 2019, arXiv:1912.09278.

\bibitem{dar2018}
S.~U.~H. Dar, M.~Yurt \emph{et~al.}, ``Synergistic reconstruction and synthesis
  via generative adversarial networks for accelerated multi-contrast {MRI},''
  2018, arXiv:1805.10704.

\bibitem{dar2017transfer}
S.~U.~H. Dar, M.~{\"O}zbey \emph{et~al.}, ``A transfer-learning approach for
  accelerated {MRI} using deep neural networks,'' \emph{Magnetic Resonance in
  Medicine}, vol.~84, no.~2, pp. 663--685, 2017.

\bibitem{zhu2018}
B.~Zhu, J.~Z. Liu \emph{et~al.}, ``Image reconstruction by domain-transform
  manifold learning,'' \emph{Nature}, vol. 555, no. 7697, pp. 487--492, 2018.

\bibitem{roth}
U.~Schmidt and S.~Roth, ``Shrinkage fields for effective image restoration,''
  in \emph{Proceedings of the IEEE Conference on Computer Vision and Pattern
  Recognition}, 2014, pp. 2774--2781.

\bibitem{admmnet}
y.~yang, J.~Sun \emph{et~al.}, ``Deep {ADMM-Net} for compressive sensing
  {MRI},'' in \emph{Advances in Neural Information Processing Systems 29},
  2016, pp. 10--18.

\bibitem{istanet}
J.~Zhang and B.~Ghanem, ``{ISTA-Net}: Interpretable optimization-inspired deep
  network for image compressive sensing,'' in \emph{Proceedings of the IEEE
  conference on computer vision and pattern recognition}, 2018, pp. 1828--1837.

\bibitem{casecadeDynamic}
J.~Schlemper, J.~Caballero \emph{et~al.}, ``A deep cascade of convolutional
  neural networks for dynamic {MR} image reconstruction,'' \emph{{IEEE} Trans.
  Med. Imag.}, vol.~37, no.~2, pp. 491--503, 2018.

\bibitem{modl}
H.~K. Aggarwal, M.~P. Mani, and M.~Jacob, ``{MoDL}: Model based deep learning
  architecture for inverse problems,'' \emph{{IEEE} Trans. Med. Imag.},
  vol.~38, no.~2, pp. 394--405, 2019.

\bibitem{hammernik}
K.~Hammernik, T.~Klatzer \emph{et~al.}, ``{Learning a Variational Network for
  Reconstruction of Accelerated {MRI} Data},'' \emph{Magnetic resonance in
  Medicine}, vol.~79, no.~6, pp. 3055--3071, 2017.

\bibitem{zhang2017magazine}
L.~Zhang and W.~Zuo, ``{Image Restoration: From Sparse and Low-Rank Priors to
  Deep Priors},'' \emph{{IEEE} Signal Process. Mag.}, vol.~34, no.~5, pp.
  172--179, 2017.

\bibitem{omodl}
A.~Pramanik, H.~K. Aggarwal, and M.~Jacob, ``Off-the-grid model based deep
  learning {O-MoDL},'' in \emph{IEEE 16th International Symposium on Biomedical
  Imaging (ISBI)}.\hskip 1em plus 0.5em minus 0.4em\relax IEEE, 2019, pp.
  1395--1398.

\bibitem{modlmussels}
H.~K. Aggarwal, M.~P. Mani, and M.~Jacob, ``{MoDL-MUSSELS}: Model-based deep
  learning for multishot sensitivity-encoded diffusion {MRI},'' \emph{{IEEE}
  Trans. Med. Imag.}, 2019.

\bibitem{mardaniGANCS}
M.~Mardani, E.~Gong \emph{et~al.}, ``Deep generative adversarial neural
  networks for compressive sensing {MRI},'' \emph{{IEEE} Trans. Med. Imag.},
  vol.~38, no.~1, pp. 167--179, 2018.

\bibitem{venkatakrishnan2013plug}
S.~V. Venkatakrishnan, C.~A. Bouman, and B.~Wohlberg, ``Plug-and-play priors
  for model based reconstruction,'' in \emph{2013 IEEE Global Conference on
  Signal and Information Processing}.\hskip 1em plus 0.5em minus 0.4em\relax
  IEEE, 2013, pp. 945--948.

\bibitem{ahmad2020plug}
R.~Ahmad, C.~A. Bouman \emph{et~al.}, ``Plug-and-play methods for magnetic
  resonance imaging: Using denoisers for image recovery,'' \emph{{IEEE} Signal
  Process. Mag.}, vol.~37, no.~1, pp. 105--116, 2020.

\bibitem{smash}
D.~K. Sodickson and W.~J. Manning, ``Simultaneous acquisition of spatial
  harmonics {SMASH}: fast imaging with radiofrequency coil arrays,''
  \emph{Magnetic resonance in medicine}, vol.~38, no.~4, pp. 591--603, 1997.

\bibitem{candes2007sparsity}
E.~Candes and J.~Romberg, ``Sparsity and incoherence in compressive sampling,''
  \emph{Inverse problems}, vol.~23, no.~3, pp. 969--985, 2007.

\bibitem{lustig2008compressed}
M.~Lustig, D.~L. Donoho \emph{et~al.}, ``Compressed sensing {MRI},'' \emph{IEEE
  signal processing magazine}, vol.~25, no.~2, p.~72, 2008.

\bibitem{vasnawalaPoissonDisc}
E.~Levine, B.~Daniel \emph{et~al.}, ``{3D} {C}artesian {MRI} with compressed
  sensing and variable view sharing using complementary {P}oisson-disc
  sampling,'' \emph{Magnetic resonance in medicine}, vol.~77, no.~5, pp.
  1774--1785, 2017.

\bibitem{Reeves2000}
Y.~Gao and S.~J. Reeves, ``Optimal k-space sampling in {MRSI} for images with a
  limited region of support,'' \emph{{IEEE} Trans. Med. Imag.}, vol.~19,
  no.~12, pp. 1168--1178, 2000.

\bibitem{xu}
D.~Xu, M.~Jacob, and Z.~Liang, ``Optimal sampling of k-space with {C}artesian
  grids for parallel {MR} imaging,'' in \emph{Proc Int Soc Magn Reson Med},
  vol.~13, 2005, p. 2450.

\bibitem{haldar2019oedipus}
J.~P. Haldar and D.~Kim, ``{OEDIPUS}: An experiment design framework for
  sparsity-constrained {MRI},'' \emph{{IEEE} Trans. Med. Imag.}, 2019.

\bibitem{levine2017}
E.~Levine and B.~Hargreaves, ``On-the-fly adaptive k-space sampling for linear
  {MRI} reconstruction using moment-based spectral analysis,'' \emph{{IEEE}
  Trans. Med. Imag.}, vol.~37, no.~2, pp. 557--567, 2017.

\bibitem{senel2019}
L.~K. Senel, T.~Kilic \emph{et~al.}, ``Statistically segregated k-space
  sampling for accelerating multiple-acquisition {MRI},'' \emph{{IEEE} Trans.
  Med. Imag.}, vol.~38, no.~7, pp. 1701--1714, 2019.

\bibitem{samsonov}
F.~Liu, A.~Samsonov \emph{et~al.}, ``{SANTIS}: Sampling-augmented neural
  network with incoherent structure for {MR} image reconstruction,''
  \emph{Magnetic resonance in medicine}, vol.~82, no.~5, pp. 1890--1904, 2019.

\bibitem{sherry2019}
F.~Sherry, M.~Benning \emph{et~al.}, ``Learning the sampling pattern for
  {MRI},'' 2019, arXiv:1906.08754.

\bibitem{gozcu2018learning}
B.~G{\"o}zc{\"u}, R.~K. Mahabadi \emph{et~al.}, ``Learning-based compressive
  {MRI},'' \emph{{IEEE} Trans. Med. Imag.}, vol.~37, no.~6, pp. 1394--1406,
  2018.

\bibitem{weiss}
T.~Weiss, S.~Vedula \emph{et~al.}, ``Learning fast magnetic resonance
  imaging,'' 2019, arXiv:1905.09324.

\bibitem{pilot}
T.~Weiss, O.~Senouf \emph{et~al.}, ``{PILOT}: Physics-informed learned optimal
  trajectories for accelerated {MRI},'' 2019, arXiv:1909.05773.

\bibitem{loupe}
C.~D. Bahadir, A.~V. Dalca, and M.~R. Sabuncu, ``Learning-based optimization of
  the under-sampling pattern in {MRI},'' in \emph{International Conference on
  Information Processing in Medical Imaging}.\hskip 1em plus 0.5em minus
  0.4em\relax Springer, 2019, pp. 780--792.

\bibitem{lustig2010spirit}
M.~Lustig and J.~M. Pauly, ``{SPIRiT}: iterative self-consistent parallel
  imaging reconstruction from arbitrary k-space,'' \emph{Magnetic resonance in
  medicine}, vol.~64, no.~2, pp. 457--471, 2010.

\bibitem{jin}
K.~H. Jin, M.~Unser, and K.~M. Yi, ``Self-supervised deep active accelerated
  {MRI},'' 2019, arXiv:1901.04547.

\bibitem{Zhang}
Z.~Zhang, A.~Romero \emph{et~al.}, ``Reducing uncertainty in undersampled {MRI}
  reconstruction with active acquisition,'' in \emph{Proceedings of the IEEE
  Conference on Computer Vision and Pattern Recognition}, 2019, pp. 2049--2058.

\bibitem{metzler2019deep}
C.~A. Metzler, H.~Ikoma \emph{et~al.}, ``Deep optics for single-shot
  high-dynamic-range imaging,'' 2019, arXiv:1908.00620.

\bibitem{muthumbi2019}
A.~Muthumbi, A.~Chaware \emph{et~al.}, ``Learned sensing: jointly optimized
  microscope hardware for accurate image classification,'' \emph{Biomed. Opt.
  Express}, vol.~10, no.~12, pp. 6351--6369, 2019.

\bibitem{horstmeyer2017}
R.~Horstmeyer, R.~Y. Chen \emph{et~al.}, ``Convolutional neural networks that
  teach microscopes how to image,'' 2017, arXiv:1709.07223.

\bibitem{cheng2019}
Y.~F. Cheng, M.~Strachan \emph{et~al.}, ``Illumination pattern design with deep
  learning for single-shot fourier ptychographic microscopy,'' \emph{Opt.
  Express}, vol.~27, no.~2, pp. 644--656, 2019.

\bibitem{chakrabarti2016}
A.~Chakrabarti, ``Learning sensor multiplexing design through
  back-propagation,'' in \emph{Advances in Neural Information Processing
  Systems}, 2016, pp. 3081--3089.

\bibitem{knollsampling}
F.~Knoll, C.~Clason \emph{et~al.}, ``Adapted random sampling patterns for
  accelerated mri,'' \emph{Magnetic resonance materials in physics, biology and
  medicine}, vol.~24, no.~1, pp. 43--50, 2011.

\bibitem{fastmri}
J.~Zbontar, F.~Knoll \emph{et~al.}, ``{fastMRI}: An open dataset and benchmarks
  for accelerated {MRI},'' 2018, arXiv:1811.08839.

\bibitem{doneva}
M.~Doneva, P.~B{\"o}rnert \emph{et~al.}, ``Compressed sensing reconstruction
  for magnetic resonance parameter mapping,'' \emph{Magnetic Resonance in
  Medicine}, vol.~64, no.~4, pp. 1114--1120, 2010.

\bibitem{ista2003wavelet}
M.~A.~T. Figueiredo, R.~D. Nowak \emph{et~al.}, ``{An EM Algorithm for
  Wavelet-Based Image Restoration},'' \emph{{IEEE} Trans. Image Process.},
  vol.~12, no.~8, pp. 906--916, 2003.

\bibitem{shiqianma2008}
S.~Ma, W.~Yin \emph{et~al.}, ``{An Efficient Algorithm for Compressed {MR}
  Imaging using Total Variation and Wavelets},'' in \emph{Computer Vision and
  Pattern Recognition}, 2008, pp. 1--8.

\bibitem{jacobspmag}
M.~Jacob, M.~P. Mani, and J.~C. Ye, ``Structured low-rank algorithms: Theory,
  {MR} applications, and links to machine learning,'' \emph{IEEE Signal
  Processing Magzine}, pp. 1--12, 2019.

\bibitem{ronneberger2015unet}
O.~Ronneberger, P.~Fischer, and T.~Brox, ``U-net: Convolutional networks for
  biomedical image segmentation,'' in \emph{International Conference on Medical
  Image Computing and Computer-Assisted Intervention (MICCAI)}.\hskip 1em plus
  0.5em minus 0.4em\relax Springer, 2015, pp. 234--241.

\bibitem{espirit2014}
M.~Uecker, P.~Lai \emph{et~al.}, ``{ESPIRiT - An eigenvalue approach to
  autocalibrating parallel {MRI}: Where {SENSE} meets {GRAPPA}},''
  \emph{Magnetic Resonance in Medicine}, vol.~71, no.~3, pp. 990--1001, 2014.

\bibitem{sparkling}
C.~Lazarus, P.~Weiss \emph{et~al.}, ``{SPARKLING}: variable-density k-space
  filling curves for accelerated {T2}*-weighted {MRI},'' \emph{Magnetic
  resonance in medicine}, vol.~81, no.~6, pp. 3643--3661, 2019.

\end{thebibliography}
